\documentclass{article}

\usepackage{PRIMEarxiv}
\usepackage[utf8]{inputenc} % allow utf-8 input
\usepackage[T1]{fontenc}    % use 8-bit T1 fonts
\usepackage{hyperref}       % hyperlinks
\usepackage{url}            % simple URL typesetting
\usepackage{booktabs}       % professional-quality tables
\usepackage{tabularx}       % Required for tables that span the text width
\usepackage{ltablex}        % Allows tabularx to break across pages
\usepackage{amsfonts}       % blackboard math symbols
\usepackage{amsmath}
\usepackage{nicefrac}       % compact symbols for 1/2, etc.
\usepackage{microtype}      % microtypography
\usepackage{lipsum}
\usepackage{fancyhdr}       % header
\usepackage{graphicx}       % graphics
\usepackage{siunitx} % Provides S column type
\usepackage{threeparttable}
\usepackage{ragged2e}
\usepackage{amsmath} % for math environments, if needed
\usepackage{textcomp} % provides additional symbols
\usepackage{tipa}     % for phonetic symbols
\usepackage{hyperref} % enhances the \url command and provides \href as well
\usepackage[numbers]{natbib}

\DeclareUnicodeCharacter{02BE}{\textquoteright}

\graphicspath{{./supplement-figures/}}    

\newcommand{\beginsupplement}{%
        \setcounter{table}{0}
        \renewcommand{\thetable}{S\arabic{table}}%
        \setcounter{figure}{0}
        \renewcommand{\thefigure}{S\arabic{figure}}%
        \setcounter{section}{6}
        \renewcommand{\thesection}{S\arabic{section}}%
     }

\title{The Economy and Public Diplomacy:  
An Analysis of RT's Economic Content and Context on Facebook
\thanks{\textit{\underline{Citation}}: 
\textbf{Lokmanoglu, A. D., Winkler, C. K., Damanhoury, K. E., Massignan, V., Villa-Turek, E., \& Chen, K. A. (2024). \textit{The Economy and Public Diplomacy: An Analysis of RT’s Economic Content and Context on Facebook} (arXiv:2405.01798). arXiv. http://arxiv.org/abs/2405.01798}} 
}

\author{
  Ayse D. Lokmanoglu \\
  Clemson University \\
  Clemson, SC\\
  \texttt{alokman@clemson.edu} \\
   \And
  Carol K. Winkler \\
  Georgia State University \\
  Atlanta, GA\\
  \texttt{cwinkler@gsu.edu} \\
   \AND
   Kareem El Damanhoury  \\
   University of Denver \\
   Denver, CO \\
   \texttt{kareem.eldamanhoury@du.edu} \\
   \And
   Virginia Massignan  \\
   Georgia State University \\
   Atlanta, GA \\
   \texttt{Vmassignan1@gsu.edu} \\
   \And
   Esteban Villa-Turek \\
   Northwestern University \\
   Evanston, IL \\
   \texttt{estebanvillaturek@gmail.com} \\
    \And
   Keyu Alexander Chen \\
   Georgia State University University \\
   Atlanta, GA \\
   \texttt{kchen15@student.gsu.edu} \\
}

\begin{document}
\maketitle

\begin{abstract}
With globalization's rise, economic interdependence's impacts have become a prominent factor affecting personal lives, as well as national and international dynamics. This study examines RT's public diplomacy efforts on its non-Russian Facebook accounts over the past five years to identify the prominence of economic topics across language accounts. Computational analysis, including word embeddings and statistical methods, investigates how offline economic indicators, like currency values and oil prices, correspond to RT's online economic content changes. The results demonstrate that RT uses message reinforcement associated economic topics as an audience targeting strategy and differentiates their use with changing currency and oil values.
\end{abstract}

% keywords can be removed
\keywords{social media \and economics \and computational social science \and public diplomacy}

\section{Introduction}
Increased gas prices, interrupted supply chains, rising inflation rates, shifting currency values, volatile cryptocurrencies, and bank failures have rendered the impacts of a globalized economy personal. Economic fluctuations have also produced societal ramifications such as support for populism \cite{Kates2019We, Rodrik2021Why, Schmeichel2015Individual}, selected policy alternatives \cite{Citrin1997Public, Fetzer2021Coronavirus, Peterson2019Carbon}, and electoral outcomes \cite{Alvarez1995Economics, Caselli2020Globalization, Jones2015Economic, DeSimone2016Rhetoric}. 

Perhaps as a result, economic content now forms an integral part of public diplomacy efforts. Economic messaging plays a key role in attracting the attention of global audiences \cite{Kim1996Determinants}. It also influences public attitudes on national issues, such as the war effort in Crimea \cite{Stoycheff2017Priming}. Previous examinations of Russian advertorials distributed in the United States and India \cite{Golan2014Advertorial}, as well as posts by a German Ambassador using social media to reach Pakistani audiences \cite{Khan2021Public}, document the high frequency of economic messaging in public diplomacy. Previous analyses of communications between members of the BRIC alliance go so far as to highlight economic aims as the chief objective of their public diplomacy efforts \cite{Li2016BRICS|}. The priority positioning of economics in public diplomacy prompts some scholars to even incorporate the concept into their definitions. Milam and Avery, for example, maintain that public diplomacy is the “direct or indirect engagement of foreign publics in support of national security, political, cultural, and economic objectives” \cite[p.~329]{Milam2012Apps4Africa:}. 

Some countries facing economic challenges such as declining currency values and increasing public debt have invested heavily in media platforms to expand their global influence. Russia’s RT serves as an illustration, as the media outreach platform functions as a chief component of Russia’s mediated public diplomacy strategy \cite{Borchers2011“Do, Kragh2017Russia’s}. After spending hundreds of millions of dollars to staff more than 20 international bureaus \cite{Alpert2014Kremlin, Wagnsson2023paperboys}, RT channels, websites, and social media platforms have attracted billions of views from citizens in more than 100 countries \cite{Carter2021Questioning, Knoblock2021Look}. 

This study examines five years of RT Facebook’s non-Russian language accounts to expand understandings of the economic dimensions of public diplomacy. It explores which topics are most prevalent in the posts on RT accounts and assesses how previously unexplored economic context variables interface with RT’s economic messaging strategies. To explain, we begin by recounting how the rise of social media has transformed public diplomacy. We then demonstrate how our study expands upon previous work related to RT’s economic messaging strategies and messaging contexts. We conclude by describing and discussing our study’s approach, results, and directions for future research.

\section{Social Media and Public Diplomacy}
Public diplomacy’s chief goal of engaging with foreign audiences to influence international environments has not changed over time \cite{Cull2009Public}, but the rise of social media has transformed its practice \cite{Williamson2012Kelleypd}. Certain social media platforms now capitalize on the viewership declines of standard news platforms (e.g., radio and television) by positioning themselves as both sources of news and as venues for socialization \cite{Napoli2017Why}. Whether followers actively seek out social media newsfeeds or not, online audiences have incidental exposure to such newsfeeds \cite{Fletcher2018Are}. Users also rely on established social media sites like Facebook, Twitter, and YouTube even when their primary news sources are alternative social media platforms like Truth Social, Gettr, Gab, and Bitchute \cite{Stocking2022Role}. Consequently, most governments now capitalize on the cost-effectiveness and accessibility of social media to post much of their public diplomacy content \cite{Surma2016Social}. The approach expands followers by delivering information around the globe. It also allows members of the diplomatic corps and their staff to gather data regarding the audience engagement levels with certain types of messaging content in ways that are useful for maximizing the value of public diplomacy efforts \cite{Kampf2015Digital}. 

By expanding the number and range of reachable followers, public diplomacy via social media proliferates opportunities for audience targeting and engagement \cite{Hayden2018Digital}. Multiple language accounts associated with online platforms such as RT, CGTN, and Voice of America position nation-states to deliver selected content to citizens of states efficiently, even those ruled by authoritarian regimes \cite{Luqiu2020Weibo}. They also permit outreach to groups that cross state boundaries that share language fluencies. RT, for example, has 35 different subscriber groups that diverge based on their nationalities, languages, and interests \cite{Crilley2022Understanding}. 

The affordances of particular social media platforms offer certain constraints and opportunities for public diplomacy. Each social media platform has unique characteristics that prompt new “logics of social practices” for their own networked publics \cite[p.~220]{boyd2010Social}. More specifically, the affordances of each platform prompt and often govern what constitutes appropriate discourse and the nature of audience interactions \cite{Norman2013design, wellmansocial2003}. Affordances also contribute to “sociability, sharing, interaction, homophily, social capital and power, and network effects” \cite[p.~9]{Bodle2010Assessing}. Consumers and producers of content, as a result, comply with certain communication rules and norms simply by selecting to view a particular social media platform or risk alienating their followers on such platforms.

Considered as a whole, the use of social media for public diplomacy purposes has produced mixed results. Some studies highlight the limited value of public diplomacy efforts on social media due to the followers’ lack of trust in the platform’s content \cite{Ceron2015Internet} or in the country that is posting the content \cite{Wasserman2018How}. Others, however, document that exposure to such forums produces success in undermining popular support for the existing world order \cite{Elswah2020“Anything}, changing perceptions of the US and its allies \cite{Carter2021Questioning, Fisher2020Demonizing}, and heightening frictions amongst global alliances \cite{Golovchenko2020Cross-Platform}. Simply put, public diplomacy efforts delivered via social media have the capacity within certain contexts to project soft power in ways to maximize the chances of accomplishing their objectives through dialogic interactions with foreign publics \cite{Fitzpatrick2007Advancing, Khan2021Public, Nye2005Soft}.

Beyond transforming an audience’s makeup and potential reactions, social media also transforms the expectations of messaging content of public diplomacy efforts. Followers of social media platforms demand that public diplomacy posts be more open, transparent, real-time, and engaging \cite{Fitzpatrick2007Advancing, Khan2021Public}. Further, social media’s capacity to reinforce and amplify messages through repetition increases the number of followers  \cite{Luqiu2020Weibo}, primes audiences to expect certain types of content \cite{Morrison2021“Scrounger-bashing”}, fosters emotional responses and persuasive outcomes in followers \cite{DiRusso2021Sustainability, Jovanovic2018Multimodal}, polarizes audiences \cite{Yarchi2021Political}, spreads misinformation (Kragh and Åsberg 2017), and reinforces ideologically based networks \cite{Chen2023Comparing}.

\subsection{Russian Public Diplomacy and the Economy}
While social media has recently changed public diplomacy practices, the influential role of economic messaging broadly considered has long been understood. Since Dierdre McCloskey’s \cite{McCloskey1998rhetoric} germinal foray linking communication and economic studies, research has documented intersections between economic communications and various ideological perspectives, including capitalism \cite{Aune2002Selling, Brown2002Global}, neoliberalism \cite{Chaput2018Trumponomics, Chaput2015Economic}, and socialism \cite{Avsar2011Mainstream, McCloskey1998rhetoric}. Economic messaging also has indirect media effects by attracting support for specific economic policies \cite{Avsar2011Mainstream, Citrin1997Public, Goidel1995Media, Soroka2015It's}, influencing perceptions of presidential leadership \cite{Alvarez1995Economics, Hanan2014From, Wood2004Presidential}, functioning as a needed response to economic crises \cite{Fetzer2021Coronavirus, Goodnight2010Rhetoric, Houck2000Rhetoric, Levasseur2015Not}, serving as a persuasive presidential strategy \cite{Crable1983Argumentative, Zarefsky1979great}, and functioning to build and challenge communities at both the national and global levels \cite{Aune2002Selling, Avsar2011Mainstream, Lebovics1992Economic}.

Growing recognition and acceptance of the societal impacts of economic messaging has prompted efforts to better understand Russia’s economic public diplomacy efforts. Previous studies, however, are limited in their consideration of platforms that serve as delivery mechanisms for such content. Past research projects on Russia’s economic content have focused on the use of newspapers (e.g., \cite{Golan2014Advertorial, Khalitova2020He}), television (e.g., \cite{Grincheva2016BRICS}), and Twitter (e.g., \cite{Khan2021Public}). This study expands understanding of Russia’s platform-specific messaging strategies by focusing on RT’s use of Facebook. With two billion daily active users \cite{Statista2022Number} and affordances of anonymity and network amplification, Facebook offers Russia access to a wide global audience with potential susceptibility to its RT messaging campaign. Thus, a fulsome accounting of Russia’s economic approach would not be complete without consideration of the Facebook platform, as it is the largest social media application in the contemporary global media environment.

Previous studies regarding Russia’s use of economic messaging in public diplomacy, while useful, are also limited in their examinations of narrow timespans and subject matter, as well as in their dated findings. By way of illustration, existing economically-related studies examine two months of RT programming following the fifth BRICs summit in 2013 \cite{Grincheva2016BRICS}, one year of advertorial content in Indian and US newspapers during 2011 \cite{Golan2014Advertorial}, communication efforts by the Russian and Polish governments in the immediate aftermath of a 2010 airplane crash that killed Polish President Lech Kaczynski and most of his Cabinet \cite{Khalitova2020He}, international news coverage of four superpower summit meetings held between the US and Russia from 1987 to 1990 \cite{Fortner1994Public}, and selected posts of Russian Minister of Foreign Affairs spokesperson Maria Zakharova in recent years \cite{Krasnyak2020Russian}. By examining the last five years of RT’s Facebook posts, this study updates these earlier studies and assesses the most frequently recurring and reinforced economic content topics by RT over time. Our study’s expanded time frame also permits a more nuanced examination of economically related context factors that consistently correspond to changes in the output levels of RT’s economic content of the platform’s preferred topics.

A third area of concern regarding previous studies of Russia’s use of economic messaging relates to considerations of the audience’s scope. While one study of China’s use of economic messaging encompasses more than 30 embassies’ public diplomacy \cite{Luqiu2020Weibo}, we found no comparable, large-scale study of Russia’s global economic efforts. Instead, the rare earlier work on the relationship between economic context variables and messaging content focuses on bilateral public diplomacy efforts, such as between two countries like Poland and Russia \cite{Khalitova2020He}. By examining all outward-facing languages of RT accounts apart from Serbian (which was not available until after the beginning of our study’s timespan), this study compares the use of representative economic topics across content RT conveys in English, French, German, Arabic, and Spanish. Examinations of the various accounts offer an opportunity to see if and how RT’s economic messaging differs towards language-based communities.

To help fill these gaps regarding Russia’s public diplomacy efforts, we will provide a long-term perspective of the language-targeting approaches of Russia on the largest social media app. To accomplish this task, we ask:

\quad \quad \textit{RQ1: How have RT’s official Facebook posts on its various language- specific accounts utilized economic messaging?}
 
Several studies examining the role of economic contexts establish a correspondence between environmental indicators and information campaigns \cite{Kayser2011Performance, Lewis-Beck1988Economics, Lewis-Beck2007Economic, LewisBeck2000Economic, Linn2010Economics}. Yet, previous examinations of economic content of Russian public diplomacy efforts are rare \cite{Soroka2015It's}. One study, for example, describes the formation of the BRIC economic alliances as a key factor in contemporary global power struggles (Li and Marsh 2016). Another, examining Chinese embassies in both Russia and 29 other countries, concludes that the number of the embassy’s social media followers does not always correlate with the country’s economic size or level of bilateral economic relations \cite{Luqiu2020Weibo}. A third finds that trade balances between Russia and Poland correspond to higher levels of attention to global media posts, regardless of the involved regime or culture \cite{Khalitova2020He}. This study adds to previous theories of economic public diplomacy and context variables by analyzing two previously unexplored variables —the role of currency rates and oil prices —as potential context factors that might correlate with levels of economic messaging. Identifying impactful context factors can better provide national governments and social media users alike with the ability to more accurately anticipate when and how governments use economic messaging in public diplomacy efforts. 

More specifically, the need to analyze currency values in the context of public diplomacy stems from their growing interconnectedness in a globalized world. Multiple factors affect a nation’s currency values including inflation, political stability levels, macroeconomic indicators (e.g., interest rates, public debt levels, central bank intervention, etc.), and appreciation or depreciation of powerful global currencies (such as the US Dollar, Euro, British Pound, and Kuwaiti Dinar). In 1995, for example, the Mexican government switched the Peso from a fixed to a floating rate, triggering a massive drop in its value \cite{Bolsen2008Polls—Trends:, Farhar1980Public, Vidigal2022Issue}. The 2016 Brexit referendum led to the exit of the United Kingdom from the European Union in 2019, causing a significant depreciation not only in the British Pound but also in the US, Australian, and Canadian Dollars \cite{Mustoe2019How}. In a 2024 interview between Tucker Carlson and Vladimir Putin, President of Russia, Putin surmised the key role of the currencies in global economics: “To use the dollar as a tool of foreign policy struggle is one of the biggest strategic mistakes made by the US political leadership. The dollar is the cornerstone of the United States power. I think everyone understands very well that no matter how many dollars are printed, they’re quickly dispersed all over the world” \cite[loc~01:17:47]{Putin2024Vladimir}. Further, the fact that currency rates can and do change quickly makes them optimal for discerning any corresponding relationships between swift changes in state-sponsored media messaging and situational factors. Thus, we ask:

\quad \quad \textit{RQ2: How have economic post levels across RT’s language-specific accounts corresponded to changes in currency values over time?}

The need to examine oil prices as a possible influential economic context factor emerges from the variable’s function as a critical component of global energy policies since the mid-1970s. The prominent placement of oil prices in global priorities has emerged from correlations between oil prices and levels of global peace and conflict, the role of state and international organizations in managing supply and demand, the diverse interests of various actors in energy policy ranging from governments to businesses and environmental groups, and shifts in global demand and supply structures such as the increasing demand from countries like Russia and China \cite{Hughes2013Politics}. Further, oil-producing countries’ domestic policies also influence global oil prices and impact certain oil cartels \cite{Chen2016Impacts, Giraud1995equilibrium}. Russia’s Ural Oil, for example, references its own oil prices separate from the other three bigger consortium's of Brent Oil, OPEC, and West Texas Intermediate. To ascertain if and how oil prices correlate with Russia’s economic public diplomacy efforts, we ask:			

\quad\quad \textit{RQ3: How have economic post levels across RT’s language-specific accounts corresponded to changes in oil prices over time? 
}
\subsubsection{Methodology}
We collected 513,836 Facebook posts across RT’s non-Russian language- and country-specific pages using CrowdTangle (a META search program) during the 5 years between 2018-09-01 and 2023-09-01. The pages were RT Main (in English), RT France, RT en Español, RT UK, RT Arabic, RT America, and RT DE. Our extracted data included the full text of posts in their original languages, and the dates of the posts. To narrow the corpus to economic content, our coauthors who are native speakers of RT’s language accounts, created a search word list through a compilation of economic literacy words relating to finance and macro/micro-economics in English, French, German, Spanish, and Arabic (for a total of 320 words) (see Tables \ref{tab:tables1} and \ref{tab:tables2}). The final corpus for analysis comprised 38,186 posts with economic content.  

\begin{figure}[htbp]
  \centering
  \includegraphics[width=\textwidth]{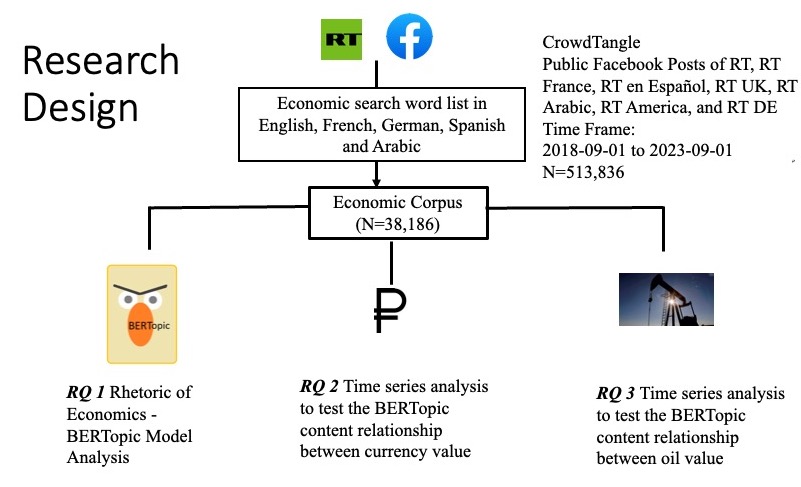} 
  \caption{Research Design. The figure represents a workflow diagram for analyzing the rhetoric of economics across various RT (Russia Today) Facebook pages in different languages, using the BERTopic model.}
  \label{fig:fig1}
\end{figure}

Figure \ref{fig:fig1} illustrates our research design. All data analyses used R (version 4.3.0 in R Studio Version 2023.03.0) and Python (3.9.16 in Visual Studio Code Version: 1.77.0). This study’s code appears in Open-Source Framework and GitHub.

We used transform-based topic modeling (BERTopic) \cite{Grootendorst2022BERTopic} with multilingual embeddings (paraphrase-multilingual-MiniLM-L12-v2) \cite{Wolf2019HuggingFace's} to reveal the topical structure of the corpus. We hyper-tuned the parameters to include at least five percent of the documents related to each topic. The generative nature of BERTopic provides one topic label for each document, and an outlier topic of -1 for documents so diffuse in their content that they defy a clear focus. We used OpenAI validated by one native speaker to relabel the topic names using documents (RT posts) with the highest probabilities. Our qualitative analysis assessed the content of topics for each unique page.

To assess what, if any, financial changes corresponded to changes in total RT Facebook economic posts, we identified the monthly (1) sum of RT posts per page and per topic, (2) mean of Ural Oil prices, and (3) mean of USD value against the Ruble’s value. The inverse representative currency exchange rates display a USD per currency unit to reveal the power of the RUB against the USD \cite{TheInternationalMonetaryFund(IMF)2023Exchange}. Thus, an inverse representative rate of 1.5 RUB to USD means 1 RUB will buy 1.5 USD, meaning that a decrease in the representative rate means the currency is becoming weaker. In contrast, an increase in the representative rate means the currency is becoming stronger in the global exchange market. 

Since the study’s timespan was 2018 to 2023, we assessed whether our findings remained constant in relation to two dominant external events as exogenous variables. The first was when the WHO declared the coronavirus pandemic on March 11th, 2021 \cite{Gehebreyesus2020WHO}, and the second corresponded to when Russia began military operations against Ukraine on February 24th, 2022 \cite{CenterforPreventiveAction2023War}. Table~\ref{tab:table1} summarizes the definitions of the independent and dependent variables.

\begin{table}[htbp]
 \caption{Explanations of Independent and Dependent Variables}
  \centering
  \begin{tabular}{p{5cm}p{10cm}}
    \toprule
    \textbf{Variable Name} & \textbf{Definition} \\
    \midrule
    \textit{x} & \\
    Inverse Russian Ruble Representative Rate & 1/ Russian Ruble “Representative exchange rates, which are reported to the Fund by the issuing central bank, are expressed in terms of currency units per US dollar, except for those indicated by (1) which are in terms of US dollars per currency unit” (The International Monetary Fund (IMF), 2023). \\
    Urals Oil monthly averages & Ural oil monthly prices acquired from OPEC (OPEC, 2023) \\
    \addlinespace
    \textit{y} & \\
    BERTopic per month per page & The 6 BERTopics per page monthly post values \\
    \addlinespace
    \textit{Exogenous Variables} & \\
    Interruption 1 – COVID-19 Pandemic & All weeks prior to March 11th, 2021 marked as 0, including March 11th, 2021 and post marked as 1 \\
    Interruption 2 – Russia \& Ukraine Military Conflict & All weeks prior to February 24th, 2022 marked as 0, including February 24th, 2022 and post marked as 1 \\
    \bottomrule
  \end{tabular}
  \label{tab:table1}
\end{table}

\begin{figure}[htbp]
  \centering
  \includegraphics[width=0.8\textwidth]{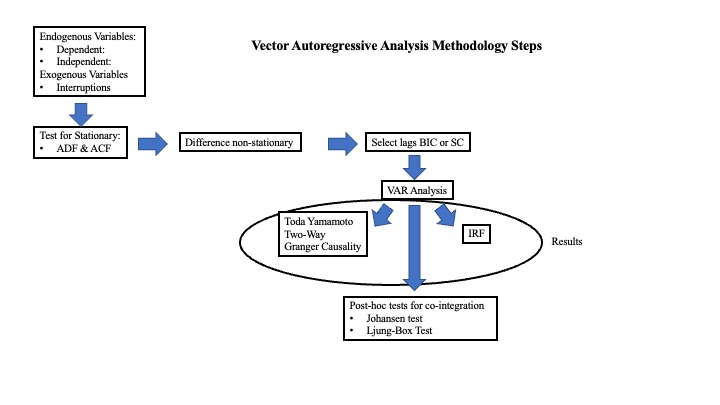} 
  \caption{Vector Autoregressive Methodology Steps. The figure outlines the steps of a Vector Autoregressive (VAR) Analysis methodology, beginning with testing for stationarity using the Augmented Dickey-Fuller (ADF) and AutoCorrelation Function (ACF) tests, proceeding to differencing non-stationary variables, selecting appropriate lag lengths using Bayesian Information Criterion (BIC) or Schwarz Criterion (SC), performing VAR Analysis, Toda-Yamamoto two-way Granger Causality tests, Impulse Response Function (IRF) and concluding with post-hoc tests for co-integration such as the Johansen test and the Ljung-Box test to arrive at results.}
  \label{fig:fig2}
\end{figure}

For our statistical analysis, we used Vector Autoregressive Models (VAR) (see Figure 2). We selected this model for three primary reasons: 1) to capture the dynamic interdependencies between variables, 2) to include endogenous and exogenous variables, and 3) to identify causal relationships between variables \cite{Lokmanoglu2023Social, Lukito2020Coordinating, Toda1995Statistical}. For a summary of our Vector Autoregressive Analysis, see Figure \ref{fig:fig2}. 

To analyze the robustness of our VAR analysis, we tested our data using Augmented Dickey–Fuller (ADF) t-statistic test for unit root to test a rejection criterion of p=0.05 (Tables \ref{tab:tableS3} and \ref{tab:tableS15}). We visually corroborated the results with Autocorrelation Function (ACF) (Figures \ref{fig:figS3} to \ref{fig:figS46}). We then selected appropriate lags with Bayesian Information Criterion (BIC) or Schwarz Criterion (SC) testing lags from one to six (Tables \ref{tab:tableS4} and \ref{tab:tableS16}) \cite{Lokmanoglu2023Social, Lukito2020Coordinating, Toda1995Statistical}. Further, to demonstrate how a dependent variable responded to the independent variable and to understand any relationship’s effect size without noise, we graphed the results using the Impulse Response Function (IRF) \cite{Lokmanoglu2023Social, Lukito2020Coordinating, Ophir2021Framing, Lutkepohl2010Impulse}. To document that the residual of all times series were random, we utilized post-hoc tests of Johansen test for co-integration with a rejection criteria of p = 0.05 (Tables \ref{tab:tables13} and \ref{tab:tableS25}) \cite{Johansen1991Estimation, Ophir2021Framing} and Ljung-Box tests for co-integration (Tables \ref{tab:tables14} and \ref{tab:tables26}) \cite{Box1970Distribution, Lukito2020Coordinating}. To interpret the results of the VAR analysis, we ran Granger Causality tests \cite{Lukito2020Coordinating, Toda1995Statistical}. 

\section{Results}
\label{sec:results}
\subsection{RQ1: Economic Content in RT Posts}
\begin{figure}[htbp]
  \centering
  \includegraphics[width=0.8\textwidth]{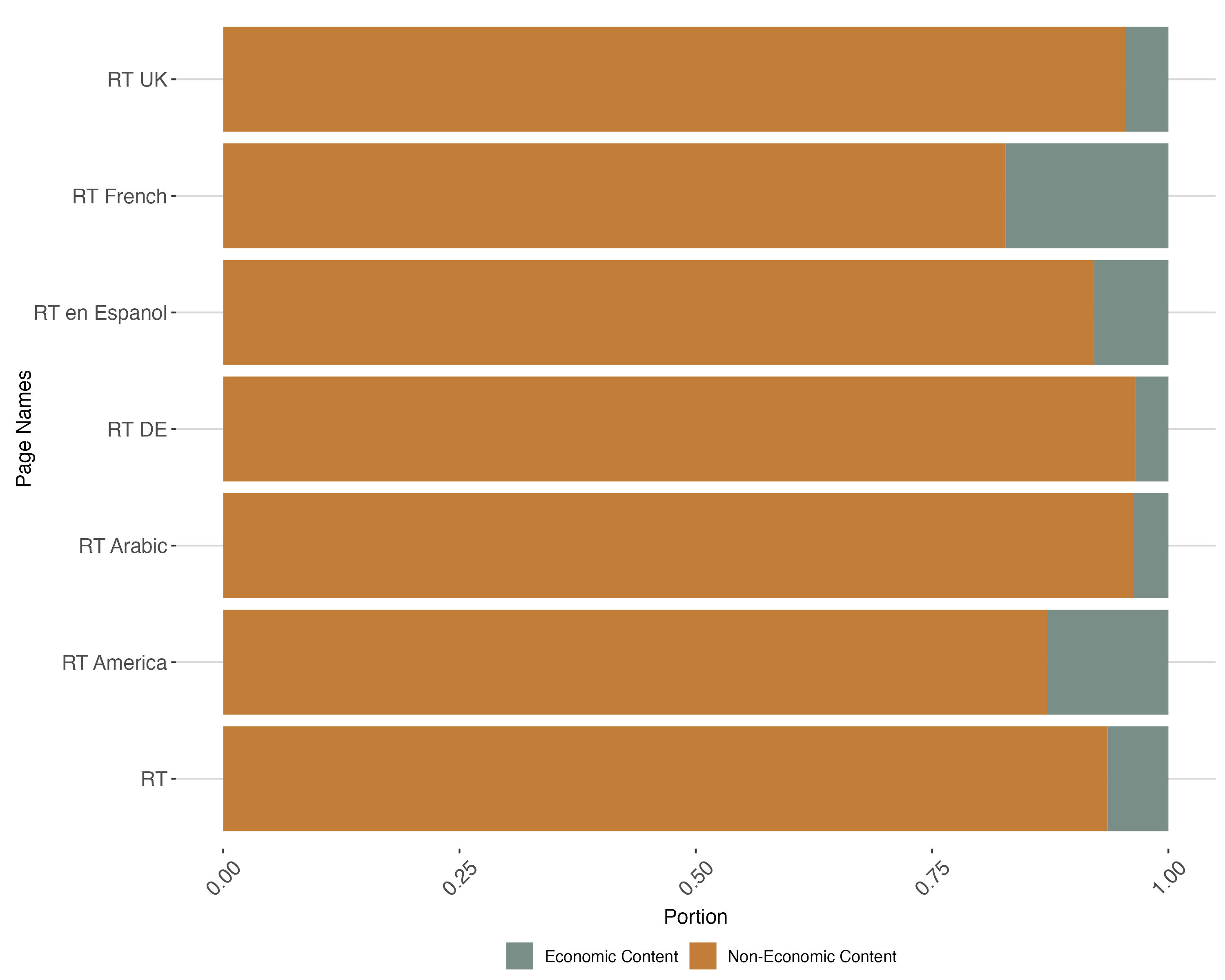} 
  \caption{RT Pages Post Volume Economic to Non-Economic Ratio.}\label{fig:fig3}
\end{figure}

Seven percent of overall RT’s Facebook posts focused on economic topics (figure \ref{fig:fig3}). Six overarching economic topics characterized the posts in RT’s multilingual corpus. \textit{Recent News Headlines} was the leading topic, followed by \textit{US-Russia-Ukraine Sanctions} and \textit{Covid Pandemic and Related Issues}. See Table~\ref{tab:table2} for topic labels and post counts across the full corpus and by language account. 

\begin{table}[htbp]
\caption{Summary of Topic Occurrences Across Different RT Channels}\label{tab:table2}\addtocounter{table}{-1}
\keepXColumns % This command is provided by ltablex to keep the X column types rather than converting them to p columns
\begin{tabularx}{\textwidth}{@{} l *{6}{>{\centering\arraybackslash}X} @{}}
\toprule
Topic Label & Bitcoin Crypto Value & City Crisis Protest & Covid Pandemic Issues & Middle East Wars & Recent News & US-Russia-Ukraine Sanctions \\
\midrule
RT          & 135 & 445 & 199 & 113 & 2343 & 2153 \\
RT America  & 54  & 13  & 103 & 17  & 520  & 420  \\
RT Arabic   & 39  & 7   & 5   & 5   & 855  & 1671 \\
RT DE       & 19  & 17  & 180 & 34  & 1432 & 843  \\
RT French   & 25  & 103 & 1480& 119 & 7515 & 2559 \\
RT Spanish  & 1051& 1115& 641 & 101 & 7588 & 3251 \\
RT UK       & 23  & 21  & 63  & 11  & 738  & 159  \\
TOTAL       & 1346& 1721& 2671& 400 & 20991& 11056 \\
\bottomrule
\end{tabularx}
\end{table}

Qualitatively each of the RT topics focused on different aspects of economic messaging. The most emphasized topic, \textit{Recent News Headlines}, spanned a spectrum of economic and societal issues globally. Posts ranged from the Yellow Vests protests in Paris and boycotts against the Turkish President to the US Federal Reserve's interest rates and China's sale of US debt (e.g., RT America on June 10, 2020). Posts delved into the Bank of England's Venezuelan gold seizure and the effects of sanctions on Venezuela's pandemic response (e.g., RT America on April 22, 2020 and RT UK May 20, 2020). The main channel of RT addressed corruption, societal concerns, and everyday challenges like debt evasion and the financial impacts of personal actions. RT DE discussed local German issues, including penalties for minor public infractions and mask mandates. RT French emphasized press freedom and civic rights amid law enforcement challenges. RT en Español focused on Latin American topics such as Colombia's defense spending during the pandemic, Venezuela's economic crisis caused by external pressures, and Ecuador's public dissent against austerity measures (e.g., RT en Español on May 12, 2020, February 13, 2019 and October 8, 2019).

The second most emphasized topic, \textit{US-Russia-Ukraine Sanctions}, focused on the conflict and sanctions across all the pages. In RT Main, RT UK, RT French, and RT DE, the posts highlighted the rising utility costs resulting from Europe EU sanctions on Russia and the importance of Russian energy. RT Arabic took a different approach, discussing Russia’s role in the Middle East brokering deals between an Indian businessmen and Saudi Arabia (e.g., RT Arabic post on November 11, 2021). RT en Español drew attention to the global consequences of sanctions, such as using posts claiming Russia and Iran had ended the use of the US dollar in bilateral economic transactions, opting instead for their own national currencies (e.g., RT en Español February 5, 2019). 

Posts representative of third most emphasized topic \textit{Covid Pandemic and Related Issues} were heavily focused on the post-pandemic period. RT generally criticized Western governments policies for causing economic problems and drew attention to price differences between Pfizer and Russian vaccines. Further, they highlighted dramatic increases in wealth for the top American billionaires, a phenomenon exacerbating concerns about wealth concentration during a notable period of widespread economic hardship. RT America primarily discussed significant stock sell-offs by top US executives and the Amazon CEO before the COVID-19 pandemic hit to raise questions about insider trading (e.g., RT America on 2020-03-24). RT UK similarly focused on criticizing UK policies regarding UK capital, COVID restrictions, and corruption. RT Arabic had only five posts related to this topic and all focused on a Russian mathematical model designed to cure Covid-19 and its resulting economic lockdowns. RT DE and RT French focused on the pandemic's effects and policy responses, with RT en Español highlighted Latin America's interactions with Russia and China amidst the pandemic.

Representative posts associated with the fourth most emphasized topic \textit{City Crisis Protest and Disaster} presented a tapestry of global events deeply interwoven with consequential economic ramifications. RT America highlighted protests in Iraq, ensuing attacks on US embassies, and Virginian debates over climate policy and public education funding (e.g., RT America on January 27th, 2020 and January 28th, 2019). RT UK highlighted the economic uncertainties stirred by Brexit and the Extinction Rebellion's protests in London. RT Arabic reported on the immediate financial costs of a possible solar storm and on broader economic trends (e.g., the decline of the Turkish lire) to expose the fragile underpinnings of national economies in the face of global challenges. Meanwhile, RT DE, RT French, and RT en Español chronicled a range of natural and man-made catastrophes across Europe to South America, with each account shedding light on the economic devastation left behind and the pressing need for economic aid and recovery initiatives.

RT channels posts associated with the fifth highest topic \textit{Bitcoin Cryptocurrency Value} encompassed the broad landscape of cryptocurrencies. The range of subject matter included RT’s detailing of significant heists, Indonesia's religious stance against crypto trading (e.g., RT on November 13, 2021), and America's apprehensions about Bitcoin's geopolitical implications (e.g., RT America on September 14, 2021). The focus of the posts spanned continents, encompassing America's scrutiny of China's burgeoning digital currency, the privacy implications of Facebook's global financial ventures (e.g., RT America September 13, 2021), and the UK's potential leap towards a national digital currency amid post-Brexit shifts (e.g., RT UK on April 4 and 21, 2021). RT Arabic highlighted the mercurial fortunes tied to digital currencies (e.g., RT Arabic on May 19, 2022) as exemplified by the dramatic losses of tech leaders and Facebook's shelving of "Diem" (e.g., RT Arabic February 2, 2022). Posts on RT De and RT French reflected European concerns over privacy, market stability, and the rise of Bitcoin. RT Spanish posts broadened the discourse to include oil market fluctuations and ethical trading issues (e.g., RT en Español March 8th, 2020). 

The smallest topic in terms of number of posts, \textit{Middle East Wars and Conflict}, often spotlighted the stark consequences of geopolitical instability and conflict, with a particular emphasis on Afghanistan. The main channel of RT covered the tumultuous withdrawal of US forces from Afghanistan and the resurgence of the Taliban, highlighting the political fallout the profound economic implications. The main channel’s posts also delved into Afghanistan's black market opium production, a multi-billion dollar industry that fueled both the global drug trade and terrorist financing. The posts also described plundered wealth by fleeing officials and the economic vacuum left by departing foreign forces (e.g., RT January 8th, 2022). RT America posts emphasized the economic costs of the US's longest war by discussing the staggering cost to American taxpayers and the fortunes made by defense contractors (e.g., RT America on August 17th, 2021). RT UK focused on the UK's military re-engagement in Kabul for evacuation operations, signaling ongoing financial and human costs of the conflict despite an official end to combat operations (e.g., RT UK on August 16th, 2021). The same channel also raised concerns about the potential rise in terror threats in the post-Afghanistan collapse and its impact on security spending. RT DE addressed the economic repercussions of war through reports on the vast amounts of US military equipment left behind and critiqued economic strategies that lhad ed to such outcomes (e.g., RT DE on September 13th, 2021). RT DE, RT French and RT en Español explored individual desperation in Afghanistan, prompting drastic measures like organ sales and the international community's response to the burgeoning humanitarian crisis (e.g., RT French October 12nd, 2021 and RT en Español August 16th, 2021). RT en Español touched on the US financial legacy in Afghanistan, critiquing the vast expenses and questioning the effectiveness of military intervention (e.g., RT en Español April 18th, 2021). In contrast, RT Arabic posts did not discuss Afghanistan and USA but did highlight the United Nations' call for the Taliban to halt punitive measures like flogging, execution, and stoning (RT Arabic on May 8th, 2023).

\begin{figure}
  \centering
  \includegraphics[width=0.8\textwidth]{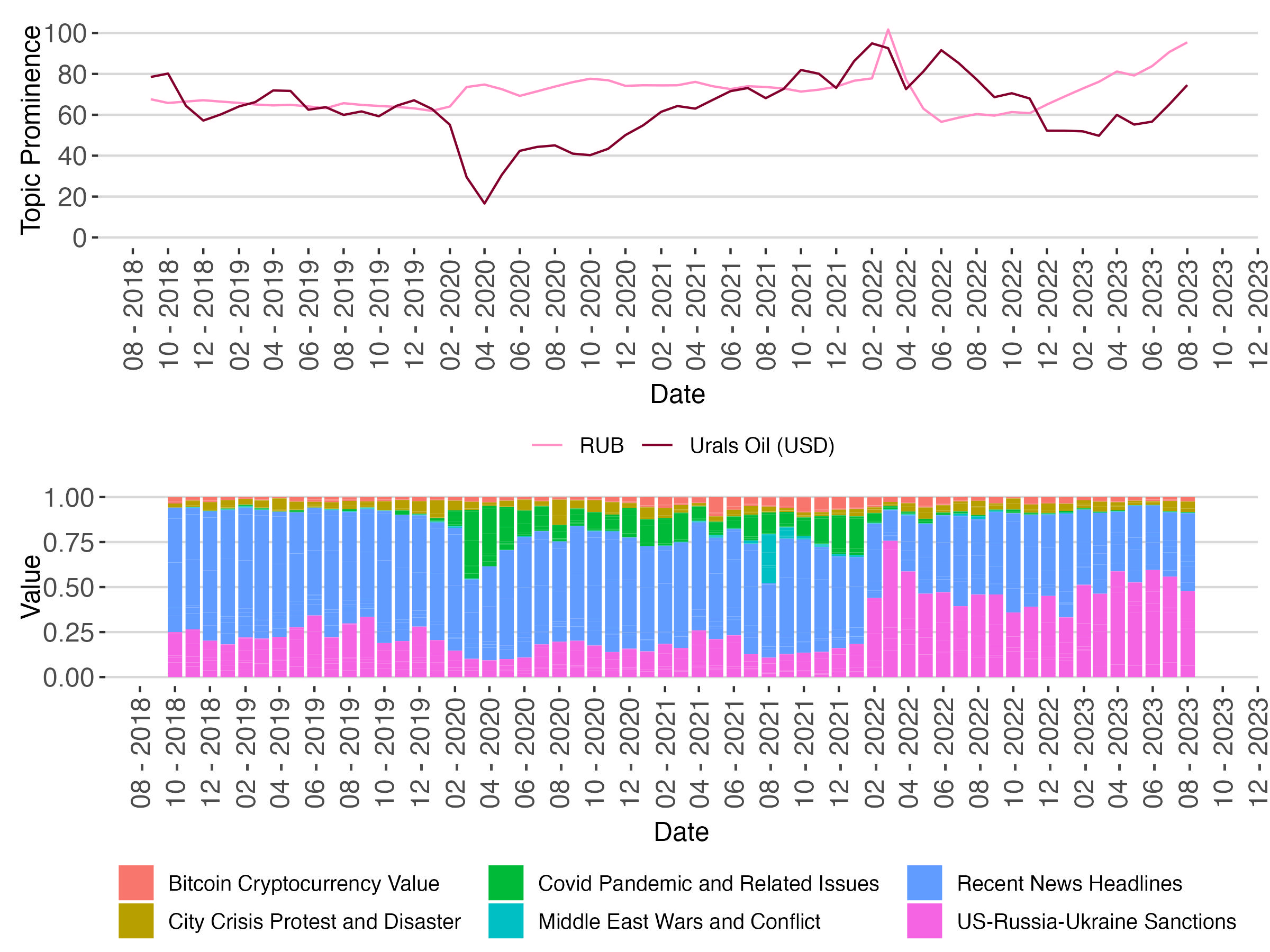} 
  \caption{Independent and Dependent Variables over Time. The figure displays two over time graphs from August 2018 to April 2023, with the top graph showing Russian Ruble (RUB) and Urals oil prices over time, and the bottom stacked bar graph depicting the normalized values of topics.}\label{fig:fig4}
\end{figure}

Overall, the RT channels highlighted a complex interplay between war, economic interests, and human costs. Collectively, they stressed the cycle where geopolitical strategies directly influenced global and local economies, often with long-lasting and far-reaching consequences. See Figure \ref{fig:fig4} for independent and dependent variables over time for the rest of the results.

\subsection{RQ2:  RT Economic Content and the Value of the Russian Ruble}
Over the past five years, the value of the Russian Ruble had a statistically significant predictive power on the level of posts related to Recent News Headlines on RT-DE and US-Russia-Ukraine Sanctions on the main RT page. Table \ref{tab:table3} displays results for the VAR estimates, and Table \ref{tab:table4} demonstrates a unidirectional relationship between the value of the Russian Ruble and the pages. For VAR results of all variables, see Tables \ref{tab:tableS5} to \ref{tab:tableS11}; for Granger Causality, see Table \ref{tab:tables12}.

\begin{table}[htbp]
  \centering
  \caption{VAR estimates for monthly topic volume per RT page on Ruble Values}\label{tab:table3}
  \begin{tabular}{lcc}
    \toprule
    & \textbf{RT DE Recent News Headlines} & \textbf{RT US-Russia-Ukraine Sanctions} \\
    \midrule
    Ruble (-1) & $4506.965^{**}$ (1551.461) & $-3755.434^{*}$ (1791.887) \\
    Interruption 1 (Covid-19) & 2.231 (3.040) & 9.476$^{*}$ (3.782) \\
    Interruption 2 (Russia-Ukraine) & -1.308 (5.549) & 16.765$^{*}$ (6.899) \\
    Constant & -0.965 (2.965) & 2.325 (3.594) \\
    Trend & 0.030 (0.149) & $-0.390^{*}$ (0.184) \\
    Num.Obs. & 58 & 58 \\
    $R^2$ & 0.320 & 0.179 \\
    $R^2$ Adj. & 0.254 & 0.100 \\
    \bottomrule
    \multicolumn{3}{p{\textwidth}}{\footnotesize $^{+} p < 0.10$, * $p < 0.05$, ** $p < 0.01$, *** $p < 0.001$} \\
  \end{tabular}
\end{table}

\begin{table}[htbp]
  \centering
  \caption{Granger Causality Test for RT page on Ruble}\label{tab:table4}
  \begin{tabular}{lc}
    \toprule
    Page & Granger Causality P Value \\
    \midrule
    RT DE Recent News Headlines & 0.0045$^{**}$ \\
    RT US-Russia-Ukraine Sanctions & 0.0385$^{*}$ \\
    \bottomrule
    \multicolumn{2}{p{\textwidth}}{\footnotesize Note: $^{*}p < 0.05$, $^{**}p < 0.01$, $^{***}p < 0.001$} \\
  \end{tabular}
\end{table}

\begin{figure}
  \centering
  \includegraphics[width=0.8\textwidth]{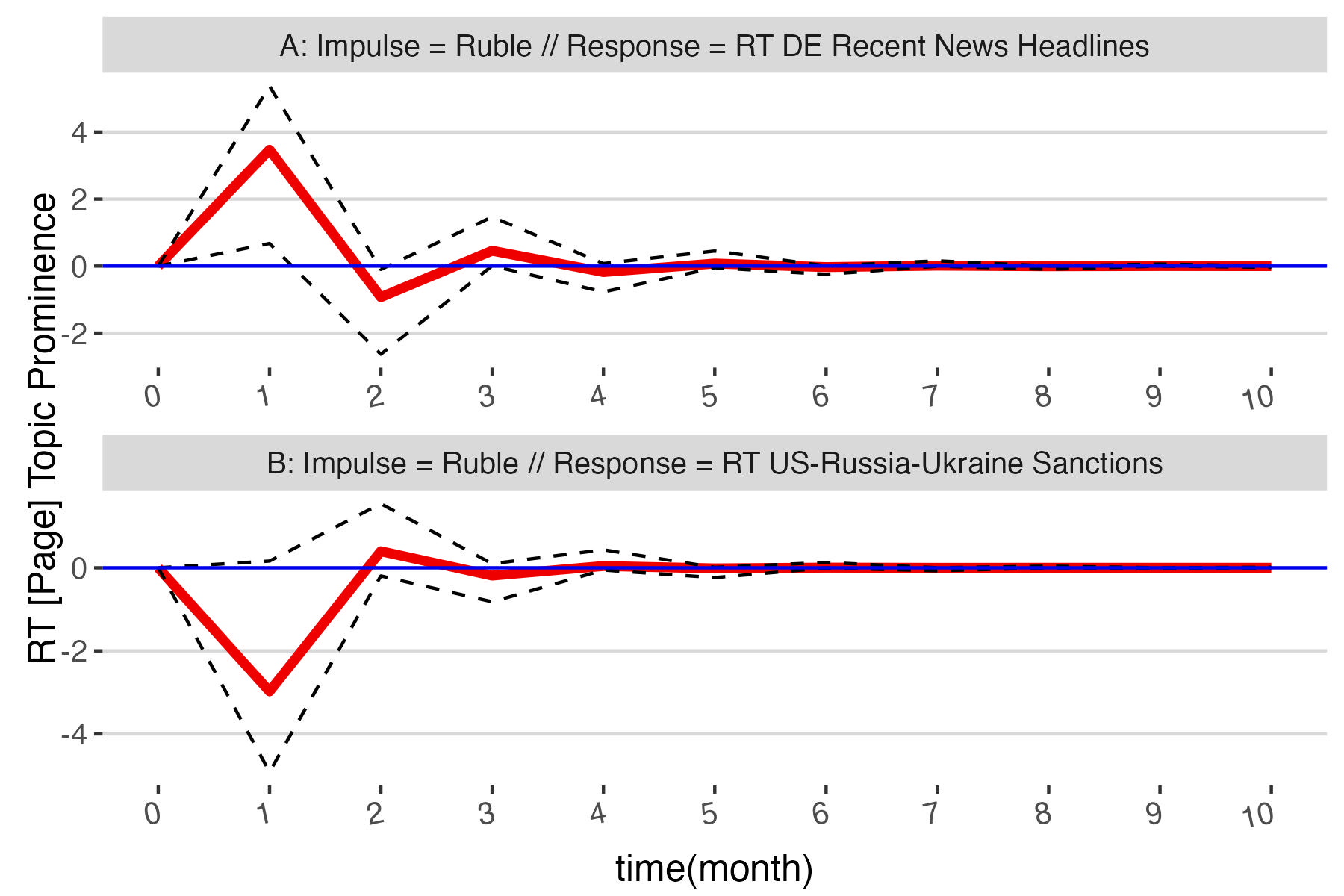} 
  \caption{IRF RT on Ruble, lags in the graph correspond to months.}
  \label{fig:fig5}
\end{figure}

The IRF graphs (Figure \ref{fig:fig5}) visualize the effect of one standard deviation change in the value of the Russian Ruble on post volumes of RT DE related to Recent News Headlines (Panel A) and RT \textit{US-Russia-Ukraine Sanctions} (Panel B) topics. Panel A shows that a positive trend in the Ruble (currency becoming stronger) led to a temporary increase in the \textit{Recent News Headlines} topic prominence in RT DE. Panel B shows that a positive trend in the Ruble (currency becoming stronger) increased the topic prominence of \textit{US-Russia-Ukraine Sanctions} in RT upon the Ruble shock, but the topic prominence sharply declined in the first month, then increased and gradually faded to the baseline level.

\subsection{RQ3:  RT Economic Content and Ural Oil Prices}

For the past five years, the Ural Oil prices (Table \ref{tab:table5}) had a statistically significant predictive power on RT Arabic’s posts on \textit{Covid Pandemic and Related Issues}, RT Main’s posts on \textit{Covid Pandemic and Related Issues} and \textit{US-Russia-Ukraine Sanctions}, RT DE’s posts on \textit{Bitcoin Cryptocurrency Value}, RT DE’s posts on \textit{Covid Pandemic and Related Issues} and RT en Español’s posts on \textit{US-Russia-Ukraine Sanctions} (for more, see Table \ref{tab:table6} for Granger causality; Tables \ref{tab:tableS17} to \ref{tab:tableS23} for VAR results of all variables; and  Table \ref{tab:tableS24} for Granger Causality). 

\begin{table}[htbp]
\caption{VAR estimates for monthly topic volume per RT page on Ural Oil Prices}\label{tab:table5}\addtocounter{table}{-1}
\begin{tabularx}{\textwidth}{@{} l *{6}{>{\centering\arraybackslash}X} @{}}
\toprule
& \textbf{RT Arabic Covid Pandemic and Related Issues} & \textbf{RT Covid Pandemic and Related Issues} & \textbf{RT DE Bitcoin Cryptocurrency Value} & \textbf{RT DE Covid Pandemic and Related Issues} & \textbf{RT en Español US-Russia-Ukraine Sanctions} & \textbf{RT US-Russia-Ukraine Sanctions} \\
\midrule
Ural Oil Price (-1) & $-0.014^*$ $(0.007)$ & $-0.163^*$ $(0.063)$ & $-0.026^*$ $(0.012)$ & $-0.153^{***}$ $(0.040)$ & $1.512^*$ $(0.654)$ & $0.383^*$ $(0.186)$ \\
Interruption 1 (Covid-19) & $-0.004$ $(0.133)$ & $-0.807$ $(1.158)$ & $0.052$ $(0.234)$ & $-1.074$ $(0.763)$ & $5.734$ $(12.716)$ & $7.716^*$ $(3.719)$ \\
Interruption 2 (Russia-Ukraine) & $-0.012$ $(0.243)$ & $-2.413$ $(2.127)$ & $-0.356$ $(0.428)$ & $-4.913^{**}$ $(1.411)$ & $20.141$ $(23.347)$ & $15.260^*$ $(6.815)$ \\
Constant & $-0.030$ $(0.132)$ & $-0.284$ $(1.144)$ & $-0.134$ $(0.229)$ & $-1.134$ $(0.755)$ & $7.117$ $(12.546)$ & $2.660$ $(3.617)$ \\
Trend & $0.001$ $(0.007)$ & $0.047$ $(0.057)$ & $0.007$ $(0.011)$ & $0.107^{**}$ $(0.038)$ & $-0.458$ $(0.624)$ & $-0.348^{+}$ $(0.181)$ \\
Num.Obs. & 58 & 58 & 58 & 58 & 58 & 58 \\
R2 & 0.404 & 0.139 & 0.252 & 0.392 & 0.140 & 0.177 \\
R2 Adj. & 0.346 & 0.056 & 0.180 & 0.334 & 0.057 & 0.098 \\
\bottomrule
\multicolumn{7}{p{\textwidth}}{\footnotesize $^{+} p < 0.10$, * $p < 0.05$, ** $p < 0.01$, *** $p < 0.001$} \\
\end{tabularx}
\end{table}

\begin{table}[htbp]
\caption{Granger Causality Test for RT page on Ural Oil Prices}\label{tab:table6}
\begin{tabularx}{\textwidth}{@{}Xl@{}}
\toprule
Page & Granger Causality P Value \\
\midrule
RT Arabic Covid Pandemic and Related Issues & 0.0383* \\
RT Covid Pandemic and Related Issues & 0.0107* \\
RT DE Bitcoin Cryptocurrency Value & 0.0318* \\
RT DE Covid Pandemic and Related Issues & 0.0002*** \\
RT en Español US-Russia-Ukraine Sanctions & 0.0228* \\
RT US-Russia-Ukraine Sanctions & 0.0418* \\
\bottomrule
\multicolumn{2}{p{\textwidth}}{\footnotesize Note: $^{*}p < 0.05$, $^{**}p < 0.01$, $^{***}p < 0.001$} \\
\end{tabularx}
\end{table}

\begin{figure}[htbp]
  \centering
  \includegraphics[width=0.8\textwidth]{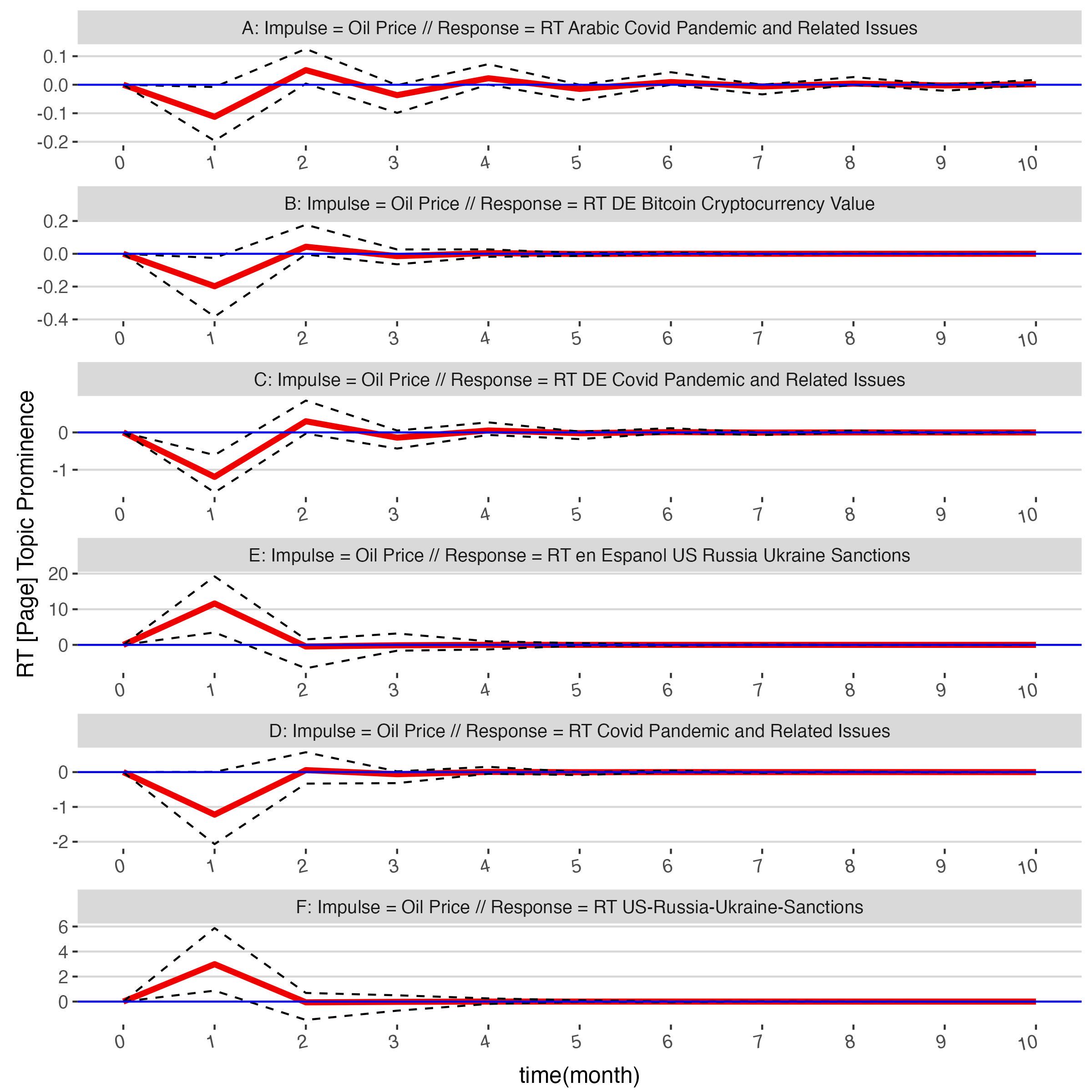} 
  \caption{IRF RT on Ural Oil Prices, lags in the graph correspond to months.}
  \label{fig:fig6}
\end{figure}

The IRF graphs (Figure \ref{fig:fig6}) depict the reaction of the RT account pages’ topic prominence following a one standard deviation increase in oil prices. Panel A documents the impact of RT Arabic posts about \textit{Covid Pandemic and Related Issues} in relation to oil prices; although modest, the response to the change in oil prices showed a consistent negative response, with the effect tapering off and returning to the baseline after approximately five months. The confidence intervals remained tight throughout the horizon. Panel B shows a pronounced negative response to post volume on \textit{Bitcoin Cryptocurrency Value} in RT DE, reaching its nadir in the second month post-increase. The topic's prominence gradually reverts to baseline levels, with the effect dissipating entirely by the tenth month. Panel C illustrates that the prominence of \textit{Covid Pandemic and Related Issues} in RT DE is initially negative, mirroring the pattern observed in RT Arabic's coverage of the same topic (Panel A), but with a quicker reversion to the mean, suggesting a transient impact of oil prices on this topic. Panel E displays a relatively stable pattern of RT en Español postings on \textit{US-Russia-Ukraine Sanctions} with minor fluctuations within the confidence interval, indicating a negligible impact of oil price shocks on the topic's prominence within the observed period. Like the Arabic and DE responses to the \textit{COVID-19 Pandemic and Related Issue}s, the same topic in RT, illustrated in Panel D, shows an evident yet negative impact, which stabilized after an initial decline, indicating a potential transient concern with oil price volatility with COVID-19 coverage. In Panel F, a unique response to US-Russia-Ukraine-Sanctions in RT was detected, characterized by a significant initial increase in topic prominence, followed by a sharp reversal and a gradual return to baseline, highlighting a complex interaction between oil price changes and the coverage of \textit{US-Russia-Ukraine Sanctions}.  

\section{Discussion}
RT Facebook posts on the economy were a pivotal component of Russia's public diplomacy efforts during the last five years. The platform posted almost forty thousand messages on economic topics, or seven percent of the full corpus during the study’s timespan. Economic posts as a percentage of Russia’s overall content on Facebook was lower than that of other Russian public diplomacy forums, such as Golan \& Viatchaninova’s \cite{Golan2014Advertorial} finding that 26\% of Russian advertorials in foreign newspapers addressed economic issues as most salient. Still, Facebook’s network amplification affordance, coupled with its wide audience, positions the social media platform to function as a key contributor to message reinforcement in Russia’s public diplomacy campaign.

RT Facebook does not treat economics as a unitary, homogenous concept. Under the \textit{Recent News Headlines}, its posts emphasize the relative strengths or weaknesses of global currencies and national economic policies. They also target local economic anxieties such as COVID-19 lockdowns, energy prices in European countries, protests arising from economic grievances in Latin America, and questions about the hegemony of the US Dollar and EU currencies. Further, Posts aligning within topics directly questioning US motivations for sanctions on Russia and the dollar’s value underscore that the United States poses an ongoing threat to the global population. On the whole, RT’s top topics all focus on economic concerns rather than the economic benefits from a Russian alliance. 

While the economic content across the full corpus consistently reiterates RT’s topics of anti-sanctions, the US dollar, the Russian Ruble, energy, and protests, the language pages often rely on unique audience strategies for the bulk of the platform’s most representative topics. Consider the topic of the US dollar, for example. RT America highlighted the US dollar's declining value and the Ruble's increasing value, while non-anglophone pages like RT DE emphasized that China and Russia would return to the gold standard to free themselves from the US Dollar’s global dominance. Such rumors and misinformation acquire power from social transmission and repeated exposure \cite{Berinsky2023Political}, two factors that further highlight Facebook’s value as a preferred platform for Russian message dissemination.  

At times, RT Facebook public diplomacy efforts emphasize distinct economic topics on their language-based pages. RT en Español, for example, placed the largest emphasis on the value of cryptocurrencies. Posts maintained that the cryptocurrencies were denigrating the US Dollar and serving as a viable alternative for the global economy. Such a targeted message has significant implications given that Latin America and the Caribbean (LAC) are leading the way in adopting digital money, led by Bukele who made Bitcoin a legal tender in El Salvador \cite{Bhattacharya2023Interest, ChainalysisTeam2023Latin, Murray2024cult}.

Mostly, however, the language accounts of RT Facebook utilize similar patterns of topic repetition when discussing economic comments.  All language accounts except RT Arabic posts are most often about \textit{Recent News Headlines}. In fact, the posts constitute almost double the number of posts associated with the second largest topic, \textit{US-Russian-Ukrainian Sanctions}. While switching the topics’ rank order, RT Arabic posts the same two topics most often. The emphasis on \textit{Recent News Headlines} suggests that RT Facebook, regardless of language account, is establishing itself as a go-to economics news source. The focus on \textit{US-Russian-Ukrainian Sanctions}, by contrast, emphasizes the global consequences of the U.S.-led, anti-Russian policy and spiked in relation to posts about recent headlines.  

Notably the dominant message factor explaining differential content across language account pages of RT Facebook are national economic contexts. In posts related to the \textit{US-Russia-Ukraine Sanction} topic, for example, the main RT page emphasized US “economic terrorism” on nations like Venezuela, China, and Russia (RT post on 5/31/2019), RT DE focused on Germany’s defense budget, and RT Arabic emphasized Western sanctions on Russia, Belarus, Turkey, Saudi Arabia, Venezuela, Syria, Hezbollah, and the Myanmar junta. The emphasis on matters of national economic concern worked to position RT as a platform capable of attracting and sustaining global viewers.

During the study’s time frame, major external shocks occurring also result in shifts in the economic content of RT Facebook pages. The onset of the COVID-19 pandemic increased the economic content on RT French and RT DE, perhaps because France and Germany were among the European epicenters of the pandemic at its start due to their close proximity to Italy \cite{Indolfi2020Outbreak, Ophir2021Framing}. The onset of Russia’s most recent invasion of Ukraine, however, increased economic content volume across all pages except for RT French. Russia’s economic costs associated with the war, heightened by the imposition of the Western sanctions, are a likely explanation for the consistent use of the global economic messaging strategy, as the prominence of the US-Russia-Ukraine Sanction topic attests. 

Adding to previous notions that trade balances between countries serve as a critical context factor at work in public diplomacy efforts \cite{Khalitova2020He}, this study documents that the frequency of RT economic content varies in relation to currency fluctuations. Changes in the Ruble's value affect the topic prominence of \textit{Recent News Headlines} and \textit{US-Russia-Ukraine Sanctions}, hinting at a reactive content strategy to economic conditions on the ground. Such shifts, however, are not uniform across all language platforms, highlighting RT's strategic targeting of regional and linguistic audience sensitivities. The temporary nature of the correspondence between the economy topic’s prominence and the value of the currency is temporary on certain language accounts, suggesting that reactive public diplomacy responses would need to be timely to distract from RT’s heightened topic focus.

The varied responses to oil prices across different topics and languages underscore the adaptability of RT’s targeted economic messaging to external economic factors. Following increased oil prices, for example, the main channel of RT, RT Arabic, and RT DE posts about \textit{Covid Pandemic and Related Issues} experienced slight declines. However, RT DE’s post volume on \textit{Bitcoin Cryptocurrency Value} dropped dramatically in relation to rising oil prices, while the most volatile RT messaging shifts both positive and negative focused on \textit{US-Russia-Ukraine-Sanctions}. Initially, explain how sanctions cause rise in oil prices. The range of topics that experience frequency changes— online currencies, health crises, and economic sanctions — suggest that oil prices constitute a flexible context variable with wide application in Russia’s public diplomacy program. 

Overall, this study makes clear that a full accounting of Russia’s economic-based public diplomacy cannot occur without consideration of multiple variables. A nuanced contemplation of specific economic topics, explorations of language-based, audience targeting strategies, an understanding of how economic variables interact with key situational events like pandemics and wars, and attention to changes in currency valuations and oil prices over time combine to define the communicative landscape. Such a multi-factor approach likely has value beyond the Russian context, as the general factors of economic messaging, audience targeting, and economic context changes recur across public diplomacy contexts. 

\section{Future Areas of Study}
The limits of this study are suggestive about productive avenues for future research. First, instead of focusing on a single currency (the Russian Ruble), a more comprehensive analysis should compare levels of public diplomacy social media content with other leading currencies, including the US Dollar, the UK Pound, the Saudi Arabian Rial, and the Euro to build out understandings of the economic context-content relationship. Second, the approach used here (BERTopic) is a generative model where the computer assigns each post to a single topic. This precludes a post from sorting into multiple topics that would be more valid if overcome. However, alternative models such as LDA topic modeling do not have multi-lingual capacities without translation. Thus, a future study comparing LDA with BERTopic would provide valuable insights into the advantages and shortcomings of both topic modeling methods for understanding public diplomacy. 

Future studies should also focus on expanding this methodology around the implications of RT's economic content and how the disseminated information could affect voter perceptions, and global financial behaviors. For example, state-sponsored media releases of information about economic events could influence commodity markets if traders and investors adjust their strategies based on the news' perceived credibility. This impact on market sentiments could be significant, especially if the information pertains to sanctions, oil prices, currency valuations, and critical economic indicators. This could influence voter perceptions of the state of the economy in a country. 

Finally, rather than focusing on RT, economic public diplomacy strategies should be compared on other state-sponsored media entities like China's CGTN or Iran's Press. Messaging approaches and platforms may differ, even if the underlying objectives influencing international perception, reinforcing national narratives, and countering Western media perspectives remain similar. This comparison highlights a trend in global information campaigns where economic messaging may be coming increasingly weaponized.

\textbf{\textit{Data Availability:}} The data and the code that support the findings of this study are openly available in  \url{https://github.com/aysedeniz09/Multilingual_Econ GitHub}.

%Bibliography
\bibliographystyle{unsrt}  
\bibliography{references}  

\begin{thebibliography}{10}

\bibitem{Kates2019We}
Sean Kates and Joshua~A. Tucker.
\newblock We never change, do we? economic anxiety and far-right identification in a postcrisis europe*.
\newblock {\em Social Science Quarterly}, 100(2):494--523, 2019.

\bibitem{Rodrik2021Why}
Dani Rodrik.
\newblock Why does globalization fuel populism? economics, culture, and the rise of right-wing populism.
\newblock {\em Annual Review of Economics}, 13(1):133--170, 2021.

\bibitem{Schmeichel2015Individual}
Brandon~J. Schmeichel and David Tang.
\newblock Individual differences in executive functioning and their relationship to emotional processes and responses.
\newblock {\em Current Directions in Psychological Science}, 24(2):93--98, 4 2015.

\bibitem{Citrin1997Public}
Jack Citrin, Donald~P. Green, Christopher Muste, and Cara Wong.
\newblock Public opinion toward immigration reform: The role of economic motivations.
\newblock {\em The Journal of Politics}, 59(3):858--881, 8 1997.
\newblock publisher: The University of Chicago Press.

\bibitem{Fetzer2021Coronavirus}
Thiemo Fetzer, Lukas Hensel, Johannes Hermle, and Christopher Roth.
\newblock Coronavirus perceptions and economic anxiety.
\newblock {\em The Review of Economics and Statistics}, 103(5):968--978, 11 2021.

\bibitem{Peterson2019Carbon}
David A.~M. Peterson, Kristy~C. Carter, Dara~M. Wald, William Gustafson, Sidney Hartz, Jacob Donahue, Joe~R. Eilers, Anne~E. Hamilton, Kyle S.~H. Hutchings, Federico~E. Macchiavelli, Aaron~J. Mehner, Zaira P.~Pagan Cajigas, Olivia Pfeiffer, and Aaron~J. Van~Middendorp.
\newblock Carbon or cash: Evaluating the effectiveness of environmental and economic messages on attitudes about wind energy in the united states.
\newblock {\em Energy Research \& Social Science}, 51:119--128, 5 2019.

\bibitem{Alvarez1995Economics}
R.~Michael Alvarez and Jonathan Nagler.
\newblock Economics, issues and the perot candidacy: Voter choice in the 1992 presidential election.
\newblock {\em American Journal of Political Science}, 39(3):714--744, 1995.
\newblock publisher: [Midwest Political Science Association, Wiley].

\bibitem{Caselli2020Globalization}
Mauro Caselli, Andrea Fracasso, and Silvio Traverso.
\newblock Globalization and electoral outcomes: Evidence from italy.
\newblock {\em Economics \& Politics}, 32(1):68--103, 2020.

\bibitem{Jones2015Economic}
Philip~Edward Jones.
\newblock Economic voting appeals in congressional campaigns.
\newblock {\em Political Communication}, 32(2):206--228, 4 2015.
\newblock publisher: Routledge.

\bibitem{DeSimone2016Rhetoric}
Elina De~Simone and Paulo~Reis Mourao.
\newblock Rhetoric on the economy: have european parties changed their economic messages?
\newblock {\em Applied Economics}, 48(22):2022--2036, 5 2016.
\newblock publisher: Routledge.

\bibitem{Kim1996Determinants}
Kyungmo Kim and George~A. Barnett.
\newblock The determinants of international news flow: A network analysis.
\newblock {\em Communication Research}, 23(3):323--352, 6 1996.
\newblock publisher: SAGE Publications Inc.

\bibitem{Stoycheff2017Priming}
Elizabeth Stoycheff and Erik~C. Nisbet.
\newblock Priming the costs of conflict? russian public opinion about the 2014 crimean conflict.
\newblock {\em International Journal of Public Opinion Research}, 29(4):657--675, 12 2017.

\bibitem{Golan2014Advertorial}
Guy~J. Golan and Evhenia~“Zhenia” Viatchaninova.
\newblock The advertorial as a tool of mediated public diplomacy.
\newblock {\em International Journal of Communication}, 8(0):21, 4 2014.
\newblock number: 0.

\bibitem{Khan2021Public}
M.~Laeeq Khan, Muhammad Ittefaq, Yadira Ixchel~Martínez Pantoja, Muhammad~Mustafa Raziq, and Aqdas Malik.
\newblock Public engagement model to analyze digital diplomacy on twitter: A social media analytics framework.
\newblock {\em International Journal of Communication}, 15(0):29, 3 2021.
\newblock number: 0.

\bibitem{Li2016BRICS|}
Hongmei Li and Leslie~L. Marsh.
\newblock Brics{\textbar} building the brics: Media, nation branding and global citizenship — introduction.
\newblock {\em International Journal of Communication}, 10(0):16, 6 2016.
\newblock number: 0.

\bibitem{Milam2012Apps4Africa:}
Lacey Milam and Elizabeth~Johnson Avery.
\newblock Apps4africa: A new state department public diplomacy initiative.
\newblock {\em Public Relations Review}, 38(2):328--335, 6 2012.

\bibitem{Borchers2011“Do}
Nils~S. Borchers.
\newblock “do you really think russia should pay up for that?”: How the russia-based tv channel rt constructs russian-baltic relations.
\newblock {\em Javnost - The Public}, 18(4):89--106, 1 2011.
\newblock publisher: Routledge.

\bibitem{Kragh2017Russia’s}
Martin Kragh and Sebastian Åsberg.
\newblock Russia’s strategy for influence through public diplomacy and active measures: the swedish case.
\newblock {\em Journal of Strategic Studies}, 40(6):773--816, 9 2017.
\newblock publisher: Routledge.

\bibitem{Alpert2014Kremlin}
Lukas~I. Alpert.
\newblock {\em Kremlin Speak: Inside Putin's Propaganda Factory}.
\newblock Tatra Press, New York, N.Y., 11 2014.

\bibitem{Wagnsson2023paperboys}
Charlotte Wagnsson.
\newblock The paperboys of russian messaging: Rt/sputnik audiences as vehicles for malign information influence.
\newblock {\em Information, Communication \& Society}, 26(9):1849--1867, 7 2023.
\newblock publisher: Routledge.

\bibitem{Carter2021Questioning}
Erin~Baggott Carter and Brett~L. Carter.
\newblock Questioning more: Rt, outward-facing propaganda, and the post-west world order.
\newblock {\em Security Studies}, 30(1):49--78, 1 2021.
\newblock publisher: Routledge.

\bibitem{Knoblock2021Look}
Natalia Knoblock.
\newblock A look at brexit by rt, a russian news source.
\newblock {\em Critical Approaches to Discourse Analysis across Disciplines}, 13(1):107--126, 2021.

\bibitem{Cull2009Public}
Nicholas~J. Cull.
\newblock {\em Public Diplomacy: Lessons from the Past}.
\newblock CPD Perspectives on Public Diplomacy. Figueroa Press, Los Angeles, CA, 2009.

\bibitem{Williamson2012Kelleypd}
William~F. Williamson and John~Robert Kelley.
\newblock Public diplomacy 2.0 classroom.
\newblock {\em Global Media Journal}, (Fall), 2012.

\bibitem{Napoli2017Why}
Philip Napoli and Robyn Caplan.
\newblock Why media companies insist they're not media companies, why they're wrong, and why it matters.
\newblock {\em First Monday}, 5 2017.
\newblock [Online; accessed 2024-03-26].

\bibitem{Fletcher2018Are}
Richard Fletcher and Rasmus~Kleis Nielsen.
\newblock Are people incidentally exposed to news on social media? a comparative analysis.
\newblock {\em New Media \& Society}, 20(7):2450--2468, 7 2018.
\newblock publisher: SAGE Publications.

\bibitem{Stocking2022Role}
Galen Stocking, Amy Mitchell, Katerina~Eva Matsa, Regina Widjaya, Mark Jurkowitz, Shreenita Ghosh, Aaron Smith, Sarah Naseer, and Christopher~St. Aubin.
\newblock The role of alternative social media in the news and information environment.
\newblock Technical report, Washington, D.C., 10 2022.
\newblock [Online; accessed 2023-05-09].

\bibitem{Surma2016Social}
Jerzy Surma.
\newblock Social exchange in online social networks. the reciprocity phenomenon on facebook.
\newblock {\em Computer Communications}, 73:342--346, 1 2016.

\bibitem{Kampf2015Digital}
Ronit Kampf, Ilan Manor, and Elad Segev.
\newblock Digital diplomacy 2.0? a cross-national comparison of public engagement in facebook and twitter.
\newblock {\em The Hague Journal of Diplomacy}, 10(4):331--362, 10 2015.
\newblock publisher: Brill Nijhoff.

\bibitem{Hayden2018Digital}
Craig Hayden.
\newblock Digital diplomacy.
\newblock {\em The encyclopedia of diplomacy}, pages 1--13, 2018.
\newblock publisher: John Wiley \& Sons, Ltd Chichester, West Sussex, UK.

\bibitem{Luqiu2020Weibo}
Luwei~Rose Luqiu and Fan Yang.
\newblock Weibo diplomacy: Foreign embassies communicating on chinese social media.
\newblock {\em Government Information Quarterly}, 37(3):101477, 7 2020.

\bibitem{Crilley2022Understanding}
Rhys Crilley, Marie Gillespie, Bertie Vidgen, and Alistair Willis.
\newblock Understanding rt’s audiences: Exposure not endorsement for twitter followers of russian state-sponsored media.
\newblock {\em The International Journal of Press/Politics}, 27(1):220--242, 1 2022.
\newblock publisher: SAGE Publications Inc.

\bibitem{boyd2010Social}
danah boyd.
\newblock Social network sites as networked publics: Affordances, dynamics, and implications.
\newblock In {\em A Networked Self}. Routledge, 2010.
\newblock number-of-pages: 20.

\bibitem{Norman2013design}
Don Norman.
\newblock {\em The design of everyday things: Revised and expanded edition}.
\newblock Basic books, 2013.

\bibitem{wellmansocial2003}
Barry Wellman, Anabel Quan-Haase, Jeffrey Boase, Wenhong Chen, Keith Hampton, Isabel Díaz, and Kakuko Miyata.
\newblock The {Social} {Affordances} of the {Internet} for {Networked} {Individualism}.
\newblock {\em Journal of Computer-Mediated Communication}, 8(3):JCMC834, April 2003.

\bibitem{Bodle2010Assessing}
Robert Bodle.
\newblock Assessing social network sites as international platforms: Guiding principles.
\newblock {\em The Journal of International Communication}, 16(2):9--24, 1 2010.
\newblock publisher: Routledge.

\bibitem{Ceron2015Internet}
Andrea Ceron.
\newblock Internet, news, and political trust: The difference between social media and online media outlets.
\newblock {\em Journal of Computer-Mediated Communication}, 20(5):487--503, 9 2015.

\bibitem{Wasserman2018How}
Herman Wasserman and Dani Madrid-Morales.
\newblock How influential are chinese media in africa? an audience analysis in kenya and south africa.
\newblock {\em International Journal of Communication}, 12(0):20, 5 2018.
\newblock number: 0.

\bibitem{Elswah2020“Anything}
Mona Elswah and Philip~N Howard.
\newblock “anything that causes chaos”: The organizational behavior of russia today (rt).
\newblock {\em Journal of Communication}, 70(5):623--645, 10 2020.

\bibitem{Fisher2020Demonizing}
Aleksandr Fisher.
\newblock Demonizing the enemy: the influence of russian state-sponsored media on american audiences.
\newblock {\em Post-Soviet Affairs}, 36(4):281--296, 7 2020.
\newblock publisher: Routledge.

\bibitem{Golovchenko2020Cross-Platform}
Yevgeniy Golovchenko, Cody Buntain, Gregory Eady, Megan~A. Brown, and Joshua~A. Tucker.
\newblock Cross-platform state propaganda: Russian trolls on twitter and youtube during the 2016 u.s. presidential election.
\newblock {\em The International Journal of Press/Politics}, 25(3):357--389, 7 2020.

\bibitem{Fitzpatrick2007Advancing}
Kathy Fitzpatrick.
\newblock Advancing the new public diplomacy: A public relations perspective.
\newblock {\em The Hague Journal of Diplomacy}, 2(3):187--211, 1 2007.
\newblock publisher: Brill Nijhoff.

\bibitem{Nye2005Soft}
Joseph~S. Nye.
\newblock {\em Soft Power}.
\newblock PublicAffairs, New York; N.Y, first edition edition, 4 2005.

\bibitem{Morrison2021“Scrounger-bashing”}
James Morrison.
\newblock “scrounger-bashing” as national pastime: the prevalence and ferocity of anti-welfare ideology on niche-interest online forums.
\newblock {\em Social Semiotics}, 31(3):383--401, 5 2021.
\newblock publisher: Routledge.

\bibitem{DiRusso2021Sustainability}
Carlina DiRusso and Jessica~Gall Myrick.
\newblock Sustainability in csr messages on social media: How emotional framing and efficacy affect emotional response, memory and persuasion.
\newblock {\em Environmental Communication}, 15(8):1045--1060, 11 2021.
\newblock publisher: Routledge.

\bibitem{Jovanovic2018Multimodal}
Danica Jovanovic and Theo Van~Leeuwen.
\newblock Multimodal dialogue on social media.
\newblock {\em Social Semiotics}, 28(5):683--699, 10 2018.
\newblock publisher: Routledge.

\bibitem{Yarchi2021Political}
Moran Yarchi, Christian Baden, and Neta Kligler-Vilenchik.
\newblock Political polarization on the digital sphere: A cross-platform, over-time analysis of interactional, positional, and affective polarization on social media.
\newblock {\em Political Communication}, 38(1-2):98--139, 3 2021.
\newblock publisher: Routledge.

\bibitem{Chen2023Comparing}
Bin Chen, Josephine Lukito, and Gyo~Hyun Koo.
\newblock Comparing the \#stopthesteal movement across multiple platforms: Differentiating discourse on facebook, twitter, and parler.
\newblock {\em Social Media + Society}, 9(3):20563051231196879, 7 2023.
\newblock publisher: SAGE Publications Ltd.

\bibitem{McCloskey1998rhetoric}
Deirdre~N. McCloskey.
\newblock {\em The rhetoric of economics}.
\newblock Rhetoric of the human sciences. University of Wisconsin Press, Madison, Wis, 2nd ed edition, 1998.

\bibitem{Aune2002Selling}
James~Arnt Aune.
\newblock {\em Selling the free market: the rhetoric of economic correctness}.
\newblock Guilford, New York; London, 2002.
\newblock OCLC: 49593997.

\bibitem{Brown2002Global}
Richard~Harvey Brown.
\newblock Global capitalism, national sovereignty, and the decline of democratic space.
\newblock {\em Rhetoric \& Public Affairs}, 5(2):347--357, 2002.

\bibitem{Chaput2018Trumponomics}
Catherine Chaput.
\newblock Trumponomics, neoliberal branding, and the rhetorical circulation of affect.
\newblock {\em Advances in the History of Rhetoric}, 21(2):194--209, 5 2018.

\bibitem{Chaput2015Economic}
Catherine Chaput and Joshua~S. Hanan.
\newblock Economic rhetoric as \textit{Taxis}: Neoliberal governmentality and the dispositif of freakonomics.
\newblock {\em Journal of Cultural Economy}, 8(1):42--61, 1 2015.

\bibitem{Avsar2011Mainstream}
Rojhat~B. Avsar.
\newblock Mainstream economic rhetoric, ideology and institutions.
\newblock {\em Journal of Economic Issues}, 45(1):137--158, 3 2011.

\bibitem{Goidel1995Media}
Robert~K. Goidel and Ronald~E. Langley.
\newblock Media coverage of the economy and aggregate economic evaluations: Uncovering evidence of indirect media effects.
\newblock {\em Political Research Quarterly}, 48(2):313--328, 1995.
\newblock publisher: [University of Utah, Sage Publications, Inc.].

\bibitem{Soroka2015It's}
Stuart~N. Soroka, Dominik~A. Stecula, and Christopher Wlezien.
\newblock It's (change in) the (future) economy, stupid: Economic indicators, the media, and public opinion.
\newblock {\em American Journal of Political Science}, 59(2):457--474, 2015.

\bibitem{Hanan2014From}
Joshua~S. Hanan.
\newblock From economic rhetoric to economic imaginaries: A critical genealogy of economic rhetoric in u.s. communication studies.
\newblock In Joshua~S. Hanan and Mark Hayward, editors, {\em Communication and the economy: history, value, and agency}, Frontiers in political communication, pages 67--94. Peter Lang, New York, 2014.

\bibitem{Wood2004Presidential}
B.~Dan Wood.
\newblock Presidential rhetoric and economic leadership.
\newblock {\em Presidential Studies Quarterly}, 34(3):573--606, 9 2004.

\bibitem{Goodnight2010Rhetoric}
G.~Thomas Goodnight and Sandy Green.
\newblock Rhetoric, risk, and markets: The dot-com bubble.
\newblock {\em Quarterly Journal of Speech}, 96(2):115--140, 2010.

\bibitem{Houck2000Rhetoric}
Davis~W Houck.
\newblock Rhetoric as currency: Herbert hoover and the 1929 stock market crash.
\newblock {\em Rhetoric \& Public Affairs}, 3(2):155--181, 2000.

\bibitem{Levasseur2015Not}
Levasseur and Gring-Pemble.
\newblock Not all capitalist stories are created equal: Mitt romney’s bain capital narrative and the deep divide in american economic rhetoric.
\newblock {\em Rhetoric \& Public Affairs}, 18(1):1, 2015.

\bibitem{Crable1983Argumentative}
Richard~E. Crable and Steven~L. Vibbert.
\newblock Argumentative stance and political faith healing: “the dream will come true”.
\newblock {\em Quarterly Journal of Speech}, 69(3):290--301, 8 1983.

\bibitem{Zarefsky1979great}
David Zarefsky.
\newblock The great society as a rhetorical proposition.
\newblock {\em Quarterly Journal of Speech}, 65(4):364--378, 12 1979.

\bibitem{Lebovics1992Economic}
Herman Lebovics.
\newblock Economic positivism as rhetoric.
\newblock {\em International Review of Social History}, 37(2):244--251, 8 1992.

\bibitem{Khalitova2020He}
Liudmila Khalitova, Barbara Myslik, Agnieszka Turska-Kawa, Sofiya Tarasevich, and Spiro Kiousis.
\newblock He who pays the piper, calls the tune? examining russia’s and poland’s public diplomacy efforts to shape the international coverage of the smolensk crash.
\newblock {\em Public Relations Review}, 46(2):101858, 6 2020.

\bibitem{Grincheva2016BRICS}
Natalia Grincheva and Jiayi Lu.
\newblock Brics summit diplomacy: Constructing national identities through russian and chinese media coverage of the fifth brics summit in durban, south africa.
\newblock {\em Global Media and Communication}, 12(1):25--47, 4 2016.
\newblock publisher: SAGE Publications.

\bibitem{Statista2022Number}
Statista.
\newblock Number of facebook users in russia from 2018 to 2027 (in millions) [graph], 6 2022.

\bibitem{Fortner1994Public}
Robert~S. Fortner.
\newblock {\em Public Diplomacy and International Politics: The Symbolic Constructs of Summits and International Radio News}.
\newblock Praeger, Westport, Conn, 4 1994.

\bibitem{Krasnyak2020Russian}
Olga Krasnyak.
\newblock Russian science diplomacy.
\newblock {\em Diplomatica}, 2(1):118--134, 5 2020.
\newblock publisher: Brill.

\bibitem{Kayser2011Performance}
Mark~Andreas Kayser and Christopher Wlezien.
\newblock Performance pressure: Patterns of partisanship and the economic vote.
\newblock {\em European Journal of Political Research}, 50(3):365--394, 2011.

\bibitem{Lewis-Beck1988Economics}
Michael~S. Lewis-Beck.
\newblock Economics and the american voter: Past, present, future.
\newblock {\em Political Behavior}, 10(1):5--21, 1988.
\newblock publisher: Springer.

\bibitem{Lewis-Beck2007Economic}
Michael~S. Lewis-Beck and Mary Stegmaier.
\newblock Economic models of voting.
\newblock In Russell~J. Dalton and Hans‐Dieter Klingemann, editors, {\em The Oxford Handbook of Political Behavior}, page~0. Oxford University Press, 8 2007.
\newblock DOI: 10.1093/oxfordhb/9780199270125.003.0027.

\bibitem{LewisBeck2000Economic}
Michael~S. Lewis-Beck and Mary Stegmaier.
\newblock Economic determinants of electoral outcomes.
\newblock {\em Annual Review of Political Science}, 3(1):183--219, 2000.

\bibitem{Linn2010Economics}
Suzanna Linn, Jonathan Nagler, and Marco~A. Morales.
\newblock Economics, elections, and voting behavior.
\newblock In Jan~E. Leighley, editor, {\em The Oxford Handbook of American Elections and Political Behavior}, page~0. Oxford University Press, 2 2010.
\newblock DOI: 10.1093/oxfordhb/9780199235476.003.0020.

\bibitem{Bolsen2008Polls—Trends:}
Toby Bolsen and Fay~Lomax Cook.
\newblock The polls—trends: Public opinion on energy policy: 1974–2006.
\newblock {\em Public Opinion Quarterly}, 72(2):364--388, 1 2008.

\bibitem{Farhar1980Public}
B~C Farhar, C~T Unseld, R~Vories, and R~Crews.
\newblock Public opinion about energy.
\newblock {\em Annual Review of Energy}, 5(1):141--172, 1980.

\bibitem{Vidigal2022Issue}
Robert Vidigal and Jennifer Jerit.
\newblock Issue importance and the correction of misinformation.
\newblock {\em Political Communication}, 39(6):715--736, 11 2022.
\newblock publisher: Routledge.

\bibitem{Mustoe2019How}
Mustoe.
\newblock How does brexit affect the pound?
\newblock {\em BBC News}, 10 2019.
\newblock [Online; accessed 2023-09-22].

\bibitem{Putin2024Vladimir}
Vladimir Putin.
\newblock Tucker carlson vladimir putin interview, 2 2024.
\newblock [Online; accessed 2024-03-25].

\bibitem{Hughes2013Politics}
Llewelyn Hughes and Phillip~Y. Lipscy.
\newblock The politics of energy.
\newblock {\em Annual Review of Political Science}, 16(1):449--469, 2013.

\bibitem{Chen2016Impacts}
Hao Chen, Hua Liao, Bao-Jun Tang, and Yi-Ming Wei.
\newblock Impacts of opec's political risk on the international crude oil prices: An empirical analysis based on the svar models.
\newblock {\em Energy Economics}, 57:42--49, 6 2016.

\bibitem{Giraud1995equilibrium}
Pierre-Noël Giraud.
\newblock The equilibrium price range of oil: Economics, politics and uncertainty in the formation of oil prices.
\newblock {\em Energy Policy}, 23(1):35--49, 1 1995.

\bibitem{Grootendorst2022BERTopic}
Maarten Grootendorst.
\newblock Bertopic, 2022.
\newblock original-date: 2020-09-22T14:19:29Z.

\bibitem{Wolf2019HuggingFace's}
Thomas Wolf, Lysandre Debut, Victor Sanh, Julien Chaumond, Clement Delangue, Anthony Moi, Pierric Cistac, Tim Rault, Rémi Louf, Morgan Funtowicz, Joe Davison, Sam Shleifer, Patrick von Platen, Clara Ma, Yacine Jernite, Julien Plu, Canwen Xu, Teven~Le Scao, Sylvain Gugger, Mariama Drame, Quentin Lhoest, and Alexander~M. Rush.
\newblock Huggingface's transformers: State-of-the-art natural language processing.
\newblock 2019.
\newblock publisher: arXiv version: 5.

\bibitem{TheInternationalMonetaryFund(IMF)2023Exchange}
The International Monetary~Fund (IMF).
\newblock Exchange rate report wizard, 4 2023.
\newblock [Online; accessed 2023-04-11].

\bibitem{Gehebreyesus2020WHO}
Tedros~Adhanom Gehebreyesus.
\newblock Who director-general's opening remarks at the media briefing on covid-19, 3 2020.
\newblock [Online; accessed 2023-05-02].

\bibitem{CenterforPreventiveAction2023War}
Center for Preventive~Action.
\newblock War in ukraine, 3 2023.
\newblock [Online; accessed 2023-05-02].

\bibitem{Lokmanoglu2023Social}
Ayse~D. Lokmanoglu, Erik~C. Nisbet, Matthew~T. Osborne, Joseph Tien, Sam Malloy, Lourdes Cueva~Chacón, Esteban Villa~Turek, and Rod Abhari.
\newblock Social media sentiment about covid-19 vaccination predicts vaccine acceptance among peruvian social media users the next day.
\newblock {\em Vaccines}, 11(4):817, 4 2023.
\newblock number: 4 publisher: Multidisciplinary Digital Publishing Institute.

\bibitem{Lukito2020Coordinating}
Josephine Lukito.
\newblock Coordinating a multi-platform disinformation campaign: Internet research agency activity on three u.s. social media platforms, 2015 to 2017.
\newblock {\em Political Communication}, 37(2):238--255, 3 2020.
\newblock publisher: Routledge.

\bibitem{Toda1995Statistical}
Hiro~Y. Toda and Taku Yamamoto.
\newblock Statistical inference in vector autoregressions with possibly integrated processes.
\newblock {\em Journal of Econometrics}, 66(1):225--250, 3 1995.

\bibitem{Ophir2021Framing}
Yotam Ophir, Dror Walter, Daniel Arnon, Ayse~D. Lokmanoglu, Michele Tizzoni, Joëlle Carota, Lorenzo D'Antiga, and Emanuele Nicastro.
\newblock The framing of covid-19 in italian media and its relationship with community mobility: A mixed-method approach.
\newblock {\em Journal of Health Communication}, 26(3):161--173, 3 2021.

\bibitem{Lutkepohl2010Impulse}
Helmut Lütkepohl.
\newblock Impulse response function.
\newblock In Steven~N. Durlauf and Lawrence~E. Blume, editors, {\em Macroeconometrics and Time Series Analysis}, pages 145--150. Palgrave Macmillan UK, London, 2010.

\bibitem{Johansen1991Estimation}
Søren Johansen.
\newblock Estimation and hypothesis testing of cointegration vectors in gaussian vector autoregressive models.
\newblock {\em Econometrica}, 59(6):1551--1580, 1991.
\newblock publisher: [Wiley, Econometric Society].

\bibitem{Box1970Distribution}
G.~E.~P. Box and David~A. Pierce.
\newblock Distribution of residual autocorrelations in autoregressive-integrated moving average time series models.
\newblock {\em Journal of the American Statistical Association}, 65(332):1509--1526, 12 1970.
\newblock publisher: Taylor \& Francis.

\bibitem{Berinsky2023Political}
Adam~J. Berinsky.
\newblock {\em Political Rumors: Why We Accept Misinformation and How to Fight It}.
\newblock Princeton University Press, 8 2023.

\bibitem{Bhattacharya2023Interest}
Rina Bhattacharya, Dmitry Vasilyev, and Mauricio Villafuerte.
\newblock Interest in central bank digital currencies picks up in latin america and the caribbean while crypto use varies, 6 2023.
\newblock [Online; accessed 2023-12-22].

\bibitem{ChainalysisTeam2023Latin}
Chainalysis Team.
\newblock Latin america cryptocurrency adoption: Data and analysis, 10 2023.
\newblock [Online; accessed 2023-12-22].

\bibitem{Murray2024cult}
Christine Murray.
\newblock The ‘cult’ of bukele: El salvador’s bitcoin-loving strongman heads for second term, 2 2024.
\newblock [Online; accessed 2024-02-05].

\bibitem{Indolfi2020Outbreak}
Ciro Indolfi and Carmen Spaccarotella.
\newblock The outbreak of covid-19 in italy: Fighting the pandemic.
\newblock {\em JACC Case Reports}, 2(9):1414, 7 2020.
\newblock publisher: Elsevier PMID: 32835287.

\end{thebibliography}
\clearpage
\section{Supplementary Material}
\beginsupplement
\subsection{Supplement A: Economic Search Words}
\begin{table}[h!]
\begin{tabularx}{\textwidth}{lXX}
% \begin{tabularx}{\textwidth}{l >{\hsize=.5\hsize}X >{\hsize=1.5\hsize}X}
\toprule
\textbf{Language} & \textbf{Terms} \\
\midrule
Arabic  & 'ajūr, aṣl, amwāl - nuqūd, ʾūsd, istithmār, istithmārāt, iqtiṣād, iqtisādī, al-'ajūr, al-'arbāḥ, al-'uṣūl, al-amwāl - al-nuqūd, al-i'timān, al-i'timānī, al-ittiḥād al-'urūbī, al-iḥtiyāṭī al-fidirālī, al-istithmār, al-istithmārāt, al-baṭālah, al-bank al-markazī, al-būrṣah, al-taḍkhīm, al-junayh, al-junayh al-istirlīnī, al-dulārāt, al-dayn, al-duyūn, al-rātib, al-ribḥ - al-arbāḥ, al-rahn al-'aqārī, al-rawātib, al-rūbal, al-sanad al-mālī, al-sanadāt, al-siyāsah al-naqdīyah, al-ḍarībah - al-ḍarā'ib, al-'uqūbāt, al-'uqūbah, al-'amal, al-'umalāt, al-'umlah, al-'umlah al-mu'shfarah, al-qarḍ, al-qurūḍ, al-līrah al-turkīyah, al-makhzūn, al-hryvnyā, al-yūrū, baṭālah, bank markazī, tabādul, taḥṣīl al-ḍarā'ib, taḥwīl, takhfīḍ qīmat al-'umlah, taḍkhīm iqtisādī, junayh aw ratl lilwazn, dūlār, dūlār amrīkī, dūlārāt, dayn, duyūn, ra's māl, ru'ūs amwāl, ribḥ - arbāḥ, raṣīd, rūbal, rūbal rūsī, riyāl sa'ūdī, s'ir al-taḥwīl, s'ir al-ṣarf, s'ir al-fā'ida, sanad mālī, sanadāt, sūq al-'ashum, sūq al-'awrāq al-mālīyah, sūq al-ṣarf al-'ajnabī, sūq ṣarf al-'umalāt al-'ajnabīyah, siyāsat naqdīyah, ḍarībah - ḍarā'ib, 'uqūbāt, 'uqūbah, 'umalāt, 'umlah, 'umlah mu'shfarah, 'umlah m'umma'ah, qarḍ, qurūḍ, līrah - līrāt, mālī, mālīyah, makhāzin, naqdī, hryvna 'ukrānīyah, hryvnya, yūrū \\
English & Fed, Federal Reserve Bank, Saudi Riyal, asset, assets, british pound, capital, central bank, credit, crypto-currency, cryptocurrency, currencies, currency, debt, devaluation, dollar, dollars, economic, economy, euro, euros, exchange, exchange rate, finance, foreign exchange market, hryvnia, hryvnya, inflation, interest rate, investment, investments, labor, lira, loan, loans, monetary, monetary policy, money, mortgage, pound, pounds, profit, ruble, russian ruble, sanction, sanctions, stock, stock market, stocks, tax, taxation, unemployment, usd, wage, wages \\
Spanish & Banco Central, Banco de la Reserva Federal, Dólar estadounidense, Finanzas, bolsa, bolsa de Valores, capital, criptomoneda, criptomoneda, crédito, desempleo, deuda, devaluación, dinero, dólar, dólares, economía, económico, euro, euros, ganancia, grivna, hipoteca, impuesto, inflación, intercambio, inversiones, inversión, la política monetaria, libra, libra esterlina, libras, lira, lucro, mano de obra, mercado \\
French & Banque centrale, Banque de réserve fédérale, Bourse, Capitale, Imposition, bourse, bénéfice, chômage, crypto-monnaie, crédit, dette, devise, devises, dollar, dollars, dévaluation, euro, euros, finance, grivna, hryvnie, hypothèque, impôt, inflation, investissement, investissements, l'argent, l'échange, la main d'oeuvre, les salaires, les sanctions, lire, livre, livre sterlling, livres, marché des changes, monétaire, monnaie, politique monétaire, profit, prêt, prêts, riyal saoudien, rouble, rouble russe, salaire, sanction, taux d'intérêt, taux de change, travail, échange, économie, économique \\
German & Abwertung, Aktienmarkt, Anlage, Arbeit, Arbeitslosigkeit, Austausch, Besteuerung, Britisches Pfund, Bundesanleihe, Bundesanleihen, Bundesreservebank, Börse, Devisenmarkt, Entwertung, Euro, Finanzen, Geld, Geldpolitik, Gewinn, Griwna, Hypothek, Inflation, Investition, Investitionen, Kapital, Kredit, Kryptowährung, Lira, Lohn, Löhne, Pfund, Profit, Rubel, Sanktion, Sanktionen, Schuld, Steuer, Tauschrate, Vermögenswerte, Wirtschaft, Währung, Währungen, Zentralbank, Zinssatz, das Darlehen, die Darlehen, monetär, riyal saudí, russischer Rubel, wirtschaftlich \\
\bottomrule
\multicolumn{2}{p{\textwidth}}{\footnotesize[1] Due to Overleaf interface we have transliterated the Arabic words with IJMES standard, you can find the original Arabic keywords in the Github page.} \\
\caption{Multilingual Financial Terms}\label{tab:tables1}
\end{tabularx}
\end{table}
\clearpage
\subsection{Supplement B: BERTopic Results}

\begin{table}[htbp!]
\keepXColumns
\begin{tabularx}{\textwidth}{>{\hsize=.5\hsize}X>{\hsize=.1\hsize}X>{\hsize=1.4\hsize}X>{\hsize=1\hsize}X}
\toprule
\textbf{Topic Label} & \textbf{Post Count} & \textbf{Top 10 Words} & \textbf{BERTopic Label} \\
\midrule
Recent News Headlines & 20,991 & lire; dólares; france; millones; millones; dólares; plus; euros; rejoignez; euro; contre; & \texttt{0\_lire\_dólares\_france\_millones} \\
US-Russia-Ukraine Sanctions & 11,056 & sanctions; us; rusia; ukraine; sanciones; china; russie; russe; russia; ucrania; & \texttt{1\_sanctions\_us\_rusia\_ukraine} \\
Covid Pandemic and Related Issues & 2,671 & covid; coronavirus; lire; sanitaire; pass; contre; france; plus; vaccination; pass; sanitaire; & \texttt{2\_covid\_coronavirus\_lire\_sanitaire} \\
Bitcoin Cryptocurrency Value & 1,346 & dólares; bitcóin; criptomoneda; bitcoin; valor; precio; millones; millones; dólares; musk; cryptocurrency; & \texttt{3\_dólares\_bitcóin\_criptomoneda\_bitcoin} \\
City Crisis Protest and Disaster & 1,721 & capital; personas; ciudad; menos; policía; manifestantes; país; calles; heridos; centro; & \texttt{4\_capital\_personas\_ciudad\_menos} \\
Middle East Wars and Conflict & 400 & afghanistan; taliban; kabul; us; afganistán; talibanes; capital; afghan; kaboul; & \texttt{5\_afghanistan\_taliban\_kabul\_us} \\
\bottomrule
\multicolumn{4}{p{\textwidth}}{\footnotesize Note: Outlier Count = 1} \\
\caption{BERTopic Results}\label{tab:tables2}
\end{tabularx}
\end{table}

\begin{figure}
  \centering
  \includegraphics[width=0.8\textwidth]{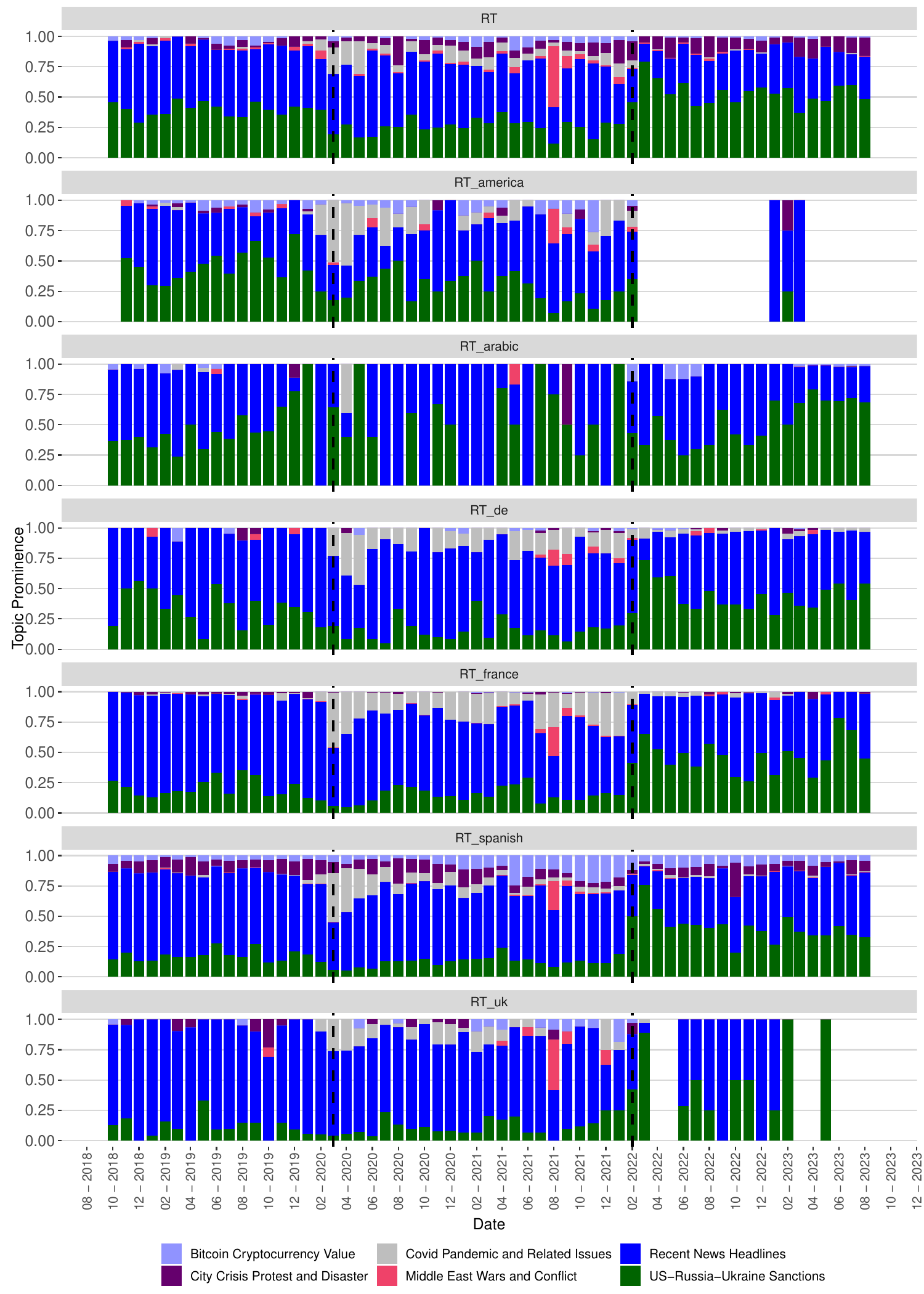} % Adjust the filename and path as necessary
  \caption{RT Pages BERTopic over Time.}
  \label{fig:figS1}
\end{figure}

\clearpage
\subsection{Supplement C: Vector Autoregressive Analysis (VAR)}
\begin{figure}[htbp]
  \centering
  \includegraphics[width=0.8\textwidth]{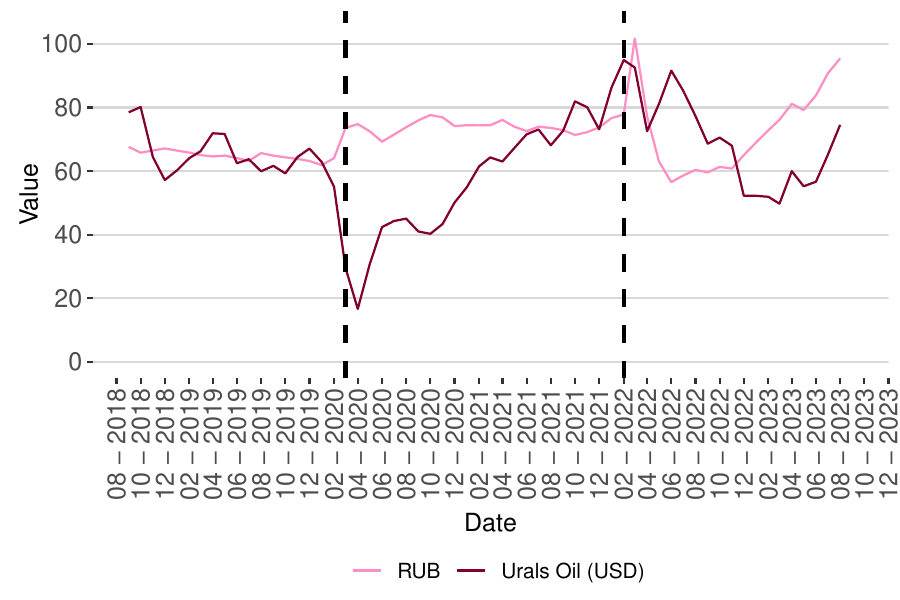} % Adjust the filename and path as necessary
  \caption{Independent Variables over Time.}
  \label{fig:figS2}
\end{figure}

\clearpage
\subsubsection{ACF \& PACF Post-Differencing Variables}

\textit{\textbf{ACF \& PACF: RT}}
\begin{figure}[htbp]
  \centering
  \includegraphics[width=0.8\textwidth]{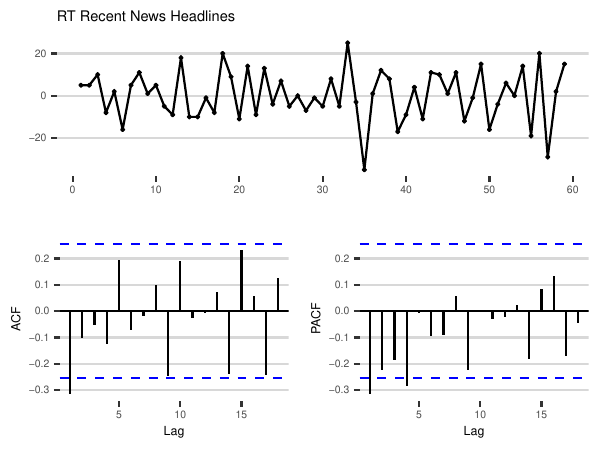} % Adjust the filename and path as necessary
  \caption{ACF RT Recent News Headlines.}
  \label{fig:figS3}
\end{figure}
\begin{figure}[htbp]
  \centering
  \includegraphics[width=0.8\textwidth]{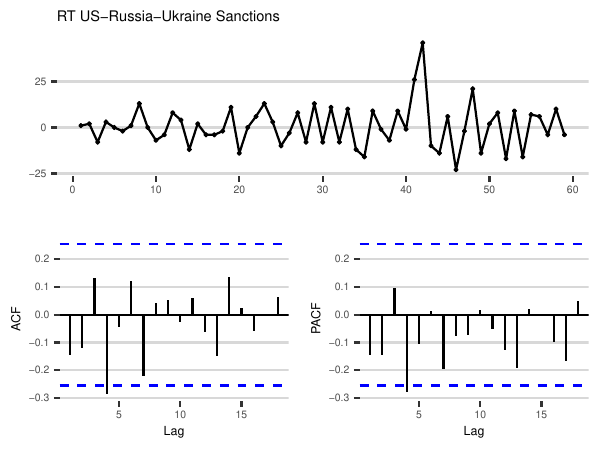} % Adjust the filename and path as necessary
  \caption{ACF RT US-Russia-Ukraine Sanctions.}
  \label{fig:figS4}
\end{figure}
\begin{figure}[htbp]
  \centering
  \includegraphics[width=0.8\textwidth]{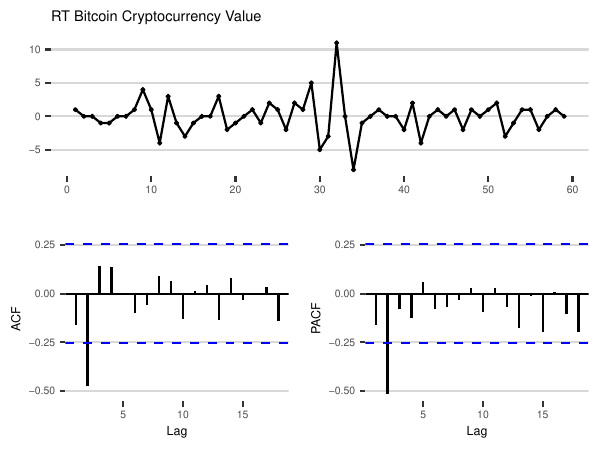} % Adjust the filename and path as necessary
  \caption{ACF RT Bitcoin Cryptocurrency Value.}
  \label{fig:figS5}
\end{figure}
\begin{figure}[htbp]
  \centering
  \includegraphics[width=0.8\textwidth]{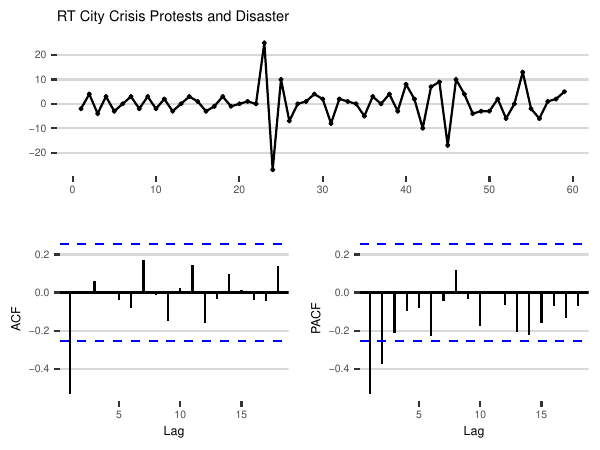} % Adjust the filename and path as necessary
  \caption{ACF RT City Crisis Protests and Disaster.}
  \label{fig:figS6}
\end{figure}
\begin{figure}[htbp]
  \centering
  \includegraphics[width=0.8\textwidth]{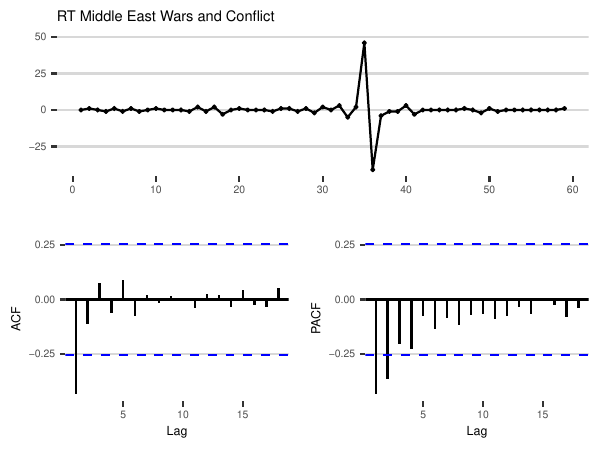} % Adjust the filename and path as necessary
  \caption{ACF RT Middle East Wars and Conflict.}
  \label{fig:figS7}
\end{figure}
\begin{figure}[htbp]
  \centering
  \includegraphics[width=0.8\textwidth]{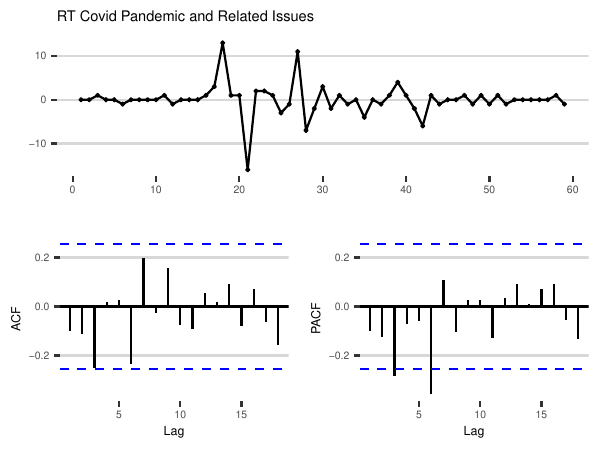} % Adjust the filename and path as necessary
  \caption{ACF RT Covid Pandemic and Related Issues.}
  \label{fig:figS8}
\end{figure}
\clearpage
\textit{\textbf{ACF \& PACF: RT America}}
\begin{figure}[htbp]
  \centering
  \includegraphics[width=0.8\textwidth]{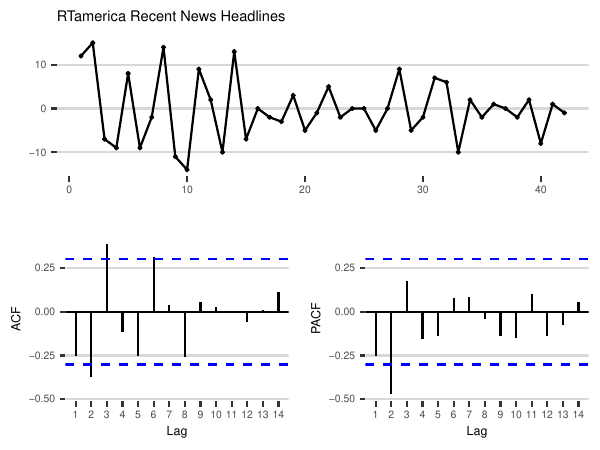} % Adjust the filename and path as necessary
  \caption{ACF RT America Recent News Headlines.}
  \label{fig:figS9}
\end{figure}
\begin{figure}[htbp]
  \centering
  \includegraphics[width=0.8\textwidth]{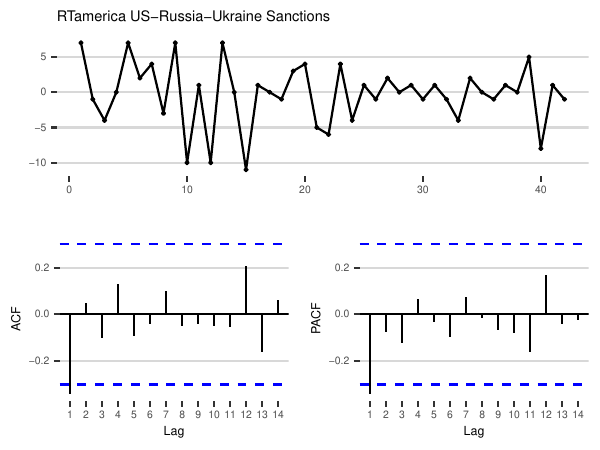} % Adjust the filename and path as necessary
  \caption{ACF RT America US-Russia-Ukraine Sanctions.}
  \label{fig:figS10}
\end{figure}
\begin{figure}[htbp]
  \centering
  \includegraphics[width=0.8\textwidth]{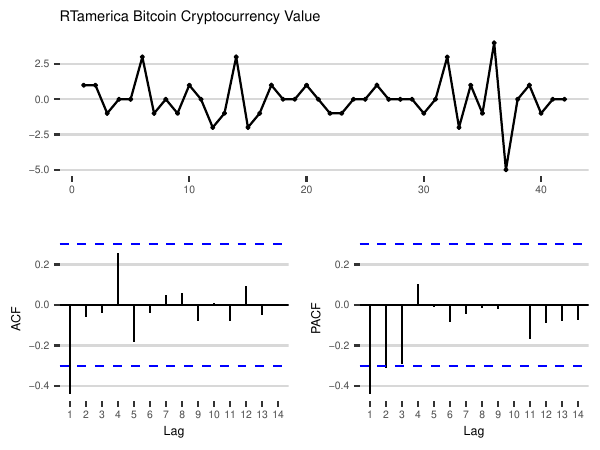} % Adjust the filename and path as necessary
  \caption{ACF RT America Bitcoin Cryptocurrency Value.}
  \label{fig:figS11}
\end{figure}
\begin{figure}[htbp]
  \centering
  \includegraphics[width=0.8\textwidth]{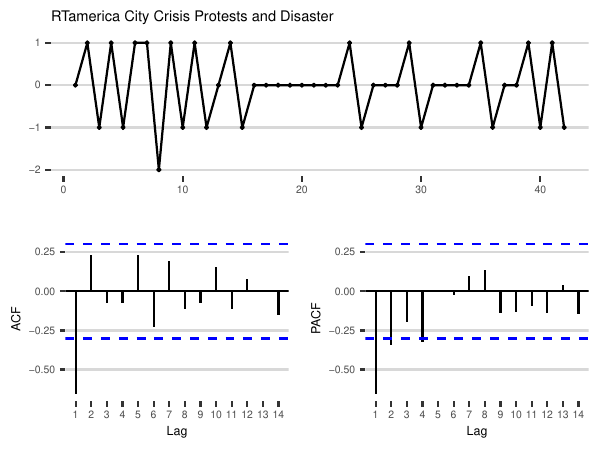} % Adjust the filename and path as necessary
  \caption{ACF RT City Crisis Protests and Disaster.}
  \label{fig:figS12}
\end{figure}
\begin{figure}[htbp]
  \centering
  \includegraphics[width=0.8\textwidth]{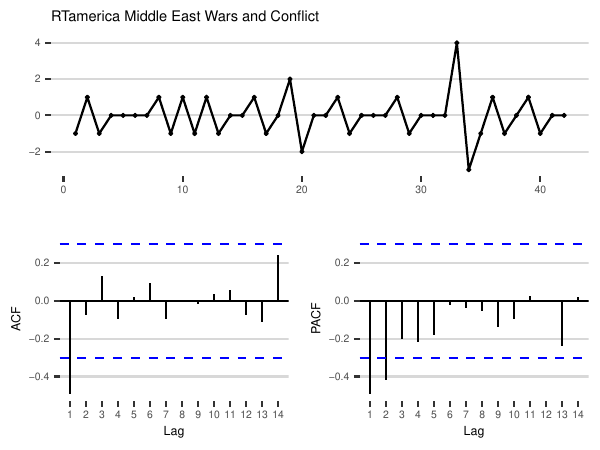} % Adjust the filename and path as necessary
  \caption{ACF RT Middle East Wars and Conflict.}
  \label{fig:figS13}
\end{figure}
\begin{figure}[htbp]
  \centering
  \includegraphics[width=0.8\textwidth]{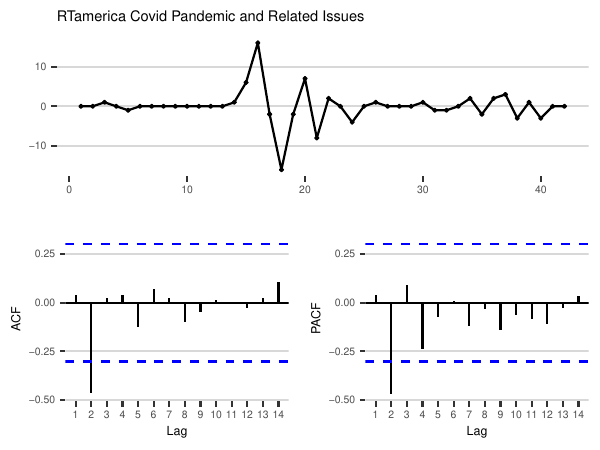} % Adjust the filename and path as necessary
  \caption{ACF RT Covid Pandemic and Related Issues.}
  \label{fig:figS14}
\end{figure}
\clearpage
\textbf{\textit{ACF \& PACF: RT Arabic}}
\begin{figure}[htbp]
  \centering
  \includegraphics[width=0.8\textwidth]{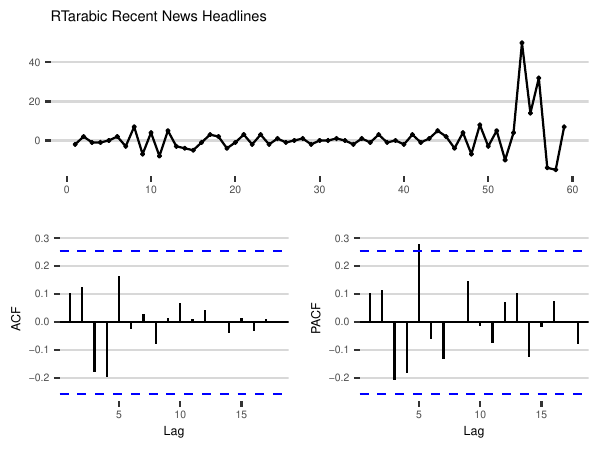} % Adjust the filename and path as necessary
  \caption{ACF RT Arabic Recent News Headlines.}
  \label{fig:figS15}
\end{figure}
\begin{figure}[htbp]
  \centering
  \includegraphics[width=0.8\textwidth]{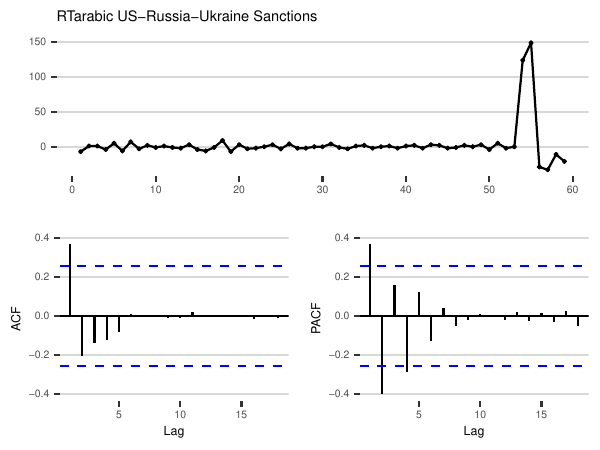} % Adjust the filename and path as necessary
  \caption{ACF RT Arabic US-Russia-Ukraine Sanctions.}
  \label{fig:figS16}
\end{figure}
\begin{figure}[htbp]
  \centering
  \includegraphics[width=0.8\textwidth]{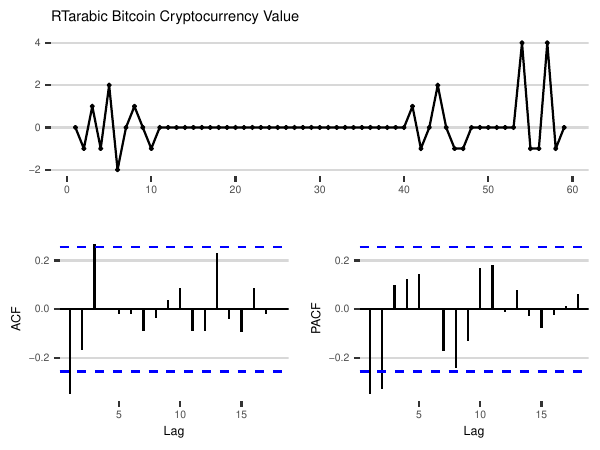} % Adjust the filename and path as necessary
  \caption{ACF RT Arabic Bitcoin Cryptocurrency Value.}
  \label{fig:figS17}
\end{figure}
\begin{figure}[htbp]
  \centering
  \includegraphics[width=0.8\textwidth]{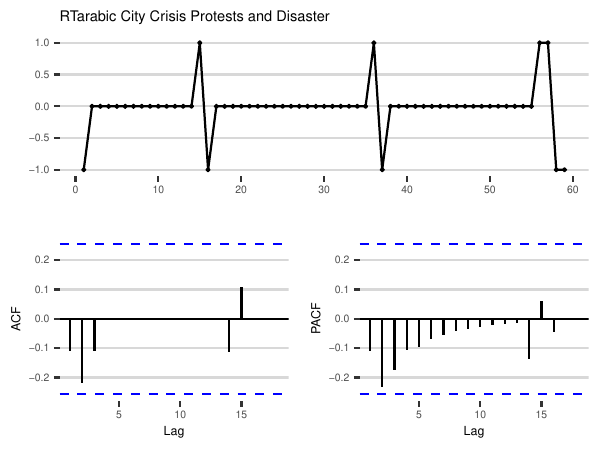} % Adjust the filename and path as necessary
  \caption{ACF RT Arabic City Crisis Protests and Disaster.}
  \label{fig:figS18}
\end{figure}
\begin{figure}[htbp]
  \centering
  \includegraphics[width=0.8\textwidth]{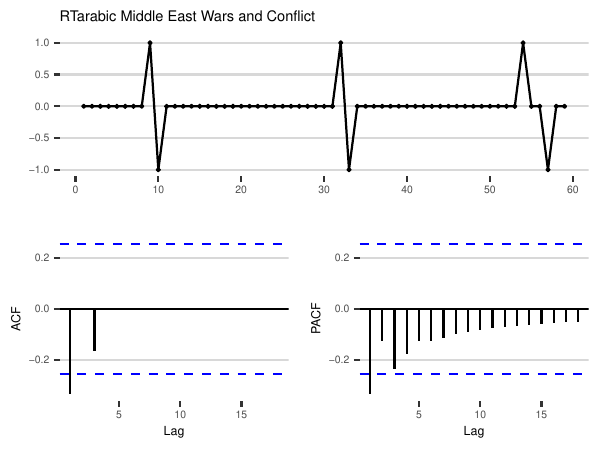} % Adjust the filename and path as necessary
  \caption{ACF RT Arabic Middle East Wars and Conflict.}
  \label{fig:figS19}
\end{figure}
\begin{figure}[htbp]
  \centering
  \includegraphics[width=0.8\textwidth]{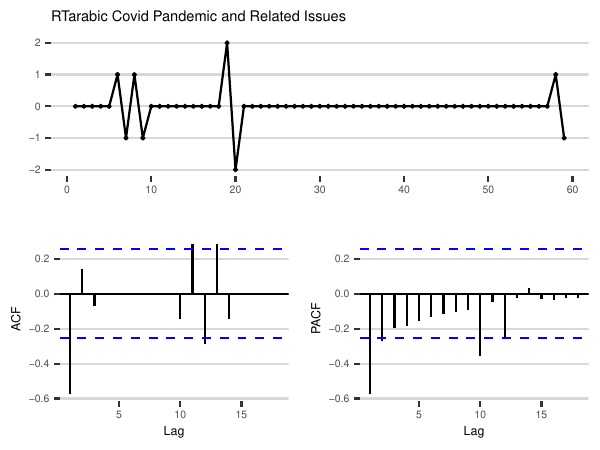} % Adjust the filename and path as necessary
  \caption{ACF RT Arabic Covid Pandemic and Related Issues.}
  \label{fig:figS20}
\end{figure}
\clearpage
\textbf{\textit{ACF \& PACF: RT DE}}
\begin{figure}[htbp]
  \centering
  \includegraphics[width=0.8\textwidth]{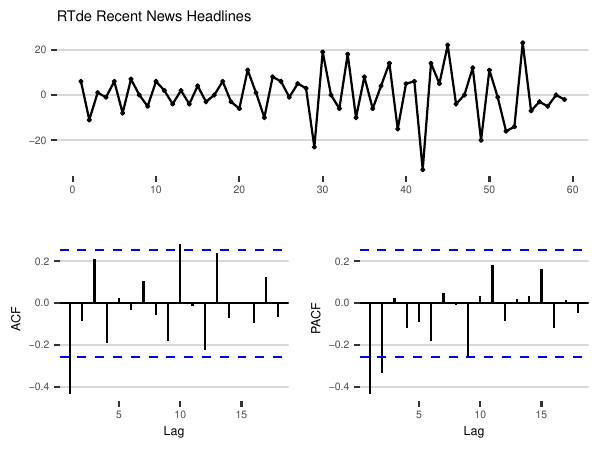} % Adjust the filename and path as necessary
  \caption{ACF RT DE Recent News Headlines.}
  \label{fig:figS21}
\end{figure}
\begin{figure}[htbp]
  \centering
  \includegraphics[width=0.8\textwidth]{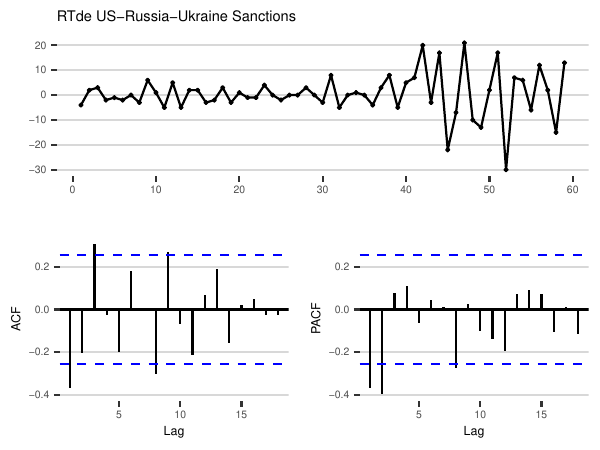} % Adjust the filename and path as necessary
  \caption{ACF RT DE US-Russia-Ukraine Sanctions.}
  \label{fig:figS22}
\end{figure}
\begin{figure}[htbp]
  \centering
  \includegraphics[width=0.8\textwidth]{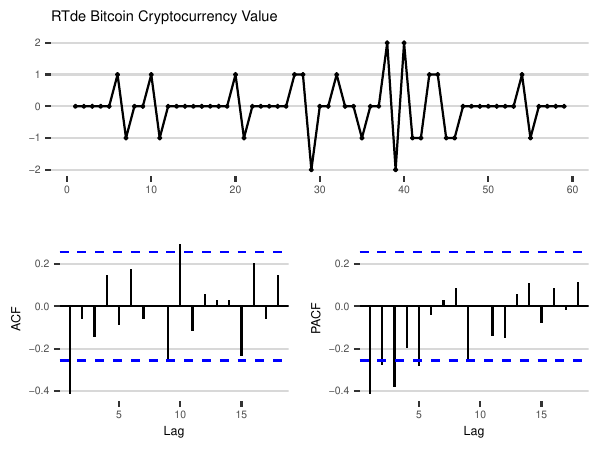} % Adjust the filename and path as necessary
  \caption{ACF RT DE Bitcoin Cryptocurrency Value.}
  \label{fig:figS23}
\end{figure}
\begin{figure}[htbp]
  \centering
  \includegraphics[width=0.8\textwidth]{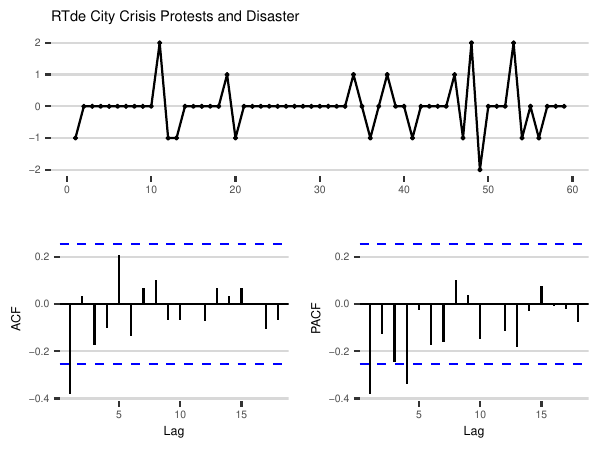} % Adjust the filename and path as necessary
  \caption{ACF RT DE City Crisis Protests and Disaster.}
  \label{fig:figS24}
\end{figure}
\begin{figure}[htbp]
  \centering
  \includegraphics[width=0.8\textwidth]{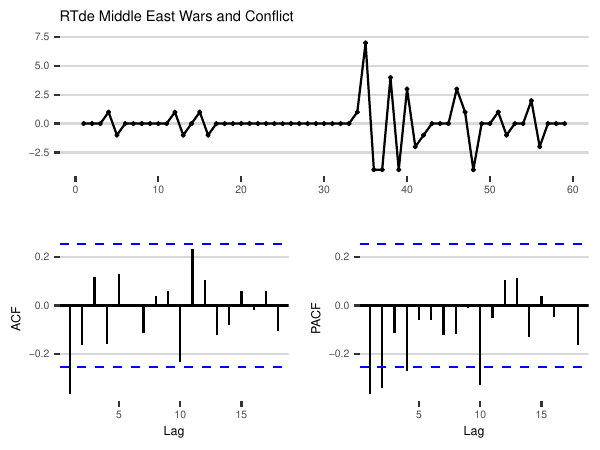} % Adjust the filename and path as necessary
  \caption{ACF RT DE Middle East Wars and Conflict.}
  \label{fig:figS25}
\end{figure}
\begin{figure}[htbp]
  \centering
  \includegraphics[width=0.8\textwidth]{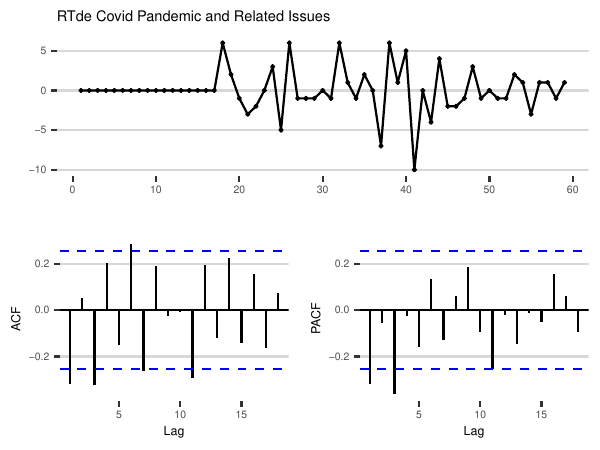} % Adjust the filename and path as necessary
  \caption{ACF RT DE Covid Pandemic and Related Issues.}
  \label{fig:figS26}
\end{figure}
\clearpage
\textbf{\textit{ACF \& PACF: RT France}}
\begin{figure}[htbp]
  \centering
  \includegraphics[width=0.8\textwidth]{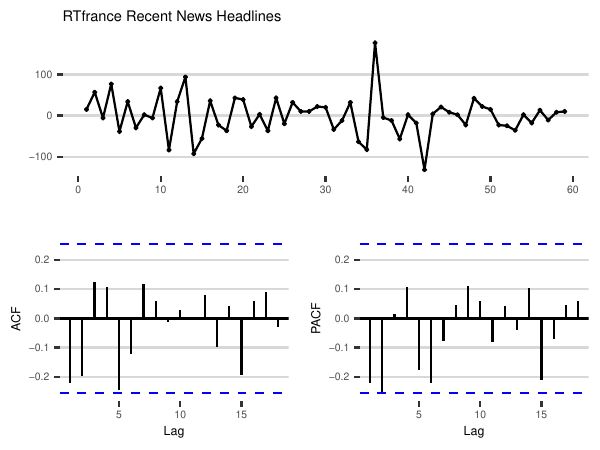} % Adjust the filename and path as necessary
  \caption{ACF RT France Recent News Headlines.}
  \label{fig:figS27}
\end{figure}
\begin{figure}[htbp]
  \centering
  \includegraphics[width=0.8\textwidth]{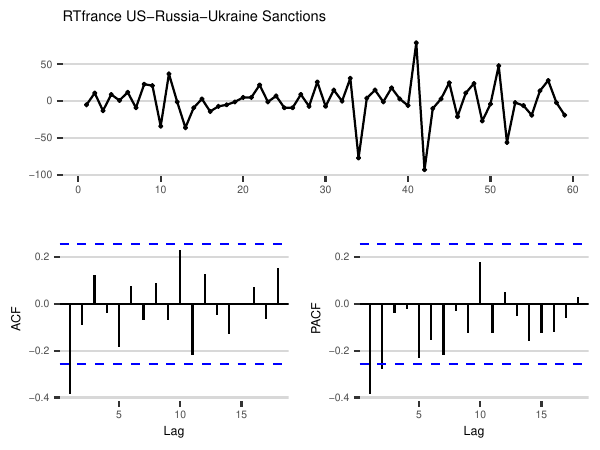} % Adjust the filename and path as necessary
  \caption{ACF RT France US-Russia-Ukraine Sanctions.}
  \label{fig:figS28}
\end{figure}
\begin{figure}[htbp]
  \centering
  \includegraphics[width=0.8\textwidth]{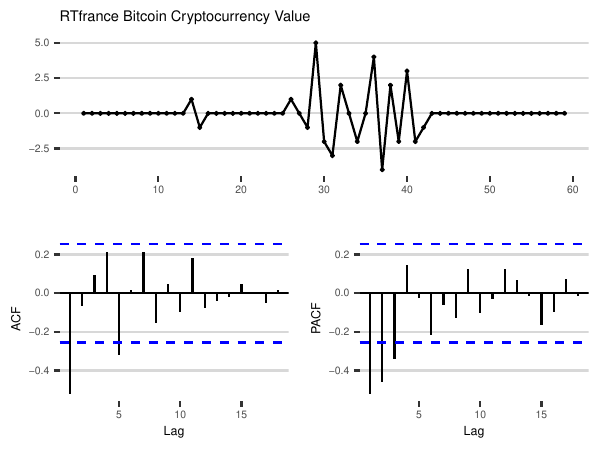} % Adjust the filename and path as necessary
  \caption{ACF RT France Bitcoin Cryptocurrency Value.}
  \label{fig:figS29}
\end{figure}
\begin{figure}[htbp]
  \centering
  \includegraphics[width=0.8\textwidth]{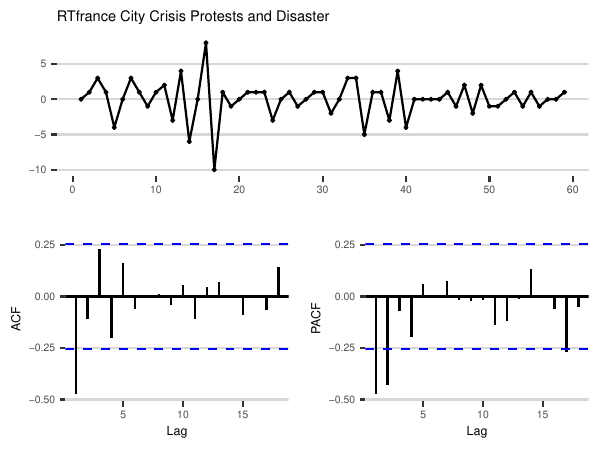} % Adjust the filename and path as necessary
  \caption{ACF RT France City Crisis Protests and Disaster.}
  \label{fig:figS30}
\end{figure}
\begin{figure}[htbp]
  \centering
  \includegraphics[width=0.8\textwidth]{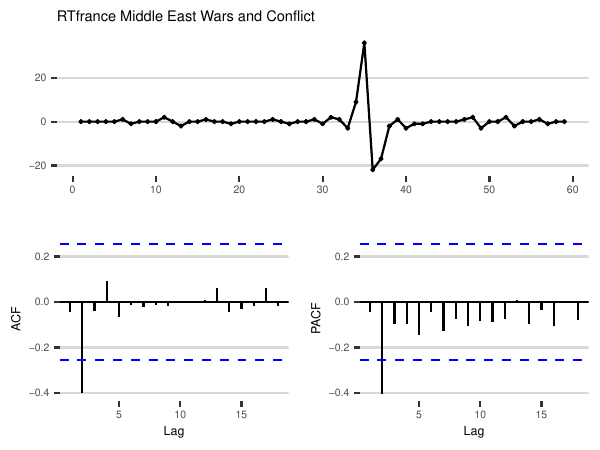} % Adjust the filename and path as necessary
  \caption{ACF RT France Middle East Wars and Conflict.}
  \label{fig:figS31}
\end{figure}
\begin{figure}[htbp]
  \centering
  \includegraphics[width=0.8\textwidth]{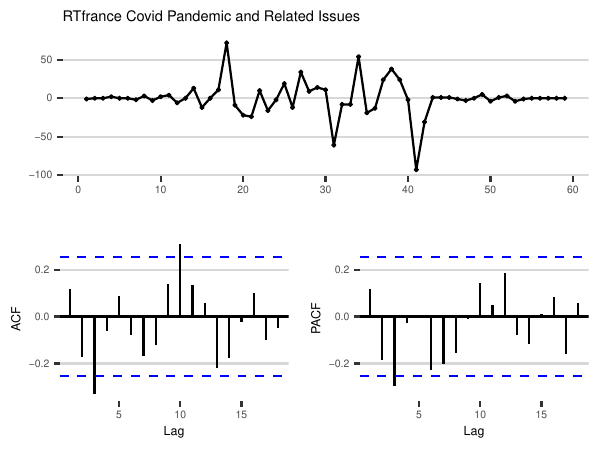} % Adjust the filename and path as necessary
  \caption{ACF RT France Covid Pandemic and Related Issues.}
  \label{fig:figS32}
\end{figure}
\clearpage
\textbf{\textit{ACF \& PACF: RT Spanish}}
\begin{figure}[htbp]
  \centering
  \includegraphics[width=0.8\textwidth]{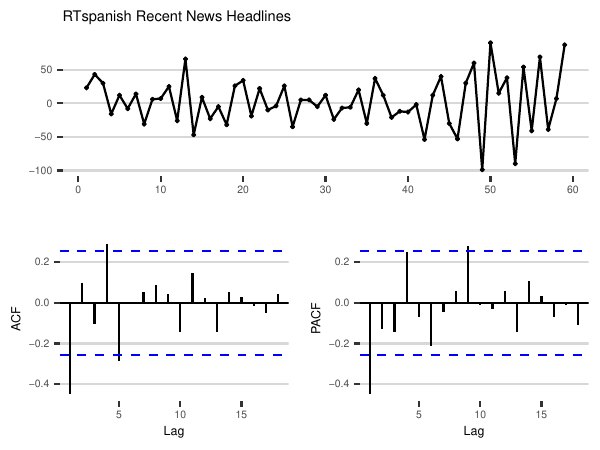} % Adjust the filename and path as necessary
  \caption{ACF RT en Espanol Recent News Headlines.}
  \label{fig:figS33}
\end{figure}
\begin{figure}[htbp]
  \centering
  \includegraphics[width=0.8\textwidth]{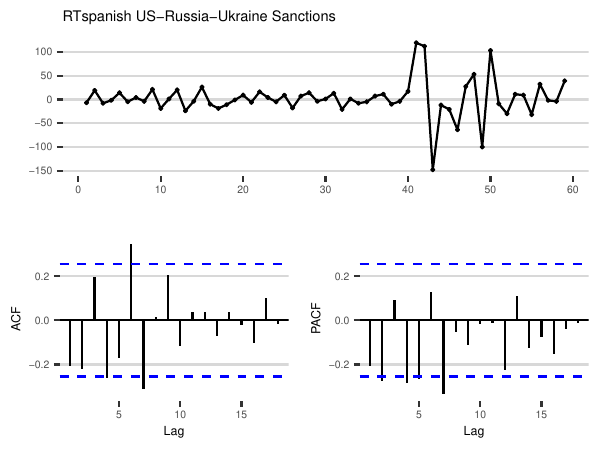} % Adjust the filename and path as necessary
  \caption{ACF RT en Espanol US-Russia-Ukraine Sanctions.}
  \label{fig:figS34}
\end{figure}
\begin{figure}[htbp]
  \centering
  \includegraphics[width=0.8\textwidth]{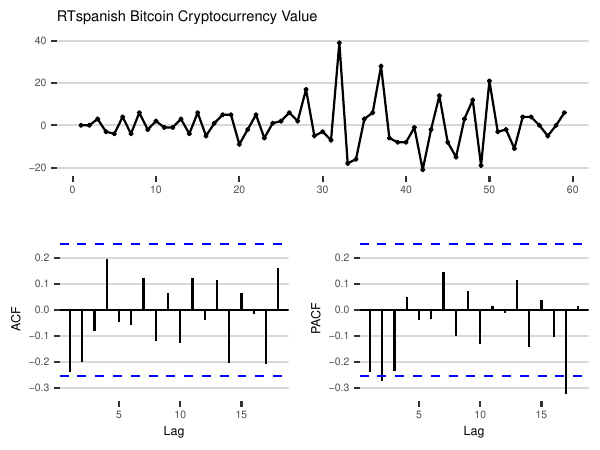} % Adjust the filename and path as necessary
  \caption{ACF RT en Espanol Bitcoin Cryptocurrency Value.}
  \label{fig:figS35}
\end{figure}
\begin{figure}[htbp]
  \centering
  \includegraphics[width=0.8\textwidth]{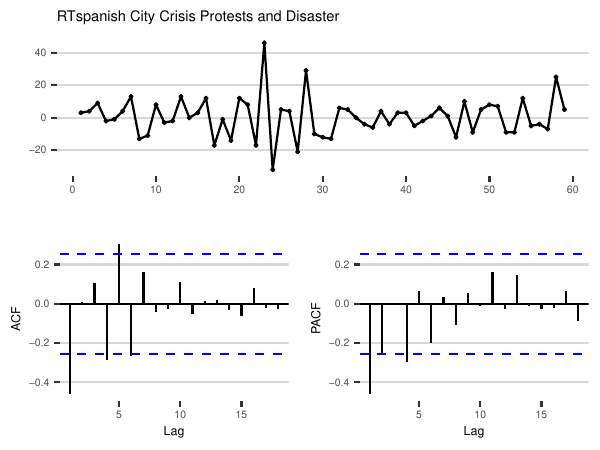} % Adjust the filename and path as necessary
  \caption{ACF RT en Espanol City Crisis Protests and Disaster.}
  \label{fig:figS36}
\end{figure}
\begin{figure}[htbp]
  \centering
  \includegraphics[width=0.8\textwidth]{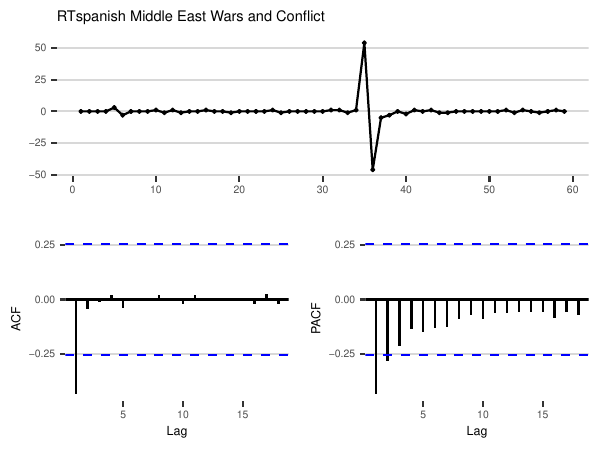} % Adjust the filename and path as necessary
  \caption{ACF RT en Espanol Middle East Wars and Conflict.}
  \label{fig:figS37}
\end{figure}
\begin{figure}[htbp]
  \centering
  \includegraphics[width=0.8\textwidth]{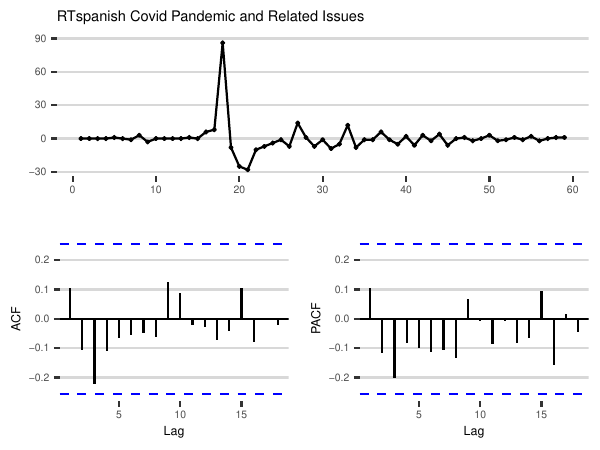} % Adjust the filename and path as necessary
  \caption{ACF RT en Espanol Covid Pandemic and Related Issues.}
  \label{fig:figS38}
\end{figure}
\clearpage
\textbf{\textit{ACF \& PACF: RT UK}}
\begin{figure}[htbp]
  \centering
  \includegraphics[width=0.8\textwidth]{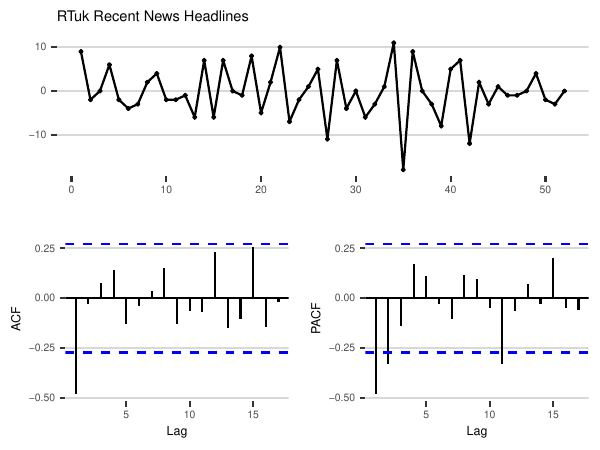} % Adjust the filename and path as necessary
  \caption{ACF RT UK Recent News Headlines.}
  \label{fig:figS39}
\end{figure}
\begin{figure}[htbp]
  \centering
  \includegraphics[width=0.8\textwidth]{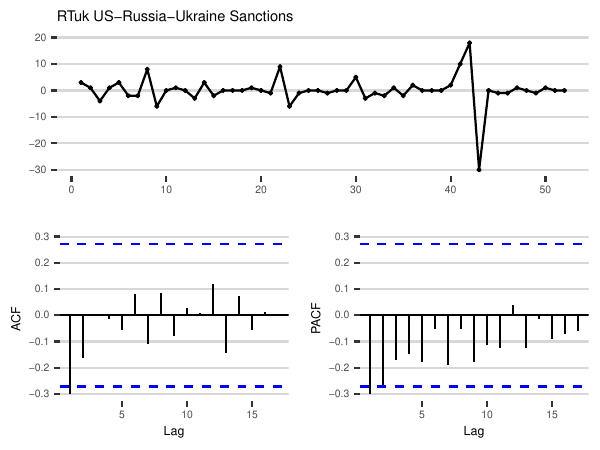} % Adjust the filename and path as necessary
  \caption{ACF RT UK US-Russia-Ukraine Sanctions.}
  \label{fig:figS40}
\end{figure}
\begin{figure}[htbp]
  \centering
  \includegraphics[width=0.8\textwidth]{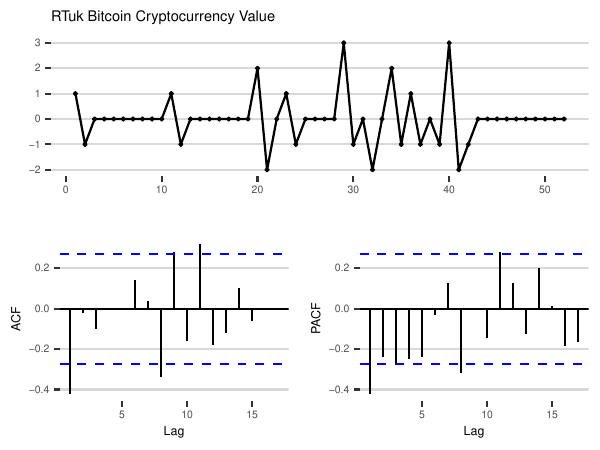} % Adjust the filename and path as necessary
  \caption{ACF RT UK Bitcoin Cryptocurrency Value.}
  \label{fig:figS41}
\end{figure}
\begin{figure}[htbp]
  \centering
  \includegraphics[width=0.8\textwidth]{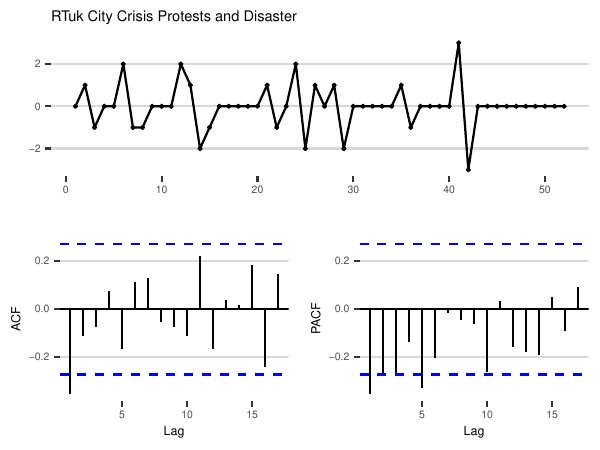} % Adjust the filename and path as necessary
  \caption{ACF RT UK City Crisis Protests and Disaster.}
  \label{fig:figS42}
\end{figure}
\begin{figure}[htbp]
  \centering
  \includegraphics[width=0.8\textwidth]{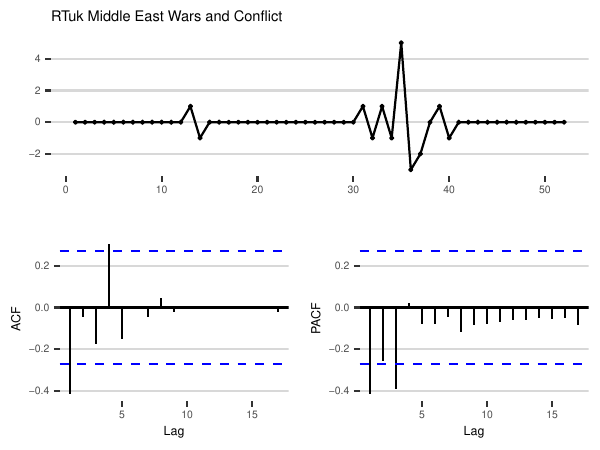} % Adjust the filename and path as necessary
  \caption{ACF RT UK Middle East Wars and Conflict.}
  \label{fig:figS43}
\end{figure}
\begin{figure}[htbp]
  \centering
  \includegraphics[width=0.8\textwidth]{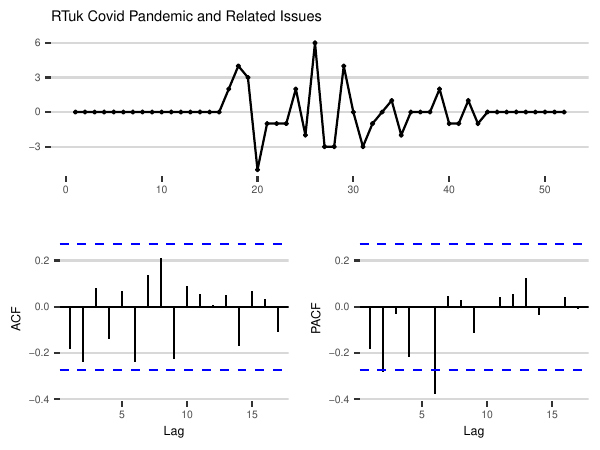} % Adjust the filename and path as necessary
  \caption{ACF RT UK Covid Pandemic and Related Issues.}
  \label{fig:figS44}
\end{figure}
\clearpage
\textbf{\textit{ACF \& PACF: Russian Ruble}}
\begin{figure}[htbp]
  \centering
  \includegraphics[width=0.8\textwidth]{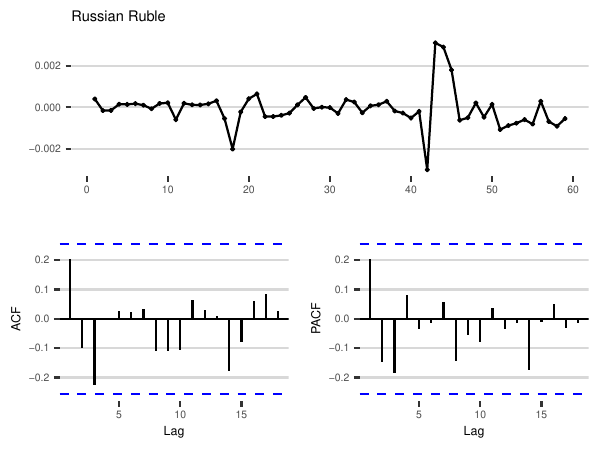} % Adjust the filename and path as necessary
  \caption{ACF Russian Rubles.}
  \label{fig:figS45}
\end{figure}
\clearpage
\textbf{\textit{ACF \& PACF: Ural Oil Prices}}
\begin{figure}[htbp]
  \centering
  \includegraphics[width=0.8\textwidth]{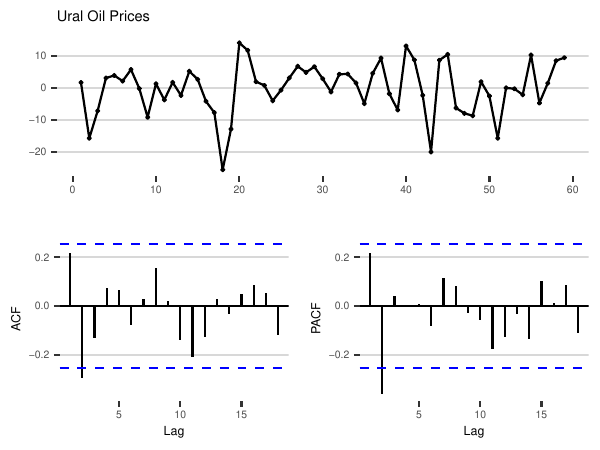} % Adjust the filename and path as necessary
  \caption{ACF Ural Oil Prices.}
  \label{fig:figS46}
\end{figure}
\clearpage
\subsubsection{Model: Russian Ruble}
\textbf{\textit{ADF Values}}
\begin{table}[htbp]
\keepXColumns
\begin{tabularx}{\textwidth}{l l X}
\toprule
\textbf{Page} & \textbf{Topic} & \textbf{P\_Value} \\
\midrule
RT & bitcoin cryptocurrency value & 0.010 \\
RT & city crisis protest and disaster & 0.010 \\
RT & covid pandemic and related issues & 0.010 \\
RT & middle east wars and conflict & 0.010 \\
RT & recent news headlines & 0.010 \\
RT & us russia ukraine sanctions & 0.010 \\
RT America & bitcoin cryptocurrency value & 0.010 \\
RT America & city crisis protest and disaster & 0.010 \\
RT America & covid pandemic and related issues & 0.018 \\
RT America & middle east wars and conflict & 0.010 \\
RT America & recent news headlines & 0.010 \\
RT America & us russia ukraine sanctions & 0.010 \\
RT Arabic & bitcoin cryptocurrency value & 0.010 \\
RT Arabic & city crisis protest and disaster & 0.010 \\
RT Arabic & covid pandemic and related issues & 0.010 \\
RT Arabic & middle east wars and conflict & 0.010 \\
RT Arabic & recent news headlines & 0.010 \\
RT Arabic & us russia ukraine sanctions & 0.010 \\
RT DE & bitcoin cryptocurrency value & 0.010 \\
RT DE & city crisis protest and disaster & 0.010 \\
RT DE & covid pandemic and related issues & 0.010 \\
RT DE & middle east wars and conflict & 0.010 \\
RT DE & recent news headlines & 0.010 \\
RT DE & us russia ukraine sanctions & 0.010 \\
RT France & bitcoin cryptocurrency value & 0.010 \\
RT France & city crisis protest and disaster & 0.010 \\
RT France & covid pandemic and related issues & 0.010 \\
RT France & middle east wars and conflict & 0.010 \\
RT France & recent news headlines & 0.010 \\
RT France & us russia ukraine sanctions & 0.010 \\
RT en Espanol & bitcoin cryptocurrency value & 0.010 \\
RT en Espanol & city crisis protest and disaster & 0.010 \\
RT en Espanol & covid pandemic and related issues & 0.010 \\
RT en Espanol & middle east wars and conflict & 0.010 \\
RT en Espanol & recent news headlines & 0.010 \\
RT en Espanol & us russia ukraine sanctions & 0.010 \\
RT UK & bitcoin cryptocurrency value & 0.010 \\
RT UK & city crisis protest and disaster & 0.010 \\
RT UK & covid pandemic and related issues & 0.010 \\
RT UK & middle east wars and conflict & 0.010 \\
RT UK & recent news headlines & 0.010 \\
RT UK & us russia ukraine sanctions & 0.010 \\
\bottomrule
\caption{ADF Values for VAR Models on Ruble}\label{tab:tableS3}
\end{tabularx}
\end{table}
\clearpage
\textbf{\textit{SC Lag Values}}
\begin{table}[htbp]
\keepXColumns
\begin{tabularx}{\textwidth}{lXl}
\toprule
\textbf{Page} & \textbf{Topic} & \textbf{SC\_Value} \\
\midrule
RT & bitcoin cryptocurrency value & 1 \\
RT & city crisis protest and disaster & 1 \\
RT & covid pandemic and related issues & 1 \\
RT & middle east wars and conflict & 1 \\
RT & recent news headlines & 1 \\
RT & us russia ukraine sanctions & 1 \\
RT America & bitcoin cryptocurrency value & 2 \\
RT America & city crisis protest and disaster & 2 \\
RT America & covid pandemic and related issues & 2 \\
RT America & middle east wars and conflict & 2 \\
RT America & recent news headlines & 2 \\
RT America & us russia ukraine sanctions & 2 \\
RT Arabic & bitcoin cryptocurrency value & 1 \\
RT Arabic & city crisis protest and disaster & 1 \\
RT Arabic & covid pandemic and related issues & 1 \\
RT Arabic & middle east wars and conflict & 1 \\
RT Arabic & recent news headlines & 1 \\
RT Arabic & us russia ukraine sanctions & 1 \\
RT DE & bitcoin cryptocurrency value & 1 \\
RT DE & city crisis protest and disaster & 1 \\
RT DE & covid pandemic and related issues & 1 \\
RT DE & middle east wars and conflict & 1 \\
RT DE & recent news headlines & 1 \\
RT DE & us russia ukraine sanctions & 1 \\
RT France & bitcoin cryptocurrency value & 1 \\
RT France & city crisis protest and disaster & 1 \\
RT France & covid pandemic and related issues & 1 \\
RT France & middle east wars and conflict & 1 \\
RT France & recent news headlines & 1 \\
RT France & us russia ukraine sanctions & 1 \\
RT en Espanol & bitcoin cryptocurrency value & 1 \\
RT en Espanol & city crisis protest and disaster & 1 \\
RT en Espanol & covid pandemic and related issues & 1 \\
RT en Espanol & middle east wars and conflict & 1 \\
RT en Espanol & recent news headlines & 1 \\
RT en Espanol & us russia ukraine sanctions & 1 \\
RT UK & bitcoin cryptocurrency value & 1 \\
RT UK & city crisis protest and disaster & 1 \\
RT UK & covid pandemic and related issues & 1 \\
RT UK & middle east wars and conflict & 1 \\
RT UK & recent news headlines & 1 \\
RT UK & us russia ukraine sanctions & 1 \\
\bottomrule
\caption{Lag Selections SC for VAR Models on Ruble}\label{tab:tableS4}
\end{tabularx}
\end{table}
\clearpage
\textbf{\textit{VAR Results}}
\begin{table}[htbp]
\begin{threeparttable}
\small % Makes the font of the table smaller
\begin{tabularx}{\textwidth}{@{}>{\RaggedRight\arraybackslash}X *{6}{>{\RaggedRight\arraybackslash}X}@{}} 
\toprule 
& \textbf{{RT bitcoin cryptocurrency value}} & \textbf{{RT city crisis protest and disaster}} & \textbf{{RT covid pandemic and related issues}} & \textbf{{RT middle east wars and conflict}} & \textbf{{RT recent news headlines}} & \textbf{{RT us russia ukraine sanctions}} \\
\midrule % Line after heading
Ruble (-1) & $164.868$ $(413.429)$ & $-217.355$ $(944.555)$ & $-421.848$ $(565.995)$ & $-356.851$ $(1194.054)$ & $155.024$ $(1794.439)$ & $-3755.434^*$ $(1791.887)$ \\
Interruption 1 (Covid-19) & $0.164$ $(0.895)$ & $0.083$ $(2.077)$ & $-0.768$ $(1.232)$ & $-0.252$ $(2.614)$ & $4.840$ $(3.902)$ & $9.476^*$ $(3.782)$ \\
Interruption 2 (Russia-Ukraine) & $0.187$ $(1.634)$ & $-0.444$ $(3.779)$ & $-1.661$ $(2.248)$ & $-0.411$ $(4.766)$ & $10.553$ $(7.122)$ & $16.765^*$ $(6.899)$ \\
Constant & $0.072$ $(0.873)$ & $0.038$ $(2.015)$ & $0.172$ $(1.201)$ & $0.252$ $(2.550)$ & $3.059$ $(3.807)$ & $2.325$ $(3.594)$ \\
Trend & $-0.007$ $(0.044)$ & $0.015$ $(0.101)$ & $0.024$ $(0.060)$ & $0.000$ $(0.128)$ & $-0.282$ $(0.192)$ & $-0.390^*$ $(0.184)$ \\
Num.Obs. & 58 & 58 & 58 & 58 & 58 & 58 \\
$R^2$ & 0.030 & 0.286 & 0.038 & 0.188 & 0.143 & 0.179 \\
$R^2$ Adj. & $-0.063$ & 0.217 & $-0.055$ & 0.110 & 0.061 & 0.100 \\
\bottomrule % Bottom line
\multicolumn{7}{p{\textwidth}}{\footnotesize $^{+} p < 0.10$, * $p < 0.05$, ** $p < 0.01$, *** $p < 0.001$} \\
\caption{VAR Models RT on Ruble}\label{tab:tableS5}
\end{tabularx}
\end{threeparttable}
\end{table}
\clearpage
\begin{table}[htbp]
\begin{threeparttable}
\small % Makes the font of the table smaller
\begin{tabularx}{\textwidth}{@{}>{\RaggedRight\arraybackslash}X *{6}{>{\RaggedRight\arraybackslash}X}@{}} 
\toprule % Top line
& \textbf{RT Arabic bitcoin cryptocurrency value} & \textbf{RT Arabic city crisis protest and disaster} & \textbf{RT Arabic covid pandemic and related issues} & \textbf{RT Arabic middle east wars and conflict} & \textbf{RT Arabic recent news headlines} & \textbf{RT Arabic us russia ukraine sanctions} \\
\midrule % Line after heading
Ruble (-1) & $183.672$ $(143.648)$ & $21.261$ $(59.302)$ & $-108.234^{+}$ $(62.151)$ & $-47.387$ $(48.379)$ & $-743.153$ $(1416.287)$ & $-1372.261$ $(3916.704)$ \\
Interruption 1 (Covid-19) & $0.006$ $(0.312)$ & $0.008$ $(0.130)$ & $0.015$ $(0.136)$ & $0.032$ $(0.106)$ & $-1.301$ $(3.102)$ & $-3.116$ $(8.573)$ \\
Interruption 2 (Russia-Ukraine) & $0.134$ $(0.568)$ & $0.032$ $(0.237)$ & $0.075$ $(0.247)$ & $0.078$ $(0.193)$ & $1.645$ $(5.641)$ & $1.855$ $(15.547)$ \\
Constant & $-0.245$ $(0.305)$ & $0.005$ $(0.128)$ & $0.022$ $(0.132)$ & $0.032$ $(0.103)$ & $-1.183$ $(3.028)$ & $-0.468$ $(8.332)$ \\
Trend & $0.010$ $(0.015)$ & $-0.001$ $(0.006)$ & $-0.002$ $(0.007)$ & $-0.002$ $(0.005)$ & $0.077$ $(0.152)$ & $0.105$ $(0.420)$ \\
Num.Obs. & 58 & 58 & 58 & 58 & 58 & 58 \\
$R^2$ & 0.188 & 0.019 & 0.389 & 0.129 & 0.071 & 0.154 \\
$R^2$ Adj. & 0.110 & $-0.076$ & 0.330 & 0.045 & $-0.018$ & 0.073 \\
\bottomrule 
\multicolumn{7}{p{\textwidth}}{\footnotesize $^{+} p < 0.10$, * $p < 0.05$, ** $p < 0.01$, *** $p < 0.001$} \\
\caption{VAR Models RT Arabic on Ruble}\label{tab:tableS6}
\end{tabularx}
\end{threeparttable}
\end{table}
\clearpage
\begin{table}[htbp]
\begin{threeparttable}
\small % Makes the font of the table smaller
\begin{tabularx}{\textwidth}{@{}>{\RaggedRight\arraybackslash}X *{6}{>{\RaggedRight\arraybackslash}X}@{}} 
\toprule % Top line
& \textbf{RT America bitcoin cryptocurrency value} & \textbf{RT America city crisis protest and disaster} & \textbf{RT America covid pandemic and related issues} & \textbf{RT America middle east wars and conflict} & \textbf{RT Arabic recent news headlines} & \textbf{RT America us russia ukraine sanctions} \\
\midrule % Line after heading
\midrule
Ruble (-1) & $688.200$ $(467.755)$ & $226.141$ $(210.095)$ & $926.738$ $(1738.873)$ & $-92.343$ $(323.263)$ & $2114.651$ $(1643.863)$ & $2211.312$ $(1527.198)$ \\
Ruble (-2) & $358.703$ $(467.507)$ & $-185.709$ $(194.163)$ & $3182.236^{+}$ $(1665.129)$ & $178.482$ $(329.766)$ & $363.159$ $(1689.415)$ & $1118.256$ $(1433.435)$ \\
Interruption 1 (Covid-19) & $0.554$ $(0.865)$ & $0.288$ $(0.359)$ & $2.128$ $(2.587)$ & $0.544$ $(0.603)$ & $3.333$ $(3.095)$ & $4.755^{+}$ $(2.614)$ \\
Interruption 2 (Russia-Ukraine) & $0.127$ $(1.348)$ & $0.654$ $(0.532)$ & $1.621$ $(3.796)$ & $0.529$ $(0.904)$ & $-0.271$ $(4.603)$ & $4.880$ $(3.897)$ \\
Constant & $0.203$ $(0.532)$ & $0.058$ $(0.220)$ & $0.988$ $(1.528)$ & $0.057$ $(0.375)$ & $-1.843$ $(1.915)$ & $0.453$ $(1.620)$ \\
Trend & $-0.026$ $(0.040)$ & $-0.013$ $(0.017)$ & $-0.105$ $(0.117)$ & $-0.021$ $(0.028)$ & $-0.057$ $(0.143)$ & $-0.193$ $(0.121)$ \\
Num.Obs. & 40 & 40 & 40 & 40 & 40 & 40 \\
$R^2$ & 0.353 & 0.578 & 0.305 & 0.395 & 0.513 & 0.236 \\
$R^2$ Adj. & 0.211 & 0.485 & 0.153 & 0.263 & 0.406 & 0.069 \\
\bottomrule
\multicolumn{7}{p{\textwidth}}{\footnotesize $^{+} p < 0.10$, * $p < 0.05$, ** $p < 0.01$, *** $p < 0.001$} \\
\caption{VAR Models RT America on Ruble}\label{tab:tableS7}
\end{tabularx}
\end{threeparttable}
\end{table}
\clearpage
\begin{table}[htbp]
\begin{threeparttable}
\small % Makes the font of the table smaller
\begin{tabularx}{\textwidth}{@{}>{\RaggedRight\arraybackslash}X *{6}{>{\RaggedRight\arraybackslash}X}@{}} 
\toprule % Top line
& \textbf{RT DE bitcoin cryptocurrency value} & \textbf{RT DE city crisis protest and disaster} & \textbf{RT DE covid pandemic and related issues} & \textbf{RT DE middle east wars and conflict} & \textbf{RT DE recent news headlines} & \textbf{RT DE US Russia Ukraine sanctions} \\
\midrule % Line after heading
Ruble (-1) & $-3.292$ $(114.048)$ & $0.042$ $(102.869)$ & $84.102$ $(399.035)$ & $141.502$ $(255.404)$ & $4506.965^{**}$ $(1551.461)$ & $-1167.060$ $(1254.928)$ \\
Interruption 1 (Covid-19) & $0.036$ $(0.246)$ & $-0.100$ $(0.225)$ & $-1.188$ $(0.870)$ & $-0.266$ $(0.558)$ & $2.231$ $(3.040)$ & $5.223^{+}$ $(2.829)$ \\
Interruption 2 (Russia-Ukraine) & $-0.280$ $(0.447)$ & $-0.344$ $(0.411)$ & $-4.364^{**}$ $(1.597)$ & $-0.765$ $(1.018)$ & $-1.308$ $(5.549)$ & $7.302$ $(5.064)$ \\
Constant & $-0.068$ $(0.238)$ & $-0.069$ $(0.219)$ & $-0.736$ $(0.849)$ & $-0.093$ $(0.545)$ & $-0.965$ $(2.965)$ & $0.621$ $(2.644)$ \\
Trend & $0.005$ $(0.012)$ & $0.007$ $(0.011)$ & $0.091^{*}$ $(0.043)$ & $0.016$ $(0.027)$ & $0.030$ $(0.149)$ & $-0.165$ $(0.136)$ \\
Num.Obs. & 58 & 58 & 58 & 58 & 58 & 58 \\
$R^2$ & 0.184 & 0.161 & 0.220 & 0.147 & 0.320 & 0.202 \\
$R^2$ Adj. & 0.106 & 0.080 & 0.145 & 0.065 & 0.254 & 0.126 \\
\bottomrule
\multicolumn{7}{p{\textwidth}}{\footnotesize $^{+} p < 0.10$, * $p < 0.05$, ** $p < 0.01$, *** $p < 0.001$} \\
\caption{VAR Models RT DE on Ruble}\label{tab:tableS8}
\end{tabularx}
\end{threeparttable}
\end{table}
\clearpage
\begin{table}[htbp]
\begin{threeparttable}
\small % Makes the font of the table smaller
\begin{tabularx}{\textwidth}{@{}>{\RaggedRight\arraybackslash}X *{6}{>{\RaggedRight\arraybackslash}X}@{}} 
\toprule % Top line
& \textbf{RT France bitcoin cryptocurrency value} & \textbf{RT France city crisis protest and disaster} & \textbf{RT France covid pandemic and related issues} & \textbf{RT France middle east wars and conflict} & \textbf{RT France recent news headlines} & \textbf{RT France US Russia Ukraine sanctions} \\
\midrule % Line after heading
Ruble (-1) & $38.442$ $(182.501)$ & $231.923$ $(370.050)$ & $2886.726$ $(3358.811)$ & $-329.245$ $(994.870)$ & $7303.599$ $(7821.946)$ & $6336.360$ $(3912.482)$ \\
Interruption 1 (Covid-19) & $-0.264$ $(0.399)$ & $-0.331$ $(0.825)$ & $-12.435$ $(7.496)$ & $0.065$ $(2.178)$ & $-13.073$ $(16.196)$ & $1.827$ $(8.302)$ \\
Interruption 2 (Russia-Ukraine) & $-0.792$ $(0.728)$ & $-0.055$ $(1.492)$ & $-35.160^{*}$ $(14.168)$ & $0.253$ $(3.972)$ & $-21.459$ $(29.437)$ & $-0.891$ $(15.148)$ \\
Constant & $-0.110$ $(0.389)$ & $0.346$ $(0.788)$ & $-5.089$ $(7.201)$ & $0.327$ $(2.124)$ & $8.274$ $(15.673)$ & $0.831$ $(8.089)$ \\
Trend & $0.016$ $(0.020)$ & $-0.004$ $(0.040)$ & $0.747^{*}$ $(0.372)$ & $-0.015$ $(0.107)$ & $0.161$ $(0.787)$ & $-0.026$ $(0.407)$ \\
Num.Obs. & 58 & 58 & 58 & 58 & 58 & 58 \\
$R^2$ & 0.287 & 0.233 & 0.127 & 0.005 & 0.090 & 0.194 \\
$R^2$ Adj. & 0.218 & 0.160 & 0.043 & $-0.091$ & 0.002 & 0.117 \\
\bottomrule
\multicolumn{7}{p{\textwidth}}{\footnotesize $^{+} p < 0.10$, * $p < 0.05$, ** $p < 0.01$, *** $p < 0.001$} \\
\caption{VAR Models RT France on Ruble}\label{tab:tableS9}
\end{tabularx}
\end{threeparttable}
\end{table}
\clearpage
\begin{table}[htbp]
\begin{threeparttable}
\small % Makes the font of the table smaller
\begin{tabularx}{\textwidth}{@{}>{\RaggedRight\arraybackslash}X *{6}{>{\RaggedRight\arraybackslash}X}@{}} 
\toprule % Top line
& \textbf{RT SP bitcoin cryptocurrency value} & \textbf{RT SP city crisis protest and disaster} & \textbf{RT SP covid pandemic and related issues} & \textbf{RT SP middle east wars and conflict} & \textbf{RT SP recent news headlines} & \textbf{RT SP US Russia Ukraine sanctions} \\
\midrule % Line after heading
Ruble (-1) & $128.040$ $(1631.263)$ & $865.049$ $(1728.603)$ & $-84.237$ $(2205.711)$ & $-723.796$ $(1360.909)$ & $-877.210$ $(5468.023)$ & $-4051.556$ $(6617.212)$ \\
Interruption 1 (Covid-19) & $-1.414$ $(3.445)$ & $-2.864$ $(3.791)$ & $-0.827$ $(4.604)$ & $0.314$ $(2.979)$ & $-10.634$ $(11.594)$ & $7.743$ $(13.487)$ \\
Interruption 2 (Russia-Ukraine) & $-7.334$ $(6.374)$ & $-0.302$ $(6.880)$ & $0.405$ $(8.406)$ & $1.003$ $(5.431)$ & $-1.952$ $(21.091)$ & $17.383$ $(24.632)$ \\
Constant & $-0.995$ $(3.361)$ & $1.657$ $(3.685)$ & $1.516$ $(4.494)$ & $0.514$ $(2.906)$ & $7.165$ $(11.307)$ & $3.778$ $(13.050)$ \\
Trend & $0.140$ $(0.170)$ & $0.017$ $(0.185)$ & $-0.040$ $(0.226)$ & $-0.033$ $(0.146)$ & $0.045$ $(0.566)$ & $-0.371$ $(0.661)$ \\
Num.Obs. & 58 & 58 & 58 & 58 & 58 & 58 \\
$R^2$ & 0.085 & 0.225 & 0.014 & 0.191 & 0.239 & 0.058 \\
$R^2$ Adj. & $-0.003$ & 0.151 & $-0.081$ & 0.113 & 0.165 & $-0.032$ \\
\bottomrule
\multicolumn{7}{p{\textwidth}}{\footnotesize $^{+} p < 0.10$, * $p < 0.05$, ** $p < 0.01$, *** $p < 0.001$} \\
\caption{VAR Models RT en Espanol on Ruble}\label{tab:tableS10}
\end{tabularx}
\end{threeparttable}
\end{table}
\clearpage
\begin{table}[htbp]
\begin{threeparttable}
\small % Makes the font of the table smaller
\begin{tabularx}{\textwidth}{@{}>{\RaggedRight\arraybackslash}X *{6}{>{\RaggedRight\arraybackslash}X}@{}} 
\toprule % Top line
& \textbf{RT UK bitcoin cryptocurrency value} & \textbf{RT UK city crisis protest and disaster} & \textbf{RT UK covid pandemic and related issues} & \textbf{RT UK middle east wars and conflict} & \textbf{RT UK recent news headlines} & \textbf{RT UK US Russia Ukraine sanctions} \\
\midrule % Line after heading
Ruble (-1) & $-10.927$ $(103.197)$ & $25.920$ $(117.098)$ & $-156.450$ $(211.156)$ & $-17.965$ $(104.176)$ & $40.092$ $(592.129)$ & $-29.105$ $(1072.489)$ \\
Interruption 1 (Covid-19) & $-0.441$ $(0.383)$ & $0.182$ $(0.430)$ & $-0.282$ $(0.766)$ & $-0.010$ $(0.386)$ & $0.770$ $(2.180)$ & $2.047$ $(2.509)$ \\
Interruption 2 (Russia-Ukraine) & $-1.149^{+}$ $(0.635)$ & $0.356$ $(0.713)$ & $-0.264$ $(1.272)$ & $-0.029$ $(0.642)$ & $2.362$ $(3.639)$ & $2.249$ $(4.140)$ \\
Constant & $-0.263$ $(0.309)$ & $0.137$ $(0.346)$ & $0.180$ $(0.619)$ & $0.031$ $(0.312)$ & $1.252$ $(1.768)$ & $0.379$ $(1.963)$ \\
Trend & $0.027$ $(0.018)$ & $-0.012$ $(0.020)$ & $0.001$ $(0.036)$ & $-0.001$ $(0.018)$ & $-0.099$ $(0.104)$ & $-0.074$ $(0.118)$ \\
Num.Obs. & 51 & 51 & 51 & 51 & 51 & 51 \\
$R^2$ & 0.236 & 0.131 & 0.049 & 0.172 & 0.259 & 0.104 \\
$R^2$ Adj. & 0.152 & 0.034 & $-0.056$ & 0.080 & 0.177 & 0.004 \\
\bottomrule
\multicolumn{7}{p{\textwidth}}{\footnotesize $^{+} p < 0.10$, * $p < 0.05$, ** $p < 0.01$, *** $p < 0.001$} \\
\caption{VAR Models RT UK on Ruble}\label{tab:tableS11}
\end{tabularx}
\end{threeparttable}
\end{table}
\clearpage
\textbf{\textit{Granger Causality Results}}
\begin{table}[htbp]
\keepXColumns
\begin{tabularx}{\textwidth}{lXXl}
\toprule
\textbf{Page} & \textbf{Topic} & \textbf{GC.P.Value} & \textbf{Significance} \\
\midrule
RT & bitcoin cryptocurrency value & 0.6909 & ns \\
RT & city crisis protest and disaster & 0.8185 & ns \\
RT & covid pandemic and related issues & 0.4578 & ns \\
RT & middle east wars and conflict & 0.7656 & ns \\
RT & recent news headlines & 0.9313 & ns \\
RT & us russia ukraine sanctions & 0.0385 & p<0.05* \\
RT America & bitcoin cryptocurrency value & 0.2901 & ns \\
RT America & city crisis protest and disaster & 0.3477 & ns \\
RT America & covid pandemic and related issues & 0.1691 & ns \\
RT America & middle east wars and conflict & 0.8136 & ns \\
RT America & recent news headlines & 0.4404 & ns \\
RT America & us russia ukraine sanctions & 0.2740 & ns \\
RT Arabic & bitcoin cryptocurrency value & 0.2039 & ns \\
RT Arabic & city crisis protest and disaster & 0.7207 & ns \\
RT Arabic & covid pandemic and related issues & 0.0846 & ns \\
RT Arabic & middle east wars and conflict & 0.3296 & ns \\
RT Arabic & recent news headlines & 0.6009 & ns \\
RT Arabic & us russia ukraine sanctions & 0.7268 & ns \\
RT DE & bitcoin cryptocurrency value & 0.9770 & ns \\
RT DE & city crisis protest and disaster & 0.9997 & ns \\
RT DE & covid pandemic and related issues & 0.8335 & ns \\
RT DE & middle east wars and conflict & 0.5807 & ns \\
RT DE & recent news headlines & 0.0045 & p<0.01** \\
RT DE & us russia ukraine sanctions & 0.3545 & ns \\
RT France & bitcoin cryptocurrency value & 0.8336 & ns \\
RT France & city crisis protest and disaster & 0.5322 & ns \\
RT France & covid pandemic and related issues & 0.3921 & ns \\
RT France & middle east wars and conflict & 0.7414 & ns \\
RT France & recent news headlines & 0.3526 & ns \\
RT France & us russia ukraine sanctions & 0.1084 & ns \\
RT en Espanol & bitcoin cryptocurrency value & 0.9376 & ns \\
RT en Espanol & city crisis protest and disaster & 0.6178 & ns \\
RT en Espanol & covid pandemic and related issues & 0.9696 & ns \\
RT en Espanol & middle east wars and conflict & 0.5960 & ns \\
RT en Espanol & recent news headlines & 0.8729 & ns \\
RT en Espanol & us russia ukraine sanctions & 0.5417 & ns \\
RT UK & bitcoin cryptocurrency value & 0.9159 & ns \\
RT UK & city crisis protest and disaster & 0.8253 & ns \\
RT UK & covid pandemic and related issues & 0.4607 & ns \\
RT UK & middle east wars and conflict & 0.8635 & ns \\
RT UK & recent news headlines & 0.9462 & ns \\
RT UK & us russia ukraine sanctions & 0.9784 & ns \\
\bottomrule
\caption{Granger Causality Tests on VAR Models on Ruble}\label{tab:tables12}
\end{tabularx}
\end{table}
\clearpage
\textbf{\textit{Post\-Hoc Tests}}
\textbf{Johansen test for co-integration
}
\begin{table}[htbp]
\keepXColumns
\begin{tabularx}{\textwidth}{lXXl}
\toprule
\textbf{Page} & \textbf{Topic} & \textbf{P Value} \\
\midrule
RT & bitcoin cryptocurrency value & 0.01000000 \\
RT & city crisis protest and disaster & 0.01000000 \\
RT & covid pandemic and related issues & 0.01000000 \\
RT & middle east wars and conflict & 0.01000000 \\
RT & recent news headlines & 0.01000000 \\
RT & us russia ukraine sanctions & 0.01000000 \\
RT America & bitcoin cryptocurrency value & 0.01000000 \\
RT America & city crisis protest and disaster & 0.01000000 \\
RT America & covid pandemic and related issues & 0.08886854 \\
RT America & middle east wars and conflict & 0.01000000 \\
RT America & recent news headlines & 0.01000000 \\
RT America & us russia ukraine sanctions & 0.01000000 \\
RT Arabic & bitcoin cryptocurrency value & 0.01000000 \\
RT Arabic & city crisis protest and disaster & 0.01000000 \\
RT Arabic & covid pandemic and related issues & 0.01000000 \\
RT Arabic & middle east wars and conflict & 0.01000000 \\
RT Arabic & recent news headlines & 0.01000000 \\
RT Arabic & us russia ukraine sanctions & 0.01000000 \\
RT DE & bitcoin cryptocurrency value & 0.01000000 \\
RT DE & city crisis protest and disaster & 0.01000000 \\
RT DE & covid pandemic and related issues & 0.01000000 \\
RT DE & middle east wars and conflict & 0.01000000 \\
RT DE & recent news headlines & 0.01000000 \\
RT DE & us russia ukraine sanctions & 0.01000000 \\
RT France & bitcoin cryptocurrency value & 0.01000000 \\
RT France & city crisis protest and disaster & 0.01000000 \\
RT France & covid pandemic and related issues & 0.01000000 \\
RT France & middle east wars and conflict & 0.01000000 \\
RT France & recent news headlines & 0.01000000 \\
RT France & us russia ukraine sanctions & 0.01000000 \\
RT en Espanol & bitcoin cryptocurrency value & 0.01000000 \\
RT en Espanol & city crisis protest and disaster & 0.01000000 \\
RT en Espanol & covid pandemic and related issues & 0.01000000 \\
RT en Espanol & middle east wars and conflict & 0.01000000 \\
RT en Espanol & recent news headlines & 0.01000000 \\
RT en Espanol & us russia ukraine sanctions & 0.01000000 \\
RT UK & bitcoin cryptocurrency value & 0.01000000 \\
RT UK & city crisis protest and disaster & 0.01000000 \\
RT UK & covid pandemic and related issues & 0.01000000 \\
RT UK & middle east wars and conflict & 0.01000000 \\
RT UK & recent news headlines & 0.01000000 \\
RT UK & us russia ukraine sanctions & 0.01000000 \\
\bottomrule
\multicolumn{3}{p{\textwidth}}{\footnotesize Note: p-value (rejection criteria $p < 0.05$} \\
\caption{Johansen test for co-integration VAR Models on Ruble}\label{tab:tables13}
\end{tabularx}
\end{table}
\clearpage
\textbf{Ljung-Box Test}
\begin{table}[htbp]
\keepXColumns
% \begin{tabularx}{\textwidth}{lXXl}
\begin{tabularx}{\textwidth}{l >{\hsize=1.5\hsize}X >{\hsize=0.75\hsize}X >{\hsize=0.75\hsize}X}
\toprule
\textbf{Page} & \textbf{Topic} & \textbf{P Value DV} & \textbf{P Value Ruble} \\
\midrule
RT & bitcoin cryptocurrency value & 0.510 & 0.898 \\
RT & city crisis protest and disaster & 0.106 & 0.945 \\
RT & covid pandemic and related issues & 0.875 & 0.903 \\
RT & middle east wars and conflict & 0.228 & 0.956 \\
RT & recent news headlines & 0.469 & 0.983 \\
RT & us russia ukraine sanctions & 0.595 & 0.913 \\
RT America & bitcoin cryptocurrency value & 0.206 & 0.918 \\
RT America & city crisis protest and disaster & 0.793 & 0.859 \\
RT America & covid pandemic and related issues & 0.764 & 0.416 \\
RT America & middle east wars and conflict & 0.768 & 0.780 \\
RT America & recent news headlines & 0.444 & 0.721 \\
RT America & us russia ukraine sanctions & 0.970 & 0.927 \\
RT Arabic & bitcoin cryptocurrency value & 0.468 & 0.985 \\
RT Arabic & city crisis protest and disaster & 0.702 & 0.911 \\
RT Arabic & covid pandemic and related issues & 0.221 & 0.960 \\
RT Arabic & middle east wars and conflict & 0.754 & 0.953 \\
RT Arabic & recent news headlines & 0.898 & 0.998 \\
RT Arabic & us russia ukraine sanctions & 0.273 & 0.926 \\
RT DE & bitcoin cryptocurrency value & 0.342 & 0.972 \\
RT DE & city crisis protest and disaster & 0.660 & 0.945 \\
RT DE & covid pandemic and related issues & 0.944 & 0.622 \\
RT DE & middle east wars and conflict & 0.266 & 0.974 \\
RT DE & recent news headlines & 0.035 & 0.883 \\
RT DE & us russia ukraine sanctions & 0.080 & 0.856 \\
RT France & bitcoin cryptocurrency value & 0.048 & 0.953 \\
RT France & city crisis protest and disaster & 0.081 & 0.993 \\
RT France & covid pandemic and related issues & 0.706 & 0.753 \\
RT France & middle east wars and conflict & 0.908 & 0.949 \\
RT France & recent news headlines & 0.614 & 0.826 \\
RT France & us russia ukraine sanctions & 0.442 & 0.577 \\
RT en Espanol & bitcoin cryptocurrency value & 0.451 & 0.986 \\
RT en Espanol & city crisis protest and disaster & 0.277 & 0.996 \\
RT en Espanol & covid pandemic and related issues & 0.969 & 0.959 \\
RT en Espanol & middle east wars and conflict & 0.347 & 0.964 \\
RT en Espanol & recent news headlines & 0.353 & 0.994 \\
RT en Espanol & us russia ukraine sanctions & 0.623 & 0.952 \\
RT UK & bitcoin cryptocurrency value & 0.508 & 0.621 \\
RT UK & city crisis protest and disaster & 0.590 & 0.837 \\
RT UK & covid pandemic and related issues & 0.659 & 0.622 \\
RT UK & middle east wars and conflict & 0.431 & 0.625 \\
RT UK & recent news headlines & 0.158 & 0.322 \\
RT UK & us russia ukraine sanctions & 0.520 & 0.381 \\
\bottomrule
\caption{Ljung-Box Test VAR Models on Ruble}\label{tab:tables14}
\end{tabularx}
\end{table}
\clearpage
\subsubsection{Model: Ural Oil Prices}
\textbf{\textit{ADF Values}}
\begin{table}[htbp]
\begin{tabularx}{\textwidth}{lXX}
\toprule
\textbf{Page} & \textbf{Topic} & \textbf{P\_Value} \\
\midrule
RT & bitcoin cryptocurrency value & 0.010 \\
RT & city crisis protest and disaster & 0.010 \\
RT & covid pandemic and related issues & 0.010 \\
RT & middle east wars and conflict & 0.010 \\
RT & recent news headlines & 0.010 \\
RT & us russia ukraine sanctions & 0.010 \\
RT America & bitcoin cryptocurrency value & 0.010 \\
RT America & city crisis protest and disaster & 0.010 \\
RT America & covid pandemic and related issues & 0.018 \\
RT America & middle east wars and conflict & 0.010 \\
RT America & recent news headlines & 0.010 \\
RT America & us russia ukraine sanctions & 0.010 \\
RT Arabic & bitcoin cryptocurrency value & 0.010 \\
RT Arabic & city crisis protest and disaster & 0.010 \\
RT Arabic & covid pandemic and related issues & 0.010 \\
RT Arabic & middle east wars and conflict & 0.010 \\
RT Arabic & recent news headlines & 0.010 \\
RT Arabic & us russia ukraine sanctions & 0.010 \\
RT DE & bitcoin cryptocurrency value & 0.010 \\
RT DE & city crisis protest and disaster & 0.010 \\
RT DE & covid pandemic and related issues & 0.010 \\
RT DE & middle east wars and conflict & 0.010 \\
RT DE & recent news headlines & 0.010 \\
RT DE & us russia ukraine sanctions & 0.010 \\
RT France & bitcoin cryptocurrency value & 0.010 \\
RT France & city crisis protest and disaster & 0.010 \\
RT France & covid pandemic and related issues & 0.010 \\
RT France & middle east wars and conflict & 0.010 \\
RT France & recent news headlines & 0.010 \\
RT France & us russia ukraine sanctions & 0.010 \\
RT en Espanol & bitcoin cryptocurrency value & 0.010 \\
RT en Espanol & city crisis protest and disaster & 0.010 \\
RT en Espanol & covid pandemic and related issues & 0.010 \\
RT en Espanol & middle east wars and conflict & 0.010 \\
RT en Espanol & recent news headlines & 0.010 \\
RT en Espanol & us russia ukraine sanctions & 0.010 \\
RT UK & bitcoin cryptocurrency value & 0.010 \\
RT UK & city crisis protest and disaster & 0.010 \\
RT UK & covid pandemic and related issues & 0.010 \\
RT UK & middle east wars and conflict & 0.010 \\
RT UK & recent news headlines & 0.010 \\
RT UK & us russia ukraine sanctions & 0.010 \\
\bottomrule
\caption{ADF Values for VAR Models on Ural Oil Prices}\label{tab:tableS15}
\end{tabularx}
\end{table}
\clearpage
\textbf{\textit{SC Lag Values}}
\begin{table}[htbp]
\begin{tabularx}{\textwidth}{lXl}
\toprule
\textbf{Page} & \textbf{Topic} & \textbf{SC\_Value} \\
\midrule
RT & bitcoin cryptocurrency value & 1 \\
RT & city crisis protest and disaster & 1 \\
RT & covid pandemic and related issues & 1 \\
RT & middle east wars and conflict & 1 \\
RT & recent news headlines & 1 \\
RT & us russia ukraine sanctions & 1 \\
RT America & bitcoin cryptocurrency value & 2 \\
RT America & city crisis protest and disaster & 2 \\
RT America & covid pandemic and related issues & 2 \\
RT America & middle east wars and conflict & 2 \\
RT America & recent news headlines & 2 \\
RT America & us russia ukraine sanctions & 2 \\
RT Arabic & bitcoin cryptocurrency value & 1 \\
RT Arabic & city crisis protest and disaster & 1 \\
RT Arabic & covid pandemic and related issues & 1 \\
RT Arabic & middle east wars and conflict & 1 \\
RT Arabic & recent news headlines & 1 \\
RT Arabic & us russia ukraine sanctions & 1 \\
RT DE & bitcoin cryptocurrency value & 1 \\
RT DE & city crisis protest and disaster & 1 \\
RT DE & covid pandemic and related issues & 1 \\
RT DE & middle east wars and conflict & 1 \\
RT DE & recent news headlines & 1 \\
RT DE & us russia ukraine sanctions & 1 \\
RT France & bitcoin cryptocurrency value & 1 \\
RT France & city crisis protest and disaster & 1 \\
RT France & covid pandemic and related issues & 1 \\
RT France & middle east wars and conflict & 1 \\
RT France & recent news headlines & 1 \\
RT France & us russia ukraine sanctions & 1 \\
RT en Espanol & bitcoin cryptocurrency value & 1 \\
RT en Espanol & city crisis protest and disaster & 1 \\
RT en Espanol & covid pandemic and related issues & 1 \\
RT en Espanol & middle east wars and conflict & 1 \\
RT en Espanol & recent news headlines & 1 \\
RT en Espanol & us russia ukraine sanctions & 1 \\
RT UK & bitcoin cryptocurrency value & 1 \\
RT UK & city crisis protest and disaster & 1 \\
RT UK & covid pandemic and related issues & 1 \\
RT UK & middle east wars and conflict & 1 \\
RT UK & recent news headlines & 1 \\
RT UK & us russia ukraine sanctions & 1 \\
\bottomrule
\caption{Lag Selections SC for VAR Models on Ural Oil Prices}\label{tab:tableS16}
\end{tabularx}
\end{table}
\clearpage
\textbf{\textit{VAR Results}}
\begin{table}[htbp]
\begin{threeparttable}
\small % Makes the font of the table smaller
\begin{tabularx}{\textwidth}{@{}>{\RaggedRight\arraybackslash}X *{6}{>{\RaggedRight\arraybackslash}X}@{}} 
\toprule 
& \textbf{{RT bitcoin cryptocurrency value}} & \textbf{{RT city crisis protest and disaster}} & \textbf{{RT covid pandemic and related issues}} & \textbf{{RT middle east wars and conflict}} & \textbf{{RT recent news headlines}} & \textbf{{RT us russia ukraine sanctions}} \\
\midrule % Line after heading
Ural Oil Price (-1) & $0.001$ $(0.045)$ & $-0.054$ $(0.104)$ & $-0.163^{*}$ $(0.063)$ & $0.014$ $(0.133)$ & $-0.006$ $(0.199)$ & $0.383^{*}$ $(0.186)$ \\
Interruption 1 (Covid-19) & $0.207$ $(0.891)$ & $0.070$ $(2.061)$ & $-0.807$ $(1.158)$ & $-0.356$ $(2.599)$ & $4.882$ $(3.885)$ & $7.716^{*}$ $(3.719)$ \\
Interruption 2 (Russia-Ukraine) & $0.263$ $(1.630)$ & $-0.679$ $(3.762)$ & $-2.413$ $(2.127)$ & $-0.526$ $(4.755)$ & $10.598$ $(7.105)$ & $15.260^{*}$ $(6.815)$ \\
Constant & $0.097$ $(0.880)$ & $-0.127$ $(2.024)$ & $-0.284$ $(1.144)$ & $0.239$ $(2.568)$ & $3.063$ $(3.829)$ & $2.660$ $(3.617)$ \\
Trend & $-0.010$ $(0.044)$ & $0.023$ $(0.101)$ & $0.047$ $(0.057)$ & $0.004$ $(0.127)$ & $-0.284$ $(0.191)$ & $-0.348^{+}$ $(0.181)$ \\
Num.Obs. & 58 & 58 & 58 & 58 & 58 & 58 \\
$R^2$ & 0.027 & 0.288 & 0.139 & 0.187 & 0.143 & 0.177 \\
$R^2$ Adj. & $-0.066$ & 0.220 & 0.056 & 0.108 & 0.061 & 0.098 \\
\bottomrule % Bottom line
\multicolumn{7}{p{\textwidth}}{\footnotesize $^{+} p < 0.10$, * $p < 0.05$, ** $p < 0.01$, *** $p < 0.001$} \\
\caption{VAR Models RT on Ural Oil Prices}\label{tab:tableS17}
\end{tabularx}
\end{threeparttable}
\end{table}
\clearpage
\begin{table}[htbp]
\begin{threeparttable}
\small % Makes the font of the table smaller
\begin{tabularx}{\textwidth}{@{}>{\RaggedRight\arraybackslash}X *{6}{>{\RaggedRight\arraybackslash}X}@{}} 
\toprule 
& \textbf{{RT Arabic bitcoin cryptocurrency value}} & \textbf{{RT Arabic city crisis protest and disaster}} & \textbf{{RT Arabic covid pandemic and related issues}} & \textbf{{RT Arabic middle east wars and conflict}} & \textbf{{RT Arabic recent news headlines}} & \textbf{{RT Arabic US Russia Ukraine sanctions}} \\
\midrule % Line after heading
Ural Oil Price (-1) & $-0.016$ $(0.016)$ & $-0.004$ $(0.007)$ & $-0.014^{*}$ $(0.007)$ & $0.001$ $(0.005)$ & $0.144$ $(0.155)$ & $-0.310$ $(0.431)$ \\
Interruption 1 (Covid-19) & $0.065$ $(0.311)$ & $0.016$ $(0.129)$ & $-0.004$ $(0.133)$ & $0.019$ $(0.106)$ & $-1.600$ $(3.063)$ & $-3.146$ $(8.501)$ \\
Interruption 2 (Russia-Ukraine) & $0.164$ $(0.570)$ & $0.031$ $(0.236)$ & $-0.012$ $(0.243)$ & $0.060$ $(0.194)$ & $1.722$ $(5.593)$ & $0.442$ $(15.444)$ \\
Constant & $-0.255$ $(0.308)$ & $-0.001$ $(0.128)$ & $-0.030$ $(0.132)$ & $0.028$ $(0.105)$ & $-0.918$ $(3.031)$ & $-1.391$ $(8.352)$ \\
Trend & $0.009$ $(0.015)$ & $-0.001$ $(0.006)$ & $0.001$ $(0.007)$ & $-0.002$ $(0.005)$ & $0.075$ $(0.151)$ & $0.150$ $(0.416)$ \\
Num.Obs. & 58 & 58 & 58 & 58 & 58 & 58 \\
$R^2$ & 0.179 & 0.022 & 0.404 & 0.114 & 0.082 & 0.160 \\
$R^2$ Adj. & 0.100 & $-0.072$ & 0.346 & 0.029 & $-0.007$ & 0.080 \\
\bottomrule % Bottom line
\multicolumn{7}{p{\textwidth}}{\footnotesize $^{+} p < 0.10$, * $p < 0.05$, ** $p < 0.01$, *** $p < 0.001$} \\
\caption{VAR Models RT Arabic on Ural Oil Prices}\label{tab:tableS18}
\end{tabularx}
\end{threeparttable}
\end{table}
\clearpage
\begin{table}[htbp]
\begin{threeparttable}
\small % Makes the font of the table smaller
\begin{tabularx}{\textwidth}{@{}>{\RaggedRight\arraybackslash}X *{6}{>{\RaggedRight\arraybackslash}X}@{}} 
\toprule 
& \textbf{{RT America bitcoin cryptocurrency value}} & \textbf{{RT America city crisis protest and disaster}} & \textbf{{RT America covid pandemic and related issues}} & \textbf{{RT America middle east wars and conflict}} & \textbf{{RT America recent news headlines}} & \textbf{{RT America US Russia Ukraine sanctions}} \\
\midrule % Line after heading
Ural Oil Price (-1) & $0.028$ $(0.024)$ & $0.006$ $(0.010)$ & $-0.034$ $(0.082)$ & $-0.001$ $(0.017)$ & $0.006$ $(0.086)$ & $0.094$ $(0.074)$ \\
Ural Oil Price (-2) & $0.019$ $(0.024)$ & $0.009$ $(0.010)$ & $0.036$ $(0.076)$ & $0.002$ $(0.017)$ & $0.011$ $(0.087)$ & $-0.008$ $(0.075)$ \\
Interruption 1 (Covid-19) & $-0.008$ $(0.791)$ & $0.258$ $(0.320)$ & $0.481$ $(2.321)$ & $0.516$ $(0.542)$ & $1.837$ $(2.849)$ & $2.357$ $(2.369)$ \\
Interruption 2 (Russia-Ukraine) & $0.094$ $(1.360)$ & $0.738$ $(0.531)$ & $0.359$ $(3.828)$ & $0.528$ $(0.902)$ & $-1.570$ $(4.711)$ & $3.053$ $(3.914)$ \\
Constant & $0.151$ $(0.532)$ & $0.079$ $(0.221)$ & $0.685$ $(1.589)$ & $0.047$ $(0.375)$ & $-2.086$ $(1.954)$ & $-0.009$ $(1.632)$ \\
Trend & $-0.009$ $(0.038)$ & $-0.014$ $(0.015)$ & $-0.045$ $(0.112)$ & $-0.020$ $(0.026)$ & $-0.004$ $(0.138)$ & $-0.104$ $(0.114)$ \\
Num.Obs. & 40 & 40 & 40 & 40 & 40 & 40 \\
$R^2$ & 0.344 & 0.568 & 0.236 & 0.388 & 0.488 & 0.213 \\
$R^2$ Adj. & 0.201 & 0.473 & 0.069 & 0.254 & 0.376 & 0.041 \\
\bottomrule % Bottom line
\multicolumn{7}{p{\textwidth}}{\footnotesize $^{+} p < 0.10$, * $p < 0.05$, ** $p < 0.01$, *** $p < 0.001$} \\
\caption{VAR Models RT America on Ural Oil Prices}\label{tab:tableS19}
\end{tabularx}
\end{threeparttable}
\end{table}
\clearpage
\begin{table}[htbp]
\begin{threeparttable}
\small % Makes the font of the table smaller
\begin{tabularx}{\textwidth}{@{}>{\RaggedRight\arraybackslash}X *{6}{>{\RaggedRight\arraybackslash}X}@{}} 
\toprule 
& \textbf{{RT DE bitcoin cryptocurrency value}} & \textbf{{RT DE city crisis protest and disaster}} & \textbf{{RT DE covid pandemic and related issues}} & \textbf{{RT DE middle east wars and conflict}} & \textbf{{RT DE recent news headlines}} & \textbf{{RT DE US Russia Ukraine sanctions}} \\
\midrule % Line after heading
Ural Oil Price (-1) & $-0.026^{*}$ $(0.012)$ & $-0.006$ $(0.012)$ & $-0.153^{***}$ $(0.040)$ & $0.008$ $(0.028)$ & $0.176$ $(0.164)$ & $0.042$ $(0.140)$ \\
Interruption 1 (Covid-19) & $0.052$ $(0.234)$ & $-0.097$ $(0.223)$ & $-1.074$ $(0.763)$ & $-0.235$ $(0.556)$ & $3.133$ $(3.221)$ & $4.729$ $(2.823)$ \\
Interruption 2 (Russia-Ukraine) & $-0.356$ $(0.428)$ & $-0.366$ $(0.409)$ & $-4.913^{**}$ $(1.411)$ & $-0.680$ $(1.017)$ & $1.121$ $(5.891)$ & $6.701$ $(5.055)$ \\
Constant & $-0.134$ $(0.229)$ & $-0.087$ $(0.221)$ & $-1.134$ $(0.755)$ & $-0.052$ $(0.549)$ & $0.047$ $(3.182)$ & $0.551$ $(2.678)$ \\
Trend & $0.007$ $(0.011)$ & $0.008$ $(0.011)$ & $0.107^{**}$ $(0.038)$ & $0.013$ $(0.027)$ & $-0.053$ $(0.158)$ & $-0.146$ $(0.135)$ \\
Num.Obs. & 58 & 58 & 58 & 58 & 58 & 58 \\
$R^2$ & 0.252 & 0.166 & 0.392 & 0.143 & 0.227 & 0.190 \\
$R^2$ Adj. & 0.180 & 0.086 & 0.334 & 0.061 & 0.152 & 0.113 \\
\bottomrule % Bottom line
\multicolumn{7}{p{\textwidth}}{\footnotesize $^{+} p < 0.10$, * $p < 0.05$, ** $p < 0.01$, *** $p < 0.001$} \\
\caption{VAR Models RT DE on Ural Oil Prices}\label{tab:tableS20}
\end{tabularx}
\end{threeparttable}
\end{table}
\clearpage
\begin{table}[htbp]
\begin{threeparttable}
\small % Makes the font of the table smaller
\begin{tabularx}{\textwidth}{@{}>{\RaggedRight\arraybackslash}X *{6}{>{\RaggedRight\arraybackslash}X}@{}} 
\toprule 
& \textbf{{RT FR bitcoin cryptocurrency value}} & \textbf{{RT FR city crisis protest and disaster}} & \textbf{{RT FR covid pandemic and related issues}} & \textbf{{RT FR middle east wars and conflict}} & \textbf{{RT FR recent news headlines}} & \textbf{{RT FR US Russia Ukraine sanctions}} \\
\midrule % Line after heading
Ural Oil Price (-1) & $-0.021$ $(0.020)$ & $0.028$ $(0.041)$ & $-0.459$ $(0.386)$ & $0.028$ $(0.112)$ & $-1.226$ $(0.802)$ & $0.551$ $(0.422)$ \\
Interruption 1 (Covid-19) & $-0.239$ $(0.393)$ & $-0.288$ $(0.818)$ & $-11.932$ $(7.404)$ & $-0.039$ $(2.165)$ & $-9.582$ $(15.790)$ & $3.135$ $(8.310)$ \\
Interruption 2 (Russia-Ukraine) & $-0.835$ $(0.719)$ & $0.127$ $(1.485)$ & $-36.996^{*}$ $(14.239)$ & $0.195$ $(3.965)$ & $-20.719$ $(28.823)$ & $3.556$ $(15.210)$ \\
Constant & $-0.158$ $(0.388)$ & $0.449$ $(0.793)$ & $-6.120$ $(7.241)$ & $0.352$ $(2.139)$ & $5.613$ $(15.594)$ & $3.107$ $(8.210)$ \\
Trend & $0.018$ $(0.019)$ & $-0.011$ $(0.040)$ & $0.783^{*}$ $(0.372)$ & $-0.013$ $(0.106)$ & $0.161$ $(0.770)$ & $-0.179$ $(0.407)$ \\
Num.Obs. & 58 & 58 & 58 & 58 & 58 & 58 \\
$R^2$ & 0.301 & 0.234 & 0.138 & 0.004 & 0.114 & 0.181 \\
$R^2$ Adj. & 0.233 & 0.161 & 0.056 & $-0.092$ & 0.029 & 0.102 \\
\bottomrule % Bottom line
\multicolumn{7}{p{\textwidth}}{\footnotesize $^{+} p < 0.10$, * $p < 0.05$, ** $p < 0.01$, *** $p < 0.001$} \\
\caption{VAR Models RT France on Ural Oil Prices}\label{tab:tableS21}
\end{tabularx}
\end{threeparttable}
\end{table}
\clearpage
\begin{table}[htbp]
\begin{threeparttable}
\small % Makes the font of the table smaller
\begin{tabularx}{\textwidth}{@{}>{\RaggedRight\arraybackslash}X *{6}{>{\RaggedRight\arraybackslash}X}@{}} 
\toprule 
& \textbf{{RT SP bitcoin cryptocurrency value}} & \textbf{{RT SP city crisis protest and disaster}} & \textbf{{RT SP covid pandemic and related issues}} & \textbf{{RT SP middle east wars and conflict}} & \textbf{{RT SP recent news headlines}} & \textbf{{RT SP US Russia Ukraine sanctions}} \\
\midrule % Line after heading
Ural Oil Price (-1) & $-0.169$ $(0.172)$ & $0.111$ $(0.193)$ & $-0.168$ $(0.257)$ & $0.019$ $(0.152)$ & $-0.789$ $(0.578)$ & $1.512^{*}$ $(0.654)$ \\
Interruption 1 (Covid-19) & $-1.253$ $(3.386)$ & $-2.741$ $(3.764)$ & $-0.678$ $(4.550)$ & $0.111$ $(2.967)$ & $-10.371$ $(11.300)$ & $5.734$ $(12.716)$ \\
Interruption 2 (Russia-Ukraine) & $-7.724$ $(6.261)$ & $0.373$ $(6.853)$ & $0.040$ $(8.315)$ & $0.742$ $(5.429)$ & $-4.468$ $(20.663)$ & $20.141$ $(23.347)$ \\
Constant & $-1.402$ $(3.348)$ & $2.071$ $(3.712)$ & $1.156$ $(4.489)$ & $0.461$ $(2.932)$ & $5.232$ $(11.184)$ & $7.117$ $(12.546)$ \\
Trend & $0.153$ $(0.167)$ & $-0.006$ $(0.184)$ & $-0.028$ $(0.223)$ & $-0.024$ $(0.145)$ & $0.128$ $(0.553)$ & $-0.458$ $(0.624)$ \\
Num.Obs. & 58 & 58 & 58 & 58 & 58 & 58 \\
$R^2$ & 0.102 & 0.227 & 0.022 & 0.187 & 0.265 & 0.140 \\
$R^2$ Adj. & 0.015 & 0.152 & $-0.072$ & 0.108 & 0.194 & 0.057 \\
\bottomrule % Bottom line
\multicolumn{7}{p{\textwidth}}{\footnotesize $^{+} p < 0.10$, * $p < 0.05$, ** $p < 0.01$, *** $p < 0.001$} \\
\caption{VAR Models RT en Espanol on Ural Oil Prices}\label{tab:tableS22}
\end{tabularx}
\end{threeparttable}
\end{table}
\clearpage
\begin{table}[htbp]
\begin{threeparttable}
\small % Makes the font of the table smaller
\begin{tabularx}{\textwidth}{@{}>{\RaggedRight\arraybackslash}X *{6}{>{\RaggedRight\arraybackslash}X}@{}} 
\toprule 
& \textbf{{RT UK bitcoin cryptocurrency value}} & \textbf{{RT UK city crisis protest and disaster}} & \textbf{{RT UK covid pandemic and related issues}} & \textbf{{RT UK middle east wars and conflict}} & \textbf{{RT UK recent news headlines}} & \textbf{{RT UK US Russia Ukraine sanctions}} \\
\midrule % Line after heading
Ural Oil Price (-1) & $-0.028$ $(0.018)$ & $0.014$ $(0.020)$ & $-0.061$ $(0.038)$ & $-0.004$ $(0.018)$ & $-0.031$ $(0.101)$ & $0.193^{+}$ $(0.108)$ \\
Interruption 1 (Covid-19) & $-0.337$ $(0.369)$ & $0.108$ $(0.422)$ & $0.074$ $(0.747)$ & $0.018$ $(0.382)$ & $0.844$ $(2.150)$ & $1.430$ $(2.439)$ \\
Interruption 2 (Russia-Ukraine) & $-1.144^{+}$ $(0.613)$ & $0.331$ $(0.702)$ & $-0.129$ $(1.234)$ & $-0.016$ $(0.637)$ & $2.321$ $(3.596)$ & $2.303$ $(3.969)$ \\
Constant & $-0.316$ $(0.303)$ & $0.162$ $(0.346)$ & $0.085$ $(0.609)$ & $0.024$ $(0.313)$ & $1.188$ $(1.777)$ & $0.731$ $(1.899)$ \\
Trend & $0.027$ $(0.017)$ & $-0.011$ $(0.020)$ & $-0.005$ $(0.035)$ & $-0.001$ $(0.018)$ & $-0.098$ $(0.103)$ & $-0.072$ $(0.113)$ \\
Num.Obs. & 51 & 51 & 51 & 51 & 51 & 51 \\
$R^2$ & 0.276 & 0.140 & 0.089 & 0.172 & 0.261 & 0.163 \\
$R^2$ Adj. & 0.196 & 0.044 & $-0.012$ & 0.080 & 0.179 & 0.069 \\
\bottomrule % Bottom line
\multicolumn{7}{p{\textwidth}}{\footnotesize $^{+} p < 0.10$, * $p < 0.05$, ** $p < 0.01$, *** $p < 0.001$} \\
\caption{VAR Models RT UK on Ural Oil Prices}\label{tab:tableS23}
\end{tabularx}
\end{threeparttable}
\end{table}
\clearpage
\clearpage
\textbf{\textit{Granger Causality Results}}
\begin{table}[htbp]
\keepXColumns
\begin{tabularx}{\textwidth}{lXXl}
\toprule
\textbf{Page} & \textbf{Topic} & \textbf{GC.P.Value} & \textbf{Significance} \\
\midrule
RT & bitcoin cryptocurrency value & 0.9851 & ns \\
RT & city crisis protest and disaster & 0.6083 & ns \\
RT & covid pandemic and related issues & 0.0107 & p<0.05* \\
RT & middle east wars and conflict & 0.9136 & ns \\
RT & recent news headlines & 0.9768 & ns \\
RT & us russia ukraine sanctions & 0.0418 & p<0.05* \\
RT America & bitcoin cryptocurrency value & 0.3634 & ns \\
RT America & city crisis protest and disaster & 0.5109 & ns \\
RT America & covid pandemic and related issues & 0.8058 & ns \\
RT America & middle east wars and conflict & 0.9924 & ns \\
RT America & recent news headlines & 0.9883 & ns \\
RT America & us russia ukraine sanctions & 0.4451 & ns \\
RT Arabic & bitcoin cryptocurrency value & 0.3084 & ns \\
RT Arabic & city crisis protest and disaster & 0.5873 & ns \\
RT Arabic & covid pandemic and related issues & 0.0383 & p<0.05* \\
RT Arabic & middle east wars and conflict & 0.8399 & ns \\
RT Arabic & recent news headlines & 0.3549 & ns \\
RT Arabic & us russia ukraine sanctions & 0.4731 & ns \\
RT DE & bitcoin cryptocurrency value & 0.0318 & p<0.05* \\
RT DE & city crisis protest and disaster & 0.5845 & ns \\
RT DE & covid pandemic and related issues & 0.0002 & p<0.001*** \\
RT DE & middle east wars and conflict & 0.7667 & ns \\
RT DE & recent news headlines & 0.2845 & ns \\
RT DE & us russia ukraine sanctions & 0.7621 & ns \\
RT France & bitcoin cryptocurrency value & 0.3012 & ns \\
RT France & city crisis protest and disaster & 0.4957 & ns \\
RT France & covid pandemic and related issues & 0.2375 & ns \\
RT France & middle east wars and conflict & 0.8043 & ns \\
RT France & recent news headlines & 0.1294 & ns \\
RT France & us russia ukraine sanctions & 0.1949 & ns \\
RT en Espanol & bitcoin cryptocurrency value & 0.3274 & ns \\
RT en Espanol & city crisis protest and disaster & 0.5670 & ns \\
RT en Espanol & covid pandemic and related issues & 0.5137 & ns \\
RT en Espanol & middle east wars and conflict & 0.9020 & ns \\
RT en Espanol & recent news headlines & 0.1751 & ns \\
RT en Espanol & us russia ukraine sanctions & 0.0228 & p<0.05* \\
RT UK & bitcoin cryptocurrency value & 0.1177 & ns \\
RT UK & city crisis protest and disaster & 0.4664 & ns \\
RT UK & covid pandemic and related issues & 0.1155 & ns \\
RT UK & middle east wars and conflict & 0.8324 & ns \\
RT UK & recent news headlines & 0.7618 & ns \\
RT UK & us russia ukraine sanctions & 0.0789 & ns \\
\bottomrule
\caption{Granger Causality Tests on VAR Models on Ural Oil Prices}\label{tab:tableS24}
\end{tabularx}
\end{table}
\clearpage
\textbf{\textit{Post\-Hoc Tests}}
\textbf{Johansen test for co-integration
}
\begin{table}[htbp]
\keepXColumns
\begin{tabularx}{\textwidth}{lXXl}
\toprule
\textbf{Page} & \textbf{Topic} & \textbf{P Value} \\
\midrule
RT & bitcoin cryptocurrency value & 0.01000000 \\
RT & city crisis protest and disaster & 0.01000000 \\
RT & covid pandemic and related issues & 0.01000000 \\
RT & middle east wars and conflict & 0.01000000 \\
RT & recent news headlines & 0.01000000 \\
RT & us russia ukraine sanctions & 0.01000000 \\
RT America & bitcoin cryptocurrency value & 0.01000000 \\
RT America & city crisis protest and disaster & 0.01000000 \\
RT America & covid pandemic and related issues & 0.01517434 \\
RT America & middle east wars and conflict & 0.01000000 \\
RT America & recent news headlines & 0.01000000 \\
RT America & us russia ukraine sanctions & 0.01113057 \\
RT Arabic & bitcoin cryptocurrency value & 0.01000000 \\
RT Arabic & city crisis protest and disaster & 0.01000000 \\
RT Arabic & covid pandemic and related issues & 0.01000000 \\
RT Arabic & middle east wars and conflict & 0.01000000 \\
RT Arabic & recent news headlines & 0.01000000 \\
RT Arabic & us russia ukraine sanctions & 0.01000000 \\
RT DE & bitcoin cryptocurrency value & 0.01000000 \\
RT DE & city crisis protest and disaster & 0.01000000 \\
RT DE & covid pandemic and related issues & 0.01000000 \\
RT DE & middle east wars and conflict & 0.01000000 \\
RT DE & recent news headlines & 0.01000000 \\
RT DE & us russia ukraine sanctions & 0.01000000 \\
RT France & bitcoin cryptocurrency value & 0.01000000 \\
RT France & city crisis protest and disaster & 0.01000000 \\
RT France & covid pandemic and related issues & 0.01000000 \\
RT France & middle east wars and conflict & 0.01000000 \\
RT France & recent news headlines & 0.01000000 \\
RT France & us russia ukraine sanctions & 0.01000000 \\
RT en Espanol & bitcoin cryptocurrency value & 0.01000000 \\
RT en Espanol & city crisis protest and disaster & 0.01000000 \\
RT en Espanol & covid pandemic and related issues & 0.01000000 \\
RT en Espanol & middle east wars and conflict & 0.01000000 \\
RT en Espanol & recent news headlines & 0.01000000 \\
RT en Espanol & us russia ukraine sanctions & 0.01000000 \\
RT UK & bitcoin cryptocurrency value & 0.01000000 \\
RT UK & city crisis protest and disaster & 0.01000000 \\
RT UK & covid pandemic and related issues & 0.01000000 \\
RT UK & middle east wars and conflict & 0.01000000 \\
RT UK & recent news headlines & 0.01000000 \\
RT UK & us russia ukraine sanctions & 0.01000000 \\
\bottomrule
\multicolumn{3}{p{\textwidth}}{\footnotesize Note: p-value (rejection criteria $p < 0.05$} \\
\caption{Johansen test for co-integration VAR Models on Ural Oil Prices}\label{tab:tableS25}
\end{tabularx}
\end{table}
\clearpage
\textbf{\textit{Ljung-Box Test}}
\begin{table}[htbp]
\keepXColumns
% \begin{tabularx}{\textwidth}{lXXX}
\begin{tabularx}{\textwidth}{l >{\hsize=1.5\hsize}X >{\hsize=0.75\hsize}X >{\hsize=0.75\hsize}X}
\toprule
\textbf{Page} & \textbf{Topic} & \textbf{P Value DV} & \textbf{P Value IV Ural Oil Prices} \\
\midrule
RT & bitcoin cryptocurrency value & 0.525 & 0.487 \\
RT & city crisis protest and disaster & 0.148 & 0.492 \\
RT & covid pandemic and related issues & 0.770 & 0.515 \\
RT & middle east wars and conflict & 0.222 & 0.502 \\
RT & recent news headlines & 0.463 & 0.495 \\
RT & us russia ukraine sanctions & 0.813 & 0.455 \\
RT America & bitcoin cryptocurrency value & 0.561 & 0.883 \\
RT America & city crisis protest and disaster & 0.635 & 0.625 \\
RT America & covid pandemic and related issues & 0.662 & 0.621 \\
RT America & middle east wars and conflict & 0.703 & 0.581 \\
RT America & recent news headlines & 0.305 & 0.596 \\
RT America & us russia ukraine sanctions & 0.877 & 0.729 \\
RT Arabic & bitcoin cryptocurrency value & 0.321 & 0.576 \\
RT Arabic & city crisis protest and disaster & 0.728 & 0.801 \\
RT Arabic & covid pandemic and related issues & 0.193 & 0.576 \\
RT Arabic & middle east wars and conflict & 0.731 & 0.454 \\
RT Arabic & recent news headlines & 0.999 & 0.513 \\
RT Arabic & us russia ukraine sanctions & 0.225 & 0.516 \\
RT DE & bitcoin cryptocurrency value & 0.233 & 0.708 \\
RT DE & city crisis protest and disaster & 0.609 & 0.514 \\
RT DE & covid pandemic and related issues & 0.351 & 0.697 \\
RT DE & middle east wars and conflict & 0.322 & 0.455 \\
RT DE & recent news headlines & 0.343 & 0.474 \\
RT DE & us russia ukraine sanctions & 0.144 & 0.477 \\
RT France & bitcoin cryptocurrency value & 0.054 & 0.592 \\
RT France & city crisis protest and disaster & 0.083 & 0.457 \\
RT France & covid pandemic and related issues & 0.626 & 0.402 \\
RT France & middle east wars and conflict & 0.876 & 0.525 \\
RT France & recent news headlines & 0.372 & 0.436 \\
RT France & us russia ukraine sanctions & 0.444 & 0.342 \\
RT en Espanol & bitcoin cryptocurrency value & 0.393 & 0.519 \\
RT en Espanol & city crisis protest and disaster & 0.274 & 0.518 \\
RT en Espanol & covid pandemic and related issues & 0.917 & 0.460 \\
RT en Espanol & middle east wars and conflict & 0.339 & 0.508 \\
RT en Espanol & recent news headlines & 0.305 & 0.601 \\
RT en Espanol & us russia ukraine sanctions & 0.827 & 0.493 \\
RT UK & bitcoin cryptocurrency value & 0.409 & 0.409 \\
RT UK & city crisis protest and disaster & 0.708 & 0.422 \\
RT UK & covid pandemic and related issues & 0.657 & 0.430 \\
RT UK & middle east wars and conflict & 0.438 & 0.468 \\
RT UK & recent news headlines & 0.143 & 0.489 \\
RT UK & us russia ukraine sanctions & 0.520 & 0.413 \\
\bottomrule
\caption{Ljung-Box Test VAR Models on Ural Oil Prices}\label{tab:tables26}
\end{tabularx}
\end{table}

\end{document}